\documentclass[ALICE,manyauthors]{cernphprep}

\usepackage[comma,square,numbers,sort&compress]{natbib}
\usepackage{hyperref}
\usepackage{lineno}
\usepackage{color}
\usepackage[subfigure]{tocloft} 
\usepackage[nameinlink,capitalise]{cleveref} 
\usepackage[T1]{fontenc}
\usepackage{multirow}
\usepackage{amsmath}
\usepackage{relsize}
\usepackage{rotating}

\begin{document}%
\newcommand{\mrm}[1]{\mathrm{#1}}
\newcommand{\mrmo}[1]{\mathrm{\overline{#1}}}
\newcommand{\bsb}[1]{\boldsymbol{#1}}
\newcommand{\circit}{\item[$\circ$]}
\newcommand{\etaless}[1]{\ensuremath{\left|\eta\right| < #1}\xspace}

\newcommand{\ITS}          {\rm{ITS}}
\newcommand{\TOF}          {\rm{TOF}}
\newcommand{\ZDC}          {\rm{ZDC}}
\newcommand{\ZDCs}         {\rm{ZDCs}}
\newcommand{\ZNA}          {\rm{ZNA}}
\newcommand{\ZNC}          {\rm{ZNC}}
\newcommand{\SPD}          {\rm{SPD}}
\newcommand{\SDD}          {\rm{SDD}}
\newcommand{\SSD}          {\rm{SSD}}
\newcommand{\TPC}          {\rm{TPC}}
\newcommand{\VZERO}        {\rm{VZERO}}
\newcommand{\VZEROA}       {\rm{VZERO-A}}
\newcommand{\VZEROC}       {\rm{VZERO-C}}
\newcommand{\pip}          {$\pi^{+}$}
\newcommand{\pim}          {$\pi^{-}$}
\newcommand{\kap}          {K$^{+}$}
\newcommand{\kam}          {K$^{-}$}
\newcommand{\pbar}         {$\rm\overline{p}$}
\newcommand{\kzero}        {\ensuremath{{\rm K}^{0}_{S}}}
\newcommand{\kstar}        {\ensuremath{{\rm K}^{*}}}
\newcommand{\He}           {\ensuremath{^{3}{\rm He}}}
\newcommand{\LH}           {\ensuremath{^{3}_{\Lambda}{\rm H}}}
\newcommand{\vzero}        {\ensuremath{{\rm V}^0}}
\newcommand{\lmb}          {\ensuremath{\Lambda}}
\newcommand{\almb}         {\ensuremath{\bar{\Lambda}}}
\newcommand{\allpart}      {$\pi^{\pm}$, K$^{\pm}$, \kzero, p(\pbar) and \lmb(\almb)}
\newcommand{\allpi}        {$\pi^{\pm}$}
\newcommand{\allk}         {K$^{\pm}$}
\newcommand{\allp}         {p(\pbar)}
\newcommand{\alllmb}       {\lmb(\almb)}
\newcommand{\degree}       {$^{\rm o}$}
\newcommand{\dg}           {\mbox{$^\circ$}}
\newcommand{\dndy}         {d$N$/d$y$}
\newcommand {\ee}            {\mbox{e$^+$e$^-$}}
\newcommand{\AuAu}         {\mbox{Au--Au}}
\newcommand{\pseudorap}    {\mbox{$\left | \eta \right | $}}
\newcommand{\dNdeta}       {\ensuremath{\mathrm{d}N_\mathrm{ch}/\mathrm{d}\eta}}
\newcommand{\dNdy}         {\ensuremath{\mathrm{d}N/\mathrm{d}y}}
\newcommand{\dNdptdy}      {\ensuremath{\mathrm{d}N/{\rm d}\pt\mathrm{d}y}}
\newcommand{\dNdyst}       {\ensuremath{\sqrt{\frac{dN_\pi/dy}{s_T}}}}
\newcommand{\dNdetatr}     {\mathrm{d}N_\mathrm{tracklets}/\mathrm{d}\eta}
\newcommand{\dNdetar}[1]   {\mathrm{d}N_\mathrm{ch}/\mathrm{d}\eta\left.\right|_{|\eta|<#1}}
\newcommand{\NtrkEtaIn}    {\ensuremath{N_{\rm tracklets}^{\etaless{0.8}}}}
\newcommand{\NtrkEtaOut}    {\ensuremath{N_{\rm tracklets}^{0.8< \left| \eta \right| < 1.5}}}
\newcommand{\lum}          {\, \mbox{${\rm cm}^{-2} {\rm s}^{-1}$}}
\newcommand{\barn}         {\, \mbox{${\rm barn}$}}
\newcommand{\m}            {\, \mbox{${\rm m}$}}
\newcommand{\ncls}         {\ensuremath{N_{cls}}}
\newcommand{\nsigma}       {\ensuremath{n\sigma}}
\newcommand{\dcaxy}        {\ensuremath{{\rm DCA}_{xy}}}
\newcommand{\dcaz}         {\ensuremath{{\rm DCA}_{z}}}
\newcommand{\EcrossB}      {E$\times$B}
\newcommand{\bb}           {Bethe-Bloch}
\newcommand{\s}            {\ensuremath{\sqrt{s}}}
\newcommand{\hlab}         {\ensuremath{\eta_{\rm lab}}}
\newcommand{\ynn}         {\ensuremath{y_{\rm NN}}}
\newcommand{\ycms}         {\ensuremath{y_{\rm CMS}}}
\newcommand{\ylab}         {\ensuremath{y_{\rm lab}}}
\newcommand{\ppi}          {\ensuremath{{\rm p}/\pi}}
\newcommand{\kpi}          {\ensuremath{{\rm K}/\pi}}
\newcommand{\lpi}          {\ensuremath{{\rm \Lambda}/\pi}}
\newcommand{\snn}          {\ensuremath{\sqrt{s_{\rm NN}}}}
\newcommand{\snnbf}        {\ensuremath{\mathbf{{\sqrt{s_{\mathbf NN}}}}}}
\newcommand{\sonly}        {\ensuremath{\sqrt{s}}}
\newcommand{\Npart}        {\ensuremath{N_\mathrm{part}}}
\newcommand{\avNpart}      {\ensuremath{\langle N_\mathrm{part} \rangle}}
\newcommand{\avNpartdata}  {\ensuremath{\langle N_\mathrm{part}^{\rm data} \rangle}}
\newcommand{\Ncoll}        {\ensuremath{N_\mathrm{coll}}}
\newcommand{\Dnpart}       {\ensuremath{D\left(\Npart\right)}}
\newcommand{\DnpartExp}    {\ensuremath{D_{\rm exp}\left(\Npart\right)}}
\newcommand{\dNdetapt}     {\ensuremath{\dNdeta\,/\left(0.5\Npart\right)}}
\newcommand{\dNdetaptr}[1] {\ensuremath{\dNdetar{#1}\,/\left(0.5\Npart\right)}}
\newcommand{\dNdetape}     {\left(\ensuremath{\dNdeta\right)/\left(\avNpart/2\right)}}
\newcommand{\dNdetaper}[1] {\ensuremath{\dNdetar{#1}\,/\left(\avNpart/2\right)}}
\newcommand{\dndydpt}      {\ensuremath{{\rm d}^2N/({\rm d}y {\rm d}p_{\rm t})}}
\newcommand{\abs}[1]       {\ensuremath{\left|#1\right|}}
\newcommand{\signn}        {\ensuremath{\sigma^{\rm inel.}_{\rm NN}}}
\newcommand{\vz}           {\ensuremath{V_{z}}}
\newcommand{\Tfo}          {\ensuremath{{T}_{\rm kin}}}
\newcommand{\Tch}          {\ensuremath{{T}_{\rm ch}}}
\newcommand{\bT}           {\ensuremath{\beta_{\rm T}}}
\newcommand{\avbT}         {\ensuremath{\left< \beta_{\rm T}\right>}}
\newcommand{\avpT}         {\ensuremath{\left< \pt \right>}}
\newcommand{\muB}          {\ensuremath{\mu_{B}}}
\newcommand{\stat}         {({\it stat.})}
\newcommand{\syst}         {({\it sys.})}
\newcommand{\Fig}[1]       {Fig.~\ref{#1}}
\newcommand{\Figure}[1]    {Figure~\ref{#1}}
\newcommand{\Ref}[1]       {Ref.~\cite{#1}}
\newcommand{\gevc}         {\ensuremath{{\rm GeV}/c}}
\newcommand{\mevc}         {\ensuremath{{\rm MeV}/c}}
\newcommand{\gs}           {\ensuremath{\gamma_{s}}}
\newcommand{\gq}           {\ensuremath{\gamma_{q}}}
\newcommand{\gc}           {\ensuremath{\gamma_{c}}}
\newcommand{\chindf}       {\ensuremath{\chi^{2}/{\rm NDF}}}
\newcommand{\avg}[1]       {\ensuremath{\left\langle#1\right\rangle}}
\newcommand{\etalab}       {\ensuremath{\eta_{{\rm lab}}}}
\newcommand {\gammas}			{\ensuremath{\gamma_{\mathrm{s}}}}
\newcommand{\pt}{\ensuremath{p_{\rm T}}\xspace}
\newcommand{\mt}{\ensuremath{m_{\rm T}}\xspace}
\newcommand{\sqrts}{\ensuremath{\sqrt{s}}\xspace}
\newcommand{\sqrtsNN}{\ensuremath{\sqrt{s_{\rm NN}}}\xspace}
\newcommand{\nch}{\ensuremath{N_{\rm ch}}\xspace}
\newcommand{\dnchdeta}{\ensuremath{{\rm d}N_{\rm ch}/{\rm d}\eta}\xspace}
\newcommand{\dnchdpt}{\ensuremath{{\rm d}N_{\rm ch}/{\rm d}\pt}\xspace}
\newcommand{\yless}[1]{\ensuremath{\left|y\right| < #1}\xspace}
\newcommand{\dnchdetaless}[1]{\ensuremath{{\rm d}N_{\rm ch}/{\rm d}\eta|_{\etaless{#1}}}\xspace}
\newcommand{\MeVc}{MeV/$c$\xspace}
\newcommand{\GeVc}{GeV/$c$\xspace}
\newcommand{\pp}{\ensuremath{\rm pp}\xspace}
\newcommand{\pPb}{p--Pb\xspace}
\newcommand{\PbPb}{Pb--Pb\xspace}
\newcommand{\ppbar}{\ensuremath{\rm p\bar{p}}\xspace}
\newcommand{\inelgtzero}{\ensuremath{\mathrm{INEL}>0}\xspace}
\newcommand{\asymmerr}[2]{\ensuremath{^{+#1}_{-#2}}\xspace}
\newcommand{\SPS}{\ensuremath{\rm Sp\bar{p}S}\xspace}
\newcommand{\Raa}{\ensuremath{R_{\rm AA}}\xspace}
\newcommand{\Rppb}{\ensuremath{R_{\rm pPb}}\xspace}
\newcommand{\Rpa}{\ensuremath{R_{\rm pA}}\xspace}
\newcommand{\average}[1]{\ensuremath{\langle #1 \rangle}\xspace}
\newcommand{\dedx}{\ensuremath{{\rm d}E/{\rm d}x}\xspace}

\newcommand{\pPiplus}{\ensuremath{{\pi}^{+}}\xspace}
\newcommand{\pPiminus}{\ensuremath{{\pi}^{-}}\xspace}
\newcommand{\sPi}{\ensuremath{{\pi}}\xspace}
\newcommand{\pKplus}{\ensuremath{{\rm K}^{+}}\xspace}
\newcommand{\pKminus}{\ensuremath{{\rm K}^{-}}\xspace}
\newcommand{\sProton}{\ensuremath{\rm p}\xspace}
\newcommand{\pProton}{\ensuremath{\rm p}\xspace}
\newcommand{\apProton}{\ensuremath{\overline{\rm p}}\xspace}
\newcommand{\sPr}{\ensuremath{\rm p}\xspace}
\newcommand{\sKzero}{\ensuremath{2{\rm K}^{0}_{S}}\xspace}
\newcommand{\pKzero}{\ensuremath{{\rm K}^{0}_{S}}\xspace}
\newcommand{\sLambda}{\ensuremath{\Lambda}\xspace}
\newcommand{\pLambda}{\ensuremath{\Lambda}\xspace}
\newcommand{\apLambda}{\ensuremath{\overline{\Lambda}}\xspace}
\newcommand{\sXi}{\ensuremath{\Xi}\xspace}
\newcommand{\pXi}{\ensuremath{\Xi^{-}}\xspace}
\newcommand{\apXi}{\ensuremath{\overline{\Xi}^{+}}\xspace}
\newcommand{\sOmega}{\ensuremath{\Omega}\xspace}
\newcommand{\pOmega}{\ensuremath{\Omega^{-}}\xspace}
\newcommand{\apOmega}{\ensuremath{\overline{\Omega}^{+}}\xspace}
\newcommand{\pJPsi}{\ensuremath{\rm J/\psi}}
\newcommand{\DMeson}{\ensuremath{\rm D}\xspace}
\newcommand{\itstof}{out-of-bunch pile-up track rejection}

\renewcommand{\labelitemi} {$-$}

\newcommand{\listwarnname}{List of Todos}
\newlistof[chapter]{warns}{wrn}{\listwarnname}

\newcommand{\warn}[1]      {%
 \refstepcounter{warns}
    {\small\textbf{\textcolor{red}{(!\footnote{\textbf{(!)}~#1})}}}%
    \addcontentsline{wrn}{warns}{\protect \numberline{\thewarns}#1}%
}

\newcommand{\warnin}[1]         {\textit{\textcolor{red}{(#1)}}}
\renewcommand{\warn}[1]      {}
\newcommand{\todo}[1]      {{\small\textbf{\textcolor{red}{(!\footnote{\textbf{(!)}\textcolor{red}{TODO: }(~#1})}}}}
\newcommand{\fake}[1]      {\textbf{\textcolor{red}{#1}}}
\newcommand{\final}[1]     {\textbf{\textcolor{blue}{#1}}}
\newcommand{\prelim}[1]    {\textbf{\textcolor{magenta}{#1}}}
\renewcommand{\mod}[1]       {\textbf{\textcolor{red}{#1}}}
\newcommand{\missingfigure}[1]{
  \begin{tabular}{|c|}
    \hline
    {\Large ~}\hspace{0.8\textwidth}~\\
    {\Large ~}\\
    {\Large ~}\\
    {\Large ~}\\
    {\Large ~}\\
    {\Large ~}\\
    {#1 }\\
    {\Large ~}\\
    {\Large ~}\\
    {\Large ~}\\
    {\Large ~}\\
    {\Large ~}\\
    {\Large ~}\\
    \hline
  \end{tabular}
}

\begin{titlepage}
\PHyear{2019}
\PHnumber{168}      
\PHdate{02 August}  
%

\title{Multiplicity dependence of (multi-)strange hadron production in\\ proton-proton collisions at \s~=~13 TeV}
\ShortTitle{Multiplicity dependence of strangeness in pp at \s~=~13 TeV}   

\Collaboration{ALICE Collaboration\thanks{See Appendix~\ref{app:collab} for the list of collaboration members}}
\ShortAuthor{ALICE Collaboration} 

\begin{abstract}
The production rates and the transverse momentum distribution of strange hadrons at mid-rapidity ($\left|y\right| < 0.5$) are measured in proton-proton collisions at \s~=~13~TeV as a function of the charged particle multiplicity, using the ALICE detector at the LHC.\@
The production rates of \pKzero, \sLambda, \sXi, and \sOmega\ increase with the multiplicity faster than what is reported for inclusive charged particles. The increase is found to be more pronounced for hadrons with a larger strangeness content.
Possible auto-correlations between the charged particles and the strange hadrons are evaluated by measuring the event-activity with charged particle multiplicity estimators covering different pseudorapidity regions.
%
When comparing to lower energy results, the yields of strange hadrons are found to depend only on the mid-rapidity charged particle multiplicity. 
%
Several features of the data are reproduced qualitatively by general purpose QCD Monte Carlo models that take into account the effect of densely-packed QCD strings in high multiplicity collisions. However,  none of the tested models reproduce the data quantitatively.
This work corroborates and extends the  ALICE findings on strangeness production in proton-proton collisions at 7 TeV.
%

\end{abstract}
\end{titlepage}
\setcounter{page}{2}

%
%
\section{Introduction}
\label{sec:introduction}


The production rates of strange and multi-strange hadrons in high-energy hadronic interactions are important observables for the study of the properties of Quantum Chromodynamics (QCD) in the non-perturbative regime.
In the simplest case, strange quarks ($s$) in proton-proton (pp) collisions can be produced from the excitation of the sea partons. Indeed, in the past decades a significant effort has been dedicated to the study of the actual strangeness content of the nuclear wave function~\cite{Forte:2013wc}.
In QCD-inspired Monte Carlo generators based on Parton Showers (PS)~\cite{Nagy:2017ggp}
the hard (perturbative) interactions are typically described at the Leading Order (LO). In this picture the $s$ quark can be  produced in the hard partonic scattering via flavour creation and flavour excitation processes as well as in the subsequent shower evolution via gluon splitting. 
At low transverse momentum, $s\bar{s}$ pairs can be produced via non-perturbative processes, as described for instance in string fragmentation models, where the production of strangeness 
is suppressed with respect to light quark production due to the larger strange quark mass~\cite{Fischer:2016zzs}. However, these models fail to quantitatively describe strangeness production in hadronic collisions~\cite{Abelev:2006cs,Abelev:2012jp,Aamodt:2011zza}.

An enhanced production of strange hadrons in heavy-ion collisions was suggested as a signature for the creation of a Quark-Gluon Plasma (QGP)~\cite{Rafelski:1982pu,Rafelski:1980rk}. The main argument in these early studies was that the mass of the strange quark is of the order of the QCD deconfinement temperature, allowing for thermal production in the deconfined medium. The lifetime of the QGP was then estimated to be comparable to the strangeness relaxation time in the plasma, leading to full equilibration. Strangeness enhancement in heavy-ion collisions was indeed observed at the SPS~\cite{Andersen:1998vu} and higher energies~\cite{Abelev:2007xp,ABELEV:2013zaa}. However, strangeness enhancement is no longer considered an unambiguous signature for deconfinement (see e.g.~\cite{Koch:2017pda}).
Strange hadron production in heavy-ion collisions is currently usually described in the framework of statistical-hadronisation (or thermal) models~\cite{Andronic:2008gu,Wheaton:2004qb}. In central heavy-ion collisions, the yields of strange hadrons turn out to be consistent with the expectation from a grand-canonical ensemble, i.e. the production of strange hadrons is compatible with thermal equilibrium, characterised by a common temperature. On the other hand, the strange hadron yields in elementary collisions are suppressed with respect to the predictions of the (grand-canonical) thermal models. The suppression of the relative abundance of strange hadrons with respect to lighter flavours was suggested to be, at least partially, a consequence of the finite volume, which makes the application of a grand-canonical ensemble not valid in hadron-hadron and hadron-nucleus interactions (canonical suppression)~\cite{Tounsi:2001ck,Tounsi:2002nd,Becattini:2008yn}. However, this approach cannot explain the observed particle abundances if the same volume is assumed for both strange and non-strange hadrons~\cite{Acharya:2018orn} and does not describe the system size dependence of the $\phi$ meson, a hidden-strange hadron~\cite{Kraus:2007hf, Kraus:2008fh}.

The ALICE Collaboration recently reported an enhancement in the relative production of (multi-) strange hadrons as a function of multiplicity in pp collisions at \s~=~7~TeV~\cite{ALICE:2017jyt} and in \pPb\ collisions at \snn~=~5.02~TeV~\cite{Abelev:2013haa,Adam:2015vsf}. In the case of p-Pb collisions, the yields of strange hadrons relative to pions reach values close to those observed in \PbPb\ collisions at full equilibrium. These are surprising observations, because thermal strangeness production was considered to be a defining feature of heavy-ion collisions, and because none of the commonly-used pp Monte Carlo models reproduced the existing data~\cite{ALICE:2017jyt,Fischer:2016zzs}. The mechanisms at the origin of this effect need to be understood, and then implemented in the state-of-the-art Monte Carlo generators~\cite{Fischer:2016zzs}. 

In this paper, strangeness production in pp interactions is studied at the highest energy reached at the LHC, \s~=~13~TeV.  We present the measurement of the yields and transverse momentum (\pt) distributions of single-strange (\pKzero, \pLambda, \apLambda) and multi-strange (\pXi, \apXi, \pOmega, \apOmega) particles at mid-rapidity, \yless{0.5}, with the ALICE detector~\cite{Aamodt:2008zz}.  
The comparison of the present results with the former ones for pp and \pPb\ interactions allows the investigation of the energy, multiplicity and system size dependence of strangeness production. 
Schematically, the multiplicity of a given pp event depends on \textit{i)} the number of Multiple Parton Interactions (MPI), \textit{ii)} the momentum transfer of those interactions, \textit{iii)} fluctuations in the fragmentation process. 
A systematic study of the biases induced by the choice of the multiplicity estimator along with the specific connections to the underlying MPI are also discussed in this paper.

The paper is organised as follows. In Sec.~\ref{sec:experimental-setup} we discuss the data set and detectors used for the measurement; in Sec.~\ref{sec:analysis-details} we describe the analysis techniques; in Sec.~\ref{sec:syst-uncert} we cover the evaluation of the systematic uncertainties; in Sec.~\ref{sec:results} we present and discuss the results; in Sec.~\ref{sec:summary} we report our conclusions.







\section{Experimental setup and data selection}
\label{sec:experimental-setup}

A detailed description of the ALICE detector and its performance can be found in~\cite{Aamodt:2008zz,Abelev:2014ffa}. In this section, we briefly outline the main detectors used for the measurements presented in this paper.
The ALICE apparatus comprises a central barrel used for vertex reconstruction, track reconstruction and charged-hadron identification, complemented by specialised forward detectors.
The central barrel covers the pseudorapidity region \etaless{0.9}. 
The main central-barrel tracking devices used for this analysis are the Inner Tracking System (ITS) and the Time-Projection Chamber (TPC), which are located inside a solenoidal magnet providing a 0.5~T magnetic field.
The ITS is composed of six cylindrical layers of high-resolution silicon tracking detectors. The innermost layers consist of two arrays of hybrid Silicon Pixel Detectors (SPD), located at an average radial distance $r$ of 3.9 and 7.6~cm from
 the beam axis and covering \etaless{2.0} and \etaless{1.4}, respectively. The SPD is also used to reconstruct tracklets, short two-point track segments covering the pseudorapidity region \etaless{1.4}.
The outer layers of the ITS are composed of silicon strips and drift detectors, with the outermost layer having a radius $r~=~43~\mathrm{cm}$.
The TPC is a large cylindrical drift detector of radial and longitudinal sizes of about $85~<~r~<~250$~cm and $-250~<~z~<~250$~cm, respectively. It is segmented in radial ``pad rows'', providing up to 159 tracking points. It also provides charged-hadron identification information via specific ionisation energy  loss in the gas filling the detector volume. The measurement of strange hadrons is based on ``global tracks'',  reconstructed using information from the TPC as well as from the ITS, if the latter is available.
{Further outwards in radial direction from the beam-pipe and located at a radius of approximately 4 m, the
Time of Flight (TOF) detector measures the time-of-flight of the particles. It is a large-area array of multigap resistive plate chambers with an intrinsic time resolution of 50~ps.
The V0 detectors are two forward scintillator hodoscopes employed for triggering, background suppression and event-class determination. They are placed on either side of the interaction region at $z~=~-0.9$~m and  $z~=~3.3$~m, covering the pseudorapidity regions $-3.7~<~\eta~<~-1.7$ and $2.8~<~\eta~<~5.1$, respectively.

The data considered in the analysis presented in this paper were collected in 2015, at the beginning of Run 2 operations of the LHC\@ at \s~=~13~TeV. The sample consists of 50~M events collected with a minimum bias trigger requiring a hit in both V0 scintillators in coincidence with the arrival of proton bunches from both directions.\@ The interaction probability per single bunch crossing ranges between 2\% and 14\%. 

The contamination from beam-induced background is removed offline by using the timing information in the V0 detectors and taking into account the correlation between tracklets and clusters in the SPD detector, as discussed in detail in~\cite{Abelev:2014ffa}.  The contamination from in-bunch pile-up events is removed offline excluding events with multiple vertices reconstructed in the SPD\@. 
Part of the data used in this paper were collected in periods in which the LHC collided ``trains'' of bunches each separated by 50~ns from its neighbours. In these beam conditions most of the ALICE detectors have a readout window wider than a single bunch spacing and are therefore sensitive to events produced in bunch crossings different from those triggering the collision. In particular, the SPD has a readout window of 300 ns. 
The drift speed in the TPC is about 2.5~cm$/\mathsmaller{\mu}$s, which implies that events produced less than about 0.5~$\mathsmaller{\mu}$s apart cannot be resolved. 
Pile-up events produced in different bunch crossings are removed exploiting  multiplicity correlations in detectors having different readout windows.




\section{Analysis details} 
\label{sec:analysis-details}

The results are presented for primary strange hadrons\footnote{A primary particle~\cite{ALICE-PUBLIC-2017-005} is defined as a particle with a mean proper decay length $ c \tau$ larger than 1~cm, which is either \textit{a)} produced directly in the interaction, or \textit{b)} from decays of particles with $ c \tau$  smaller than 1~cm, excluding particles produced in interactions with material.}.
The measurements reported here have been obtained for events having at least one charged particle produced with \pt~$>$~0 in the pseudorapidity interval \etaless{1} (\inelgtzero), corresponding to about 75\% of the total inelastic cross-section.  In order to study the multiplicity dependence of strange and multi-strange hadrons, for each multiplicity estimator the sample is divided into event classes based on the total charge deposited in the V0 detectors (V0M amplitude) or on the number of tracklets in two pseudorapidity regions: \etaless{0.8} and $0.8 < \left| \eta \right| < 1.5$. The event classes are summarised in Table~\ref{tab:multi}. Since the measurement of strange hadrons is performed in the region \yless{0.5}, the usage of these three estimators allows one to evaluate potential biases on particle production, arising from measuring the multiplicity in a pseudorapidity region partially overlapping with the one of the reconstructed strange hadrons (\NtrkEtaIn), or disjoint from it (V0M and \NtrkEtaOut). 
In particular, the effect of fluctuations can be expected to be stronger if the multiplicity estimator and the observable of interest are measured in the same pseudorapidity region. The usage of two different non-overlapping estimators allows the study of the effect of a rapidity gap between the multiplicity estimator and the measurement of interest.

The events used for the data analysis are required to have a reconstructed vertex in the fiducial region $\left|z\right|~<$~10~cm. As mentioned in the previous section, events containing more than one distinct vertex are tagged as pile-up and discarded. 
For each event class and each multiplicity estimator, the average pseudorapidity density of primary charged-particles \average{\dnchdeta} is measured at mid-rapidity (\etaless{0.5}) using the technique described in~\cite{ALICE:2012xs}. The \average{\dnchdeta} values, corrected for acceptance and efficiency as well as for contamination from secondary particles and combinatorial background, are listed in Table~\ref{tab:multi}. When multiplicity event classes are selected outside the \etaless{0.8} region, the corresponding charged particle multiplicity at mid-rapidity  
is characterized by large a variance. In the case of the V0M estimator, the variance ranges between 30\% and 70\% of the mean \dNdeta\ for the highest and lowest multiplicity classes, respectively.

\begin{sidewaystable}[tp]
\centering
   \caption{Event classes selected according to different multiplicity estimators (see text for details). For each estimator, the second column summarises the relative multiplicity w.r.t. the INEL $>$ 0 event class and the third column represents the corresponding fraction of the INEL $>$ 0 cross-section. The average charged particle multiplicity density is reported in the last column for all event classes. For all the multiplicity estimators, the charged particle multiplicity density is quoted in \etaless{0.5}.} 
  \begin{tabular}{c|ccc|ccc|ccc}
  \hline
  \hline
      &  &   &  & & & & \multicolumn{3}{c}{}                          \\
      &  &   &  & \multicolumn{3}{c|}{$\sigma/\sigma_{\mathrm INEL>0}$ }  & \multicolumn{3}{c}{$\langle \dNdeta \rangle$ }                          \\
    Event Class   & $\frac{\langle \mathrm{V0M} \rangle_{\mathrm{mult}}}{\langle \mathrm{V0M} \rangle_{\mathrm{INEL>0}}}$  & $\frac{\langle \NtrkEtaIn \rangle_{\mathrm{mult}}}{\langle \NtrkEtaIn \rangle_{\mathrm{INEL>0}}}$ & $\frac{\langle\NtrkEtaOut \rangle_{\mathrm{mult}}}{\langle\NtrkEtaOut \rangle_{\mathrm{INEL>0}}}$ & V0M & \NtrkEtaIn &  \NtrkEtaOut & V0M & \NtrkEtaIn &  \NtrkEtaOut \\
      &  &   &  & & & & \multicolumn{3}{c}{}                          \\
\hline
   INEL$>$0                       &   \multicolumn{3}{c|}{Not Applicable}  &  \multicolumn{3}{c|}{Not Applicable}  & \multicolumn{3}{c}{6.89$\pm$0.11} \\
\hline
   I 	 	 	 &  3.7 & 4.7 & 3.8 & 0.0-0.90\% & 0.0-0.91\% &	0.0-0.91\% & 25.75$\pm$0.40  &  32.49$\pm$0.50 &  26.32$\pm$0.40 	       \\ 
\hline
   II 	 	 	 &  2.9 & 3.4 & 2.8  &0.90-4.5\% & 0.91-4.6\% &	0.91-4.6\% & 19.83$\pm$0.30  &  23.42$\pm$0.35 &  19.51$\pm$0.29 	       \\ 
\hline
   III 	 	 	 &  2.3 & 2.7 & 2.2  &4.5-8.9\% & 4.6-9.1\% & 4.6-9.1\%	 & 16.12$\pm$0.24  &  18.29$\pm$0.28 &  15.45$\pm$0.23 	       \\ 
\hline
   IV 	 	 	 &  2.0 & 2.2 & 1.9  &8.9-13.5\% &9.1-13.7\% & 9.1-13.7\%	 & 13.76$\pm$0.21  &  14.90$\pm$0.23 &  13.14$\pm$0.20 	       \\ 
\hline
   V 	 	 	 &  1.8 & 1.9 & 1.7  & 13.5-18.0\% &13.7-18.3\% & 13.7-18.2\%	 & 12.06$\pm$0.18  &  12.90$\pm$0.19 &  11.63$\pm$0.17 	       \\ 
\hline
   VI 	 	 	 &  1.5 & 1.6 & 1.4  & 18.0-27.0\% & 18.3-27.4\% & 18.2-27.4\%	 & 10.11$\pm$0.15  &  10.72$\pm$0.16 &   9.50$\pm$0.14 	       \\ 
\hline
   VII 	 	 	 &  1.2 & 1.2 & 1.1  & 27.0-36.1\% &  27.4-36.6\% & 27.4-36.6\%	 &  8.07$\pm$0.12  &   8.14$\pm$0.12 &   7.68$\pm$0.11 	       \\ 
\hline
   VIII 	 	 &  0.95 & 0.85 & 0.91  & 36.1-45.3\% & 36.6-45.9\% & 36.6-45.9\%	 &  6.48$\pm$0.10  &   5.95$\pm$0.09 &   6.35$\pm$0.10 	       \\ 
\hline
   IX 	 	 	 &  0.68 & 0.55 & 0.62 	 & 45.3-64.5\% & 45.9-65.0\% & 45.9-65.6\% &  4.64$\pm$0.07  &   3.82$\pm$0.06 &   4.36$\pm$0.06 	       \\ 
\hline
   X 	 	 	 &  0.37 & 0.25 & 0.38 	  & 64.5-100.0\% & 65.0-100.0\% & 65.6-100\% &  2.52$\pm$0.04  &   1.76$\pm$0.03 &   2.67$\pm$0.04 	       \\ 
\hline
\hline
  \end{tabular}
  \label{tab:multi}
\end{sidewaystable}

The strange hadrons \pKzero, \pLambda, \apLambda, \pXi, \apXi, \pOmega and \apOmega are reconstructed at mid-rapidity (\yless{0.5}) with an invariant mass analysis, exploiting their specific weak decay topology. The following decay channels are studied~\cite{Tanabashi:2018oca}:
\begin{center}
  \centering
  \begin{tabular*}{\linewidth}{rll@{\extracolsep{2cm}}l}
    \pKzero & $\to$ & \pPiplus + \pPiminus & {\footnotesize B.R. = (69.20 $\pm$ 0.05) \%} \\
    \pLambda(\apLambda) & $\to$ & \pProton (\apProton) + \pPiminus (\pPiplus) & {\footnotesize B.R. = (63.9 $\pm$ 0.5) \%} \\
    \pXi (\apXi) & $\to$ & \pLambda (\apLambda) + \pPiminus (\pPiplus) & {\footnotesize B.R. = (99.887 $\pm$ 0.035) \%} \\ 
    \pOmega (\apOmega) & $\to$ & \pLambda (\apLambda) + \pKminus (\pKplus) & {\footnotesize B.R. = (67.8 $\pm$ 0.7) \%} \\
  \end{tabular*}
\end{center}

In the following, we refer to the sum of particles and anti-particles, $\pLambda+\apLambda$, $\pXi+\apXi$ and $\pOmega+\apOmega$,  simply as \sLambda, \sXi and \sOmega.

The details of the analysis have been discussed in earlier ALICE publications~\cite{Abelev:2013haa, Aamodt:2011zza, Abelev:2012jp, Acharya:2018orn}.
The tracks retained in the analysis are required to cross at least 70 TPC readout pads out of a maximum of 159. Tracks are also required not to have large gaps in the number of expected tracking points in the radial direction. This is achieved by checking 
that the number of clusters expected based on the reconstructed trajectory and the measurements in neighbouring TPC pad rows do not differ by more than 20\%. 

\begin{table}[t!]
\begin{center}
\caption{Track, topological and candidate selection criteria applied to \pKzero, \pLambda and \apLambda candidates. DCA stands for ``distance of closest approach'', PV represents the ``primary event vertex'' and $\theta$ is the angle between the momentum vector of the reconstructed $V^0$ and the
displacement vector between the decay and primary vertices. The selection on DCA between $V^0$ daughter tracks takes into account the corresponding experimental resolution.
} 
\begin{tabular}{ l | l }
\hline
\hline
\bf{Toplogical Variable}    &  \bf{\pKzero (\pLambda and \apLambda) Selection criteria}  \\
\hline
$V^0$ transverse decay radius & $>$ 0.50~cm \\
\hline
DCA (Negative / Positive Track - PV) & $>$ 0.06~cm \\
\hline
Cosine of $V^0$ Pointing Angle ($\theta_{V^0}$) & $>$ 0.97 (0.995) \\
\hline
DCA between $V^0$ daughter tracks & $<$ 1.0~standard deviations \\
\hline
\hline
\bf{Track selection}    &  \bf{\pKzero (\pLambda and \apLambda) Selection criteria}   \\
\hline
Daughter Track Pseudorapidity Interval & $|\eta|<0.8$ \\
\hline
Daughter Track $N_{crossed rows}$ & $\geq$ 70 \\
\hline
Daughter Track $N_{crossed}/N_{findable}$ & $\geq$ 0.8 \\
\hline
{\it p} inner wall TPC (proton only)  & $>$ 0.3 $\mathrm{GeV}/c$ \\
\hline
TPC d$E$/d$x$ & $<$ $5\sigma$ \\
\hline
\itstof & requested for at least one daughter \\
\hline
\hline
\bf{Candidate selection}                &  \bf{\pKzero (\pLambda and \apLambda) Selection criteria}  \\
\hline
Rapidity Interval & $|y|<0.5$ \\ 
\hline
Proper Lifetime ($mL/p$) & $<$ 20~cm (30~cm) \\
\hline
Competing $V^0$ Rejection & 5~$\mathrm{MeV}/c^{2}$ (10~$\mathrm{MeV}/c^{2}$) \\
\hline
MC Association (MC Only) & identity assumption for $V^0$ and for daughter tracks \\
\hline
\hline

\end{tabular}
\label{tab:cutV0}
\end{center}
\end{table}
\begin{table}[!t]
\begin{center}
\caption{Track, topological and candidate selection criteria applied to charged \pXi, \apXi, \pOmega and \apOmega candidates. DCA stands for ``distance of closest approach'', PV represents the ``primary event vertex'' and $\theta$ is the angle between the momentum vector of the reconstructed $V^0$ or cascade and the displacement vector between the decay and primary vertices. The selection on DCA between $V^0$ daughter tracks takes into account the corresponding experimental resolution.}
\begin{tabular}{ l | l }

\hline
\hline
\bf{Topological Variable}                          & \bf{$\Xi$ ($\Omega$) Selection criteria}                            \\
\hline
Cascade transverse decay radius R$_{2D}$  & $>$ 0.6 (0.5) cm                              \\
\hline
$V^0$ transverse decay radius                 & $>$  1.2 (1.1)  cm \\
\hline
DCA (bachelor - PV)                                   & $>$  0.04 cm   \\
\hline
DCA ($V^0$ - PV)                                & $>$  0.06 cm    \\
\hline
DCA (meson $V^0$ track - PV)                & $>$  0.04 cm       \\
\hline
DCA (baryon $V^0$ track - PV)                & $>$  0.03 cm      \\
\hline
DCA between $V^0$ daughter tracks                          &$<$ 1.5~standard deviations \\
\hline
DCA (bachelor - $V^0$)                            & $<$ 1.3 cm \\
\hline
Cosine of Cascade Pointing Angle  ($\theta_{\rm casc}$)          & $>$ 0.97                        \\
\hline
Cosine of $V^0$ Pointing Angle ($\theta_{V^0}$)            & $>$ 0.97                 \\
\hline
$V^0$ invariant mass window            & $\pm$ 0.008 $\mathrm{GeV}/c^2$  \\
\hline
\hline
\bf{Track selection}               & \bf{$\Xi$ ($\Omega$) Selection criteria}                            \\
\hline
Daughter Track Pseudorapidity Interval & $|\eta|<0.8$ \\
\hline
Daughter Track $N_{TPC clusters}$ & $\geq$ 70 \\
\hline
TPC d$E$/d$x$ & $<$ $5\sigma$ \\
\hline
\itstof & requested for at least one daughter \\
\hline
\hline
\bf{Candidate selection}               & \bf{$\Xi$ ($\Omega$) Selection criteria}                            \\
\hline
Rapidity Interval & $|y|<0.5$ \\ 
\hline
Proper Lifetime ($mL/p$) & $<3 \times c \tau$ \\
\hline
Competing Cascade Rejection (only $\Omega$) & $|M(\pLambda\pi)-1321|>8~\mathrm{MeV}/c^{2}$ \\
\hline
MC Association (MC Only) & identity assumption for cascades and for daughter tracks \\
\hline
\hline

\end{tabular}
\label{tab:cutCasc}
\end{center}
\end{table}

Each decay product arising from $V^0$ (\pKzero, \pLambda, \apLambda) and cascade (\pXi, \apXi, \pOmega, \apOmega) candidates is verified 
to lie within the fiducial tracking region \etaless{0.8}. The specific energy loss (\dedx) measured in the TPC, used for the particle identification (PID) of the decay products, is also requested to be compatible within 5$\sigma$ with the one expected for the corresponding particle species' hypothesis. 
The \dedx\ is evaluated as a truncated mean using the lowest 60\% of the values out of a possible total of 159. This leads to a resolution of about 6\%. 
A set of  ``geometrical'' selections is applied in order to identify specific decay topologies (topological selection), improving the signal/background ratio. The topological variables used for $V^0$s and cascades are described in detail in~\cite{Aamodt:2011zza}. 
In addition, in order to reject the residual out-of-bunch pile-up background on the measured yields, it is requested that at least one of  the tracks from the decay products of the (multi-)strange hadron under study is matched in either the ITS or the TOF detector. The selections used in this paper are summarised in Table~\ref{tab:cutV0} for the $V^0$s and in Table~\ref{tab:cutCasc} for the cascades.

Strange hadron candidates are required to be in the rapidity window \yless{0.5}. \pKzero (\pLambda) candidates compatible with the alternative $V^0$ hypothesis are rejected if they lie within $\pm$5~$\mathrm{MeV}/c^2$ ($\pm$10~$\mathrm{MeV}/c^2$) of the nominal \pLambda (\pKzero) mass. A similar selection 
is applied to the \sOmega, where candidates compatible within  $\pm$8~$\mathrm{MeV}/c^2$ of the nominal \sXi\ mass are rejected. The width of the rejected region was determined according to the invariant mass resolution of the corresponding competing signal. Furthermore, candidates whose proper lifetimes are unusually large for their expected species 
are also rejected to avoid combinatorial background from interactions with the detector material.
The signal extraction is performed as a function of \pt. A preliminary fit is performed on the invariant mass distribution using a Gaussian plus a linear function describing the background.
This allows for the extraction of the mean ($\mu$) and width ($\sigma_{\rm G}$) of the peak. 
A ``peak''  region is defined within $\pm 6(4)\sigma_{\rm G}$ for  $V^0$s (cascades) with respect to $\mu$ for any measured \pt bin. 
Adjacent background bands, covering a combined mass interval as wide as the peak region, are defined on both sides of that central region.
The signal is then extracted with a bin counting procedure, subtracting counts in the background region from those of the signal region. Alternatively, the signal is extracted by fitting the background with a  linear function extrapolated under the signal region. This procedure  is used to compute the systematic uncertainty due to the signal extraction. Examples of the invariant mass peaks for all particles are shown in Fig.~\ref{fig:signExtraction}.

\begin{figure}[!t]
  \begin{center}
  \includegraphics[width=0.40\textwidth]{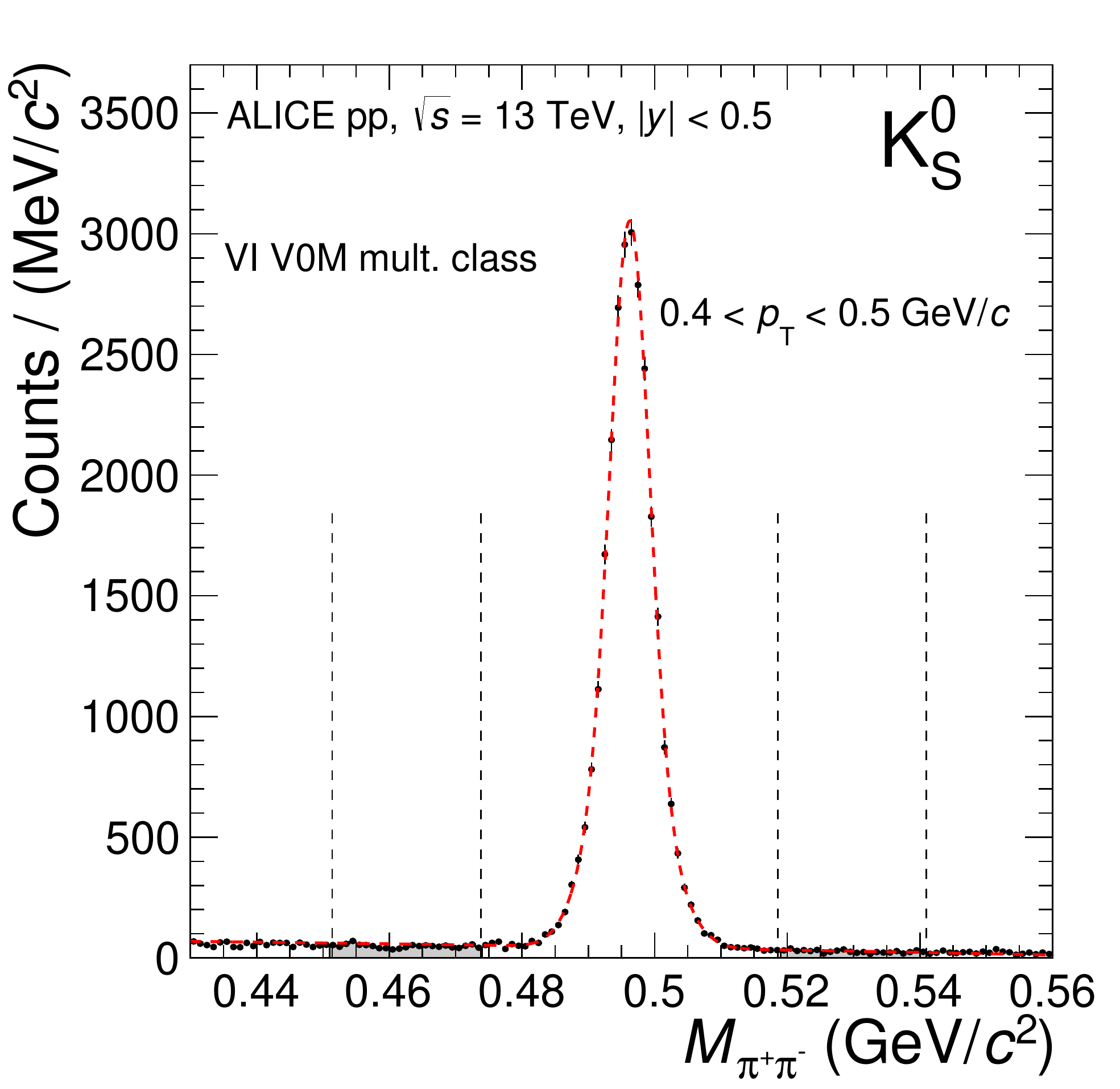}
    \includegraphics[width=0.40\textwidth]{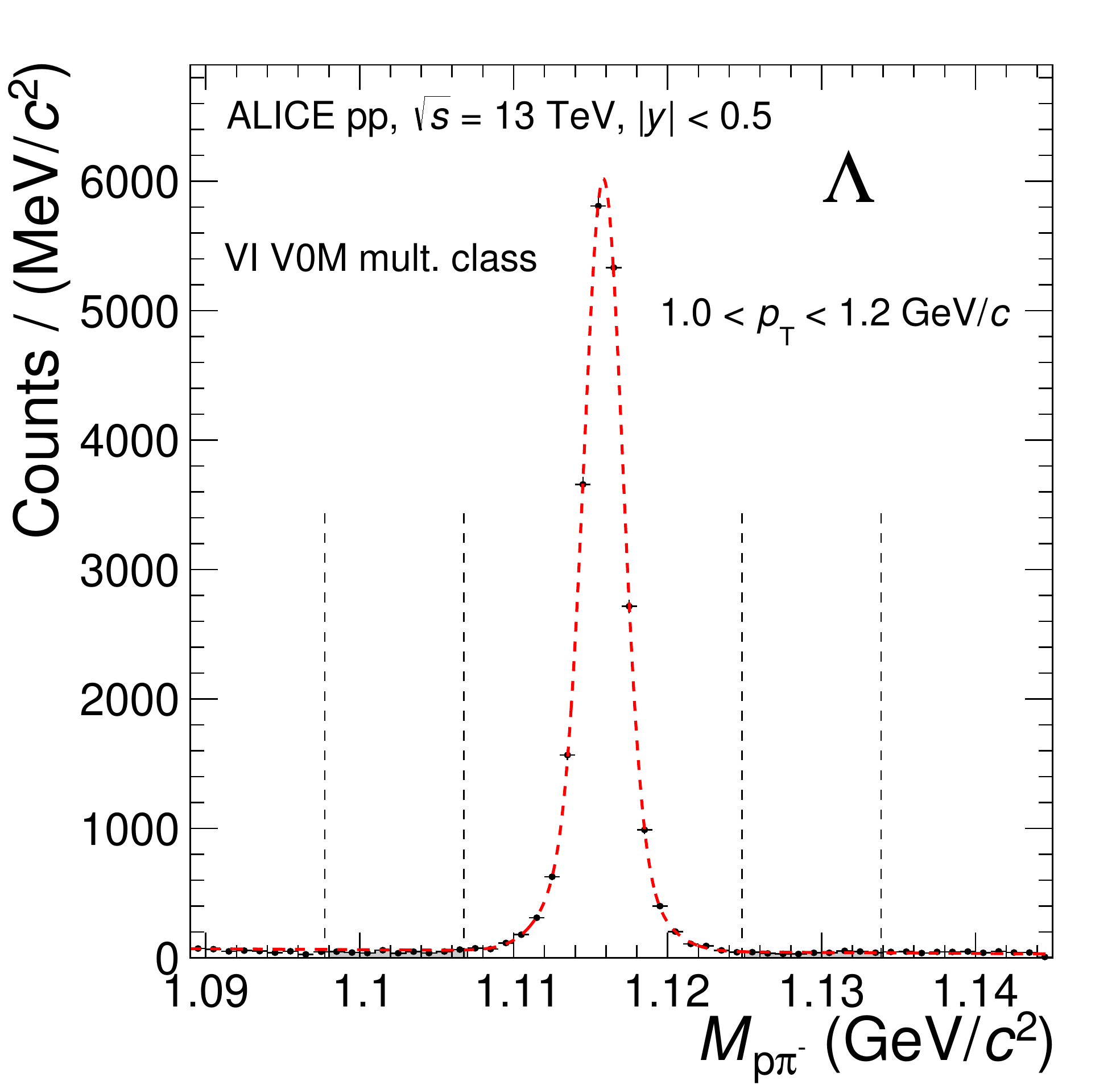}
    \includegraphics[width=0.40\textwidth]{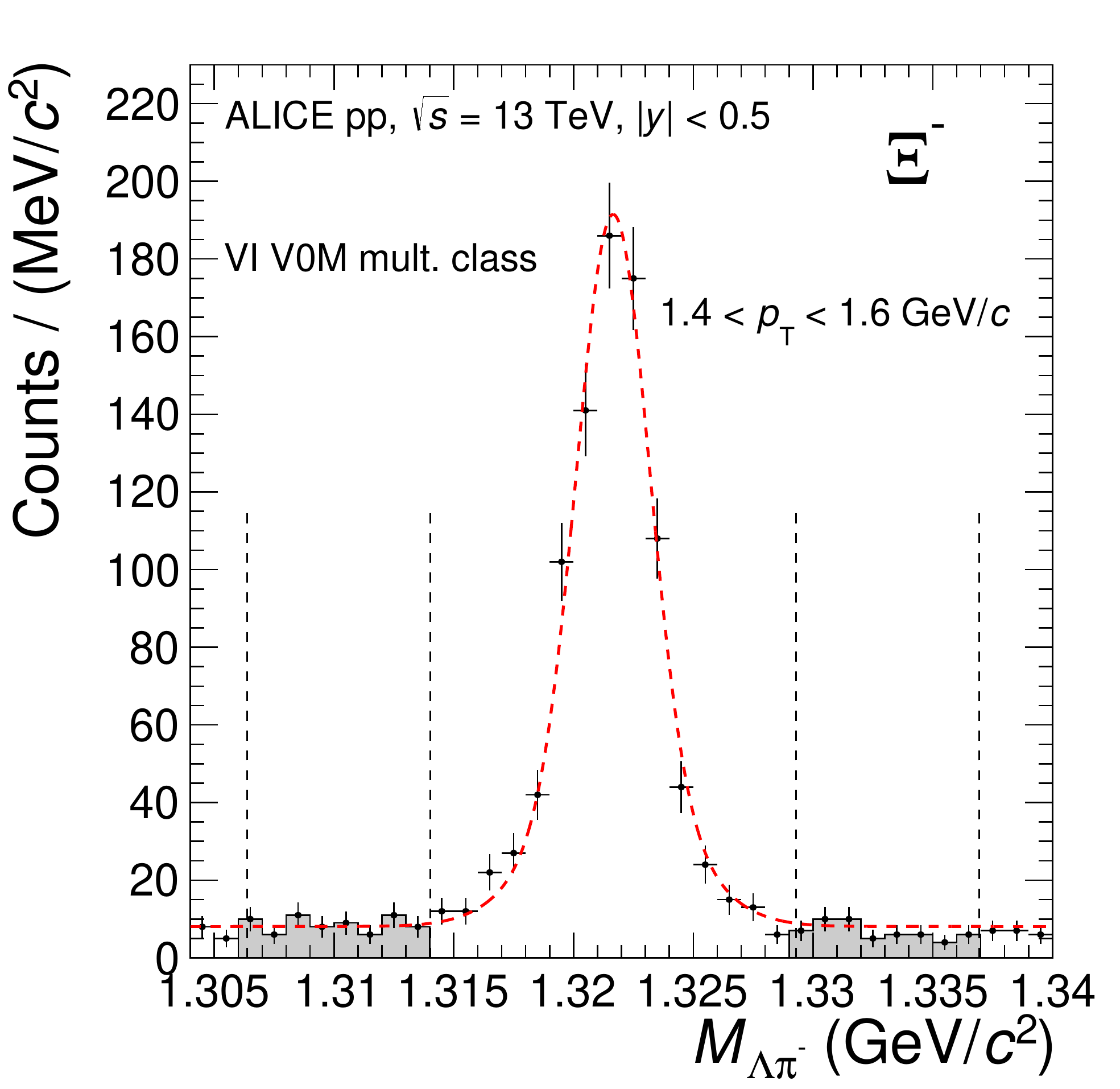}
    \includegraphics[width=0.40\textwidth]{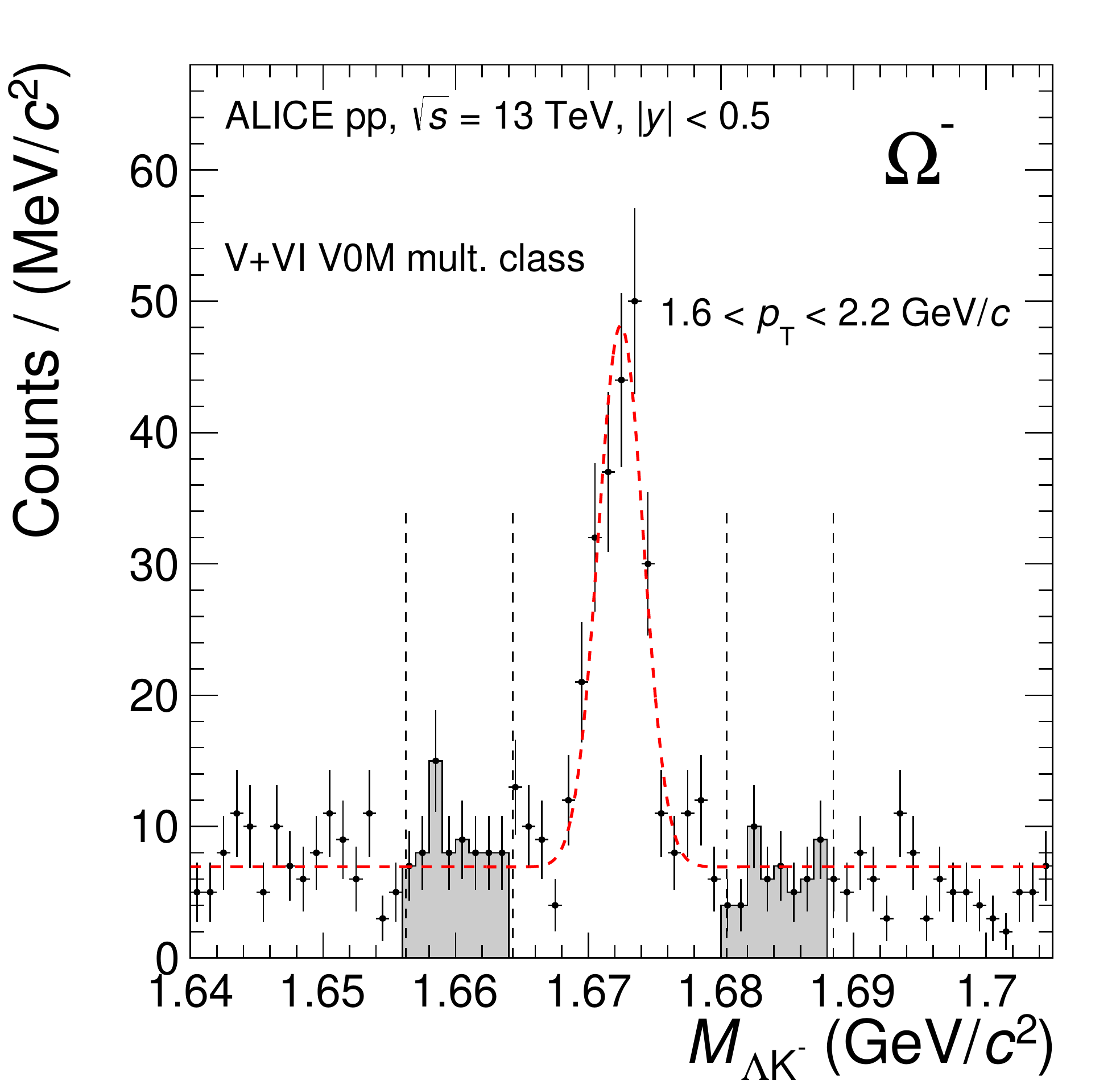}
  \end{center}
  \caption{\label{fig:signExtraction} Invariant mass distributions for \kzero, \lmb, \pXi, \pOmega
in different V0M multiplicity and \pt intervals. The candidates are reconstructed in \yless{0.5}. The grey areas delimited by the
short-dashed lines are used for signal extraction in the bin counting procedure. The red dashed lines
represent the fit to the invariant mass distributions, shown for drawing purpose only. }
\end{figure}

Only the \pLambda\ turns out to be affected by a significant contamination from secondary particles, coming from the decay of charged and neutral \sXi. In order to estimate this contribution we use the measured \pXi\ and \apXi\ spectra, folded with a \pt-binned 2D matrix describing the decay kinematics and secondary \pLambda reconstruction efficiencies.
The $\Xi \to \Lambda \pi $ decay matrix is  extracted from Monte Carlo (see below for the details on the generator settings). The fraction of secondary \pLambda\ particles in the measured spectrum varies between 10\% and 20\%, depending on \pt\ and multiplicity. Further details on the uncertainties characterising the feed-down contributions are provided in the next section.


The raw \pt\ distributions are corrected for acceptance and efficiency using Monte Carlo simulated data. Events are generated using the 
PYTHIA 6.425, (Tune Perugia 2011)~\cite{Sjostrand:2006za,Skands:2010ak} event generator, and transported through a GEANT 3~\cite{Brun:1994aa} (v2-01-1) model of the detector. 
With respect to previous GEANT 3 versions, the adopted one contains
a more realistic description of (anti)proton interactions. The quality of this description was cross-checked comparing to the results obtained with the state-of-the-art transport codes FLUKA~\cite{Ferrari:898301,BOHLEN2014211} and GEANT~4.9.5~\cite{Agostinelli:2002hh}. It was found that a correction factor  $<5\%$ is needed for the 
efficiency of \apLambda, \apXi, \apOmega\ for $\pt < 1~\gevc$, while the effect is negligible at higher \pt. 
Events generated using PYTHIA 8.210 (tune Monash 2013)~\cite{Sjostrand:2014zea,Skands:2014pea} and EPOS-LHC (CRMC package 1.5.4)~\cite{Pierog:2013ria} and transported in the same way are used for systematic studies, namely to compute the systematic uncertainties arising from the normalisation and from the closure of the correction procedure (details provided in the next sections).

%
%

The acceptance-times-efficiency changes with \pt, saturating at a value of about 40\%, 30\%, 30\% and 20\% at \pt $\simeq$ 2, 3, 3 and 4 \gevc\ for  \pKzero, \pLambda, \sXi and \sOmega, respectively. These values include the losses due to the branching ratio. 
They are found to be independent of the multiplicity class within 2\%, limited by the available Monte Carlo simulated data.\warn{change this if the conclusion is different for central estimators.}
The dependence of the efficiency on the generated \pt\ distributions was checked for all particle species. It is found to be relevant only in the case of the \sOmega, where large \pt\ bins are used. This effect is removed by reweighting the
Monte Carlo \pt\ distribution with the measured one using an iterative procedure.

In order to compute \average\pt and the \pt-integrated production yields, the spectra are fitted with a Tsallis-L\'evy~\cite{Prato:1999jj} distribution to extrapolate in the unmeasured \pt\ region. The systematic uncertainties on this extrapolation procedure are evaluated using other fit functions, as discussed in Sec.~\ref{sec:syst-uncert}.


\section{Systematic uncertainties}
\label{sec:syst-uncert}


Several sources of systematic effects on the evaluation of the \pt\ distributions were investigated. The main contributions for three representative \pt\ values are summarised in Table~\ref{tab:syst} 
for the INEL $>$ 0 data sample.

The stability of the signal extraction method was checked by varying the widths used to define the ``signal'' and 
``background'' regions, expressed in terms of number of $\sigma_{G}$ as defined in Sec.~\ref{sec:analysis-details}.
The raw counts were also extracted with a fitting procedure and compared to the standard ones computed by the bin counting technique. 
An uncertainty ranging between 0.2\% and 3.5\% depending on \pt\ is assigned to the signal extraction of the $V^0$s and cascades based on these checks.

The stability of the acceptance and efficiency corrections was verified by varying all 
track, candidate and topological selection criteria within ranges leading to a maximum variation of $\pm 10\%$ in the raw signal yield. 
The results were compared to those obtained with the default selection criteria (Sec.~\ref{sec:analysis-details}). 
Variations not compatible with statistical fluctuations (following the prescription in~\cite{Barlow:2002yb} with a 2$\sigma$ threshold) are added to the systematic uncertainty.

The resulting uncertainty from topological and track selections (except TPC d$E$/d$x$) depends on \pt\ and amounts at most to 4\%, 5\%, 4\% and 6\% for \pKzero, \pLambda, \sXi and \sOmega, respectively.

The TPC d$E$/d$x$ selection is used to reduce the combinatorial background in the strange baryon invariant mass distribution. 
The uncertainty was evaluated  varying the TPC d$E$/d$x$ selection requirements between 4 and 7$\sigma$ and was found to be at most 1\% (3\%) for \pLambda (\sXi and \sOmega).
For the \pKzero the uncertainty due to the TPC d$E$/d$x$ usage was evaluated by comparing results obtained adopting the default loose PID requirements ($5~\sigma$) with those obtained without applying any PID selection. The difference is found to be negligible ($< 1\%$).

The systematic uncertainty for the competing decay rejection was investigated by removing entirely this condition for \pLambda and \sOmega. 
It resulted  in a deviation on the \pt spectra of at most 4\% and 6\%, respectively. For the \pKzero the systematic uncertainty was evaluated by changing the width of the competing rejected mass window between 3 and 5.5~$\mathrm{MeV}/c^2$ and the corresponding deviation was found to be at most 1\%.

For the strange baryons, the systematic uncertainty related to the proper lifetime was computed by varying the selection requirements between 2.5 and 5~$c\tau$. The variation range for the \pKzero was set to 5-15~$c\tau$.
The statistically significant deviations were found to be at most 3\% for the \pLambda and negligible ($<1$\%) for all other particles.  

An uncertainty related to the absorption in the detector material was assigned to the anti-baryons, mostly due to the interactions of anti-proton daughters. It was estimated on the basis of the comparison of the different transport codes mentioned in Sec.~\ref{sec:analysis-details}. The uncertainty on the absorption cross section for baryons and \pKzero\
 was found to be negligible.

Furthermore an additional 2\% uncertainty is added to account for possible variations of the tracking efficiency with multiplicity (Sec.~\ref{sec:analysis-details}).

The uncertainty due to approximations in the description of the detector material was estimated with a Monte Carlo simulation where the material budget was varied within its uncertainty~\cite{Abelev:2014ffa}. The assigned systematic uncertainty ranges between 8\% at low \pt\ to about 1\% at high \pt.
%

The \pLambda\ \pt spectrum is affected by an uncertainty coming from the feed-down correction, due to the uncertainties on the measured \sXi\ spectrum and on the multiplicity dependence of the 
feed-down fraction. Furthermore the contribution from neutral $\Xi^{0}$ was taken into account by assuming $\Xi^{\pm}/\Xi^{0} = 1$ or using the ratio provided by the Monte Carlo (using the reference Pythia 6 sample described in the previous section). The difference between these two estimates was taken into account in the calculation of the total uncertainty due to the feed-down correction, which ranges from 2\% to 4\% depending on \pt\ and multiplicity.

%

The systematic uncertainty due to the out-of-bunch pile-up rejection was evaluated by changing the matching scheme with the relevant detectors, considering the following configurations: matching of at least one decay track with the ITS (TOF) detector below (above) 2 GeV/$c$ of the reconstructed (multi-) strange hadron; ITS matching of at least one decay track in the full \pt\ range. Half of the maximum difference between these configurations and the standard selection was taken as the systematic uncertainty, which was found to increase with transverse momentum and to saturate at high \pt, reaching a maximum value of 2.4\% (3\%) for $V^0$s (cascades).

The effect of a possible residual contamination from in-bunch pile-up events was estimated varying the pile-up rejection criteria and dividing the data sample in three groups with an average interaction probability per bunch crossing  of 3\%, 6\%, 13\%, respectively. The resulting uncertainty is larger at low multiplicity, and ranges between 1\% (3\%) for the \pKzero to 2\% (3\%) for the baryons in high-(low-)multiplicity events.

\begin{table}[tbp]
  \caption{Main sources and values of the relative systematic uncertainties (expressed in \%) of the \pt-differential yields. These values are reported for low, intermediate and high \pt. The values for the $\mathrm{INEL}>0$ data sample are shown in the table. Results as a function of multiplicity are further affected by an uncertainty originating from the multiplicity dependence of the efficiency (2\%) and, in the case of the \pLambda, of the feed-down contributions (2\%).}
\begin{center}
\resizebox{1.0\textwidth}{!}{
\begin{tabular}{| l |  c c c |  ccc  | ccc  |  ccc  | }
\hline
\hline
Hadron & \multicolumn{3}{c|}{\pKzero}  & \multicolumn{3}{c|}{\pLambda + \apLambda}  & \multicolumn{3}{c|}{\pXi + \apXi} & \multicolumn{3}{c|}{\pOmega + \apOmega}  \\
\pt (GeV/$c$)  & $\simeq$0.95  & $\simeq$4.8  & $\simeq$9.0  & $\simeq$0.50 & $\simeq$4.5 & $\simeq$7.3 & $\simeq$0.80 & $\simeq$3.2 & $\simeq$5.8 & $\simeq$1.3 & $\simeq$2.8 & $\simeq$4.7 \\
\hline
\hline
Signal extraction & 0.6 & 0.6 & 1.6 & 1.1 & 2.4 & 1.1 & 0.6 & 0.4 & negl. & 2.1 & 2.1 & 3.3 \\
Topological and track & \multirow{2}{*}{0.7} & \multirow{2}{*}{2.7}  & \multirow{2}{*}{2.5}  & \multirow{2}{*}{2.9} & \multirow{2}{*}{2.6}  & \multirow{2}{*}{2.3}  & \multirow{2}{*}{4.8} & \multirow{2}{*}{2.6}  & \multirow{2}{*}{2.0} & \multirow{2}{*}{5.6} & \multirow{2}{*}{6.0}  &  \multirow{2}{*}{5.5}  \\
selection (but TPC d$E$/d$x$) &               &   &   &   &    &   &  &   & &  &   &    \\
TPC d$E$/d$x$ selection & 0.1 & negl. & negl. & 0.7 & 0.3 & 1.7 & 0.4 & 0.3 & 0.3 & 1.7 & 2.2 & 1.6 \\
Competing decay rejection & 0.1 & 0.2 & 1.0 & negl. & 0.9 & 5.7 & \multicolumn{3}{c|}{not applied} & 1.2 & 3.1 & 5.6 \\
Proper lifetime & 0.1 & negl. & negl. & 1.1 & negl. & negl. & 0.5 & 0.8 & 0.8 & 3.1 & 1.8 & 1.0 \\
Transport code (for anti-particles) &  \multicolumn{3}{c|}{not applied}  & 1.8 & negl.  & negl.  & 1.1 & negl.  & negl. & 0.6 & negl.  &  negl. \\
Material budget &  1.1 & 0.5  & 0.5  & 8.3 & 0.8  & 0.8  & 5.1 & 1.2  & 0.6 & 3.3 & 1.5  &  1.5 \\
Feed-down correction  & \multicolumn{3}{c|}{not applied} & 2.2 & 1.3 & 2.1 & \multicolumn{3}{c|}{not applied} & \multicolumn{3}{c|}{not applied} \\
out-of-bunch pile-up track rejection  & 1.1  & 1.5  & 1.5  & 1.2  & 2.4   & 2.4  & 1.0 & 2.5  & 2.5 & 3.0 & 3.0  & 3.0   \\
Residual in-bunch pile-up  & 1.6 & 2.5 & 2.5 & 2.0 & 2.9 & 2.9 & 2.0 & 2.0 & 2.9 & 2.0 & 2.0 & 2.0 \\
\hline
\hline
Total  & 2.4 & 4.1 & 4.3 & 9.6 & 5.5 & 7.8 & 7.5 & 4.4 & 4.4 & 8.6 & 8.6 & 9.6 \\
Common ($N_{\rm ch}$-independent) & 2.3 & 4.0 & 3.8 & 9.2 & 5.3 & 7.0 & 6.8 & 4.2 & 4.2 & 7.7 & 8.4 & 7.5 \\
\hline
\hline
\end{tabular}
}
\end{center}
  \label{tab:syst}
\end{table}

Several additional consistency checks have been performed which are described in the following.
The analysis has been repeated separately for events with positive and negative $z$-vertex position, as well as considering candidates reconstructed in positive and negative rapidity windows. The resulting \pt-spectra were found to be statistically compatible with the standard analysis.

In order to ensure that the estimated systematic uncertainties are not affected by statistical fluctuations, two checks were performed. First of all, the threshold used to consider a variation statistically significant was varied between one and three standard deviations. Then, the analysis was repeated with wider \pt\ bins. These checks showed that statistical fluctuations in the systematic uncertainty analysis are well under control.

The results related to the \average\pt\ and \pt-integrated yields for all particles but the \pKzero\ are further affected by an uncertainty coming from the extrapolation to zero \pt.
The default extrapolation is performed using a L\'evy-Tsallis distribution. As confirmed by a $\chi^{2}$ analysis, this function describes the \pt\ spectra well for all the examined strange hadrons over the measured \pt\ range.

The uncertainty on the extrapolated fraction was estimated repeating the fit to the spectra with five alternative functions (Blast-Wave, Boltzmann, Bose-Einstein, $m_{\rm T}$-exponential, Fermi-Dirac). Since these alternative functions do not, in general, describe the full \pt -distribution, the fit range was limited to a small number of data points in order to obtain a good description of the fitted part of the spectrum. 
This procedure was repeated separately for the low and for the high \pt\ extrapolation, with the final uncertainties being dominated by the low \pt\ contribution.
The reliability of the extrapolation uncertainty estimate was checked using a simple linear extrapolation to \pt $= 0$ as an extreme case and in a full Monte Carlo closure test where the EPOS model was used with data and PYTHIA was used to estimate the corrections.
 The resulting uncertainty on the integrated yields is around 2.5\% for the \pLambda and ranges between 3\% (4\%) at high multiplicity to  19\% (12\%) at low multiplicity for the \sXi (\sOmega).
The extrapolation uncertainty on \average\pt\ is $\sim$2\% for the \pLambda and ranges between 2\% (3\%) at high multiplicity to 12\% (7\%) at low multiplicity  for the \sXi (\sOmega).

The main focus of this paper is the study of the multiplicity dependence of strangeness production. 
In this light, the different systematic uncertainties can be categorised in the following way:

\begin{enumerate}
\item Fully uncorrelated uncertainties: the change in the data is completely uncorrelated across multiplicity classes. These sources are rare, as most systematic effects have a smooth evolution with multiplicity.
\item Fully correlated uncertainties: lead to a correlated shift of the data in the same direction, independently of the multiplicity class being studied. 
The common part of this shift has to be considered separately, since it does not affect the shape of the multiplicity-dependent observable. 
\item Anti-correlated uncertainties: the effect is opposite in low and high multiplicity events.
\end{enumerate}

In the following we quote separately uncertainties that affect the trends as a function of multiplicity and those that lead to a constant fractional shift across all multiplicity classes.
The effect of every systematic variation was evaluated simultaneously in each multiplicity class and in minimum bias events to separate the fully correlated part of the uncertainty. 
Sources leading to a global, fully-correlated shift in the spectra were subtracted from the total uncertainty while the remaining contribution could in principle belong to any of the three aforementioned categories.
However, since different (independent) sources are combined in the final uncertainties, it is a reasonable assumption to consider them as uncorrelated.
For the figures shown in Sec.~\ref{sec:results}, total uncertainties are shown as boxes, while shadowed boxes represent uncertainties uncorrelated across different multiplicity classes.
Statistical uncertainties are depicted as error bars.



\section{Results and discussion}
\label{sec:results}


Particles and anti-particles turn out to have compatible transverse momentum distributions within uncertainties, consistently with previous results at the LHC. In the following, unless specified otherwise, we present results for their sum. 
The \pt\ distributions of strange hadrons, measured in \yless{0.5}, are shown in 
\crefrange{fig:spectra-V0M}{fig:spectra-tracklets0815}
for the different multiplicity classes, selected using the estimators V0M, tracklets in \etaless{0.8} and tracklets in $0.8 < \left| \eta \right| < 1.5$, as summarised in Table~\ref{tab:multi}. 
The L\'evy-Tsallis fit to the \pt\  distributions, used for the extrapolation, are also displayed. The bottom panels depict the ratio to the minimum bias (\inelgtzero) \pt\ distribution. 
%
%
The \pt\ spectra become harder as the multiplicity increases, as also shown in Fig.~\ref{fig:avpt-vs-mult}, which shows the average \pt\ (\avpT) as a function of the mid-rapidity charged particle multiplicity. While the V0M and \NtrkEtaOut\ results show the same trend, the spectra obtained with the \NtrkEtaIn\ estimator are systematically softer for comparable \dNdeta\ values (though still compatible within uncertainty in the case of the strange baryons). 

The increase of the \avpT\ as a function of the charged-particle multiplicity (Fig.~\ref{fig:avpt-vs-mult}) is compatible, within uncertainties, for all particle species. 
The hardening of \pt\ spectra with the charged-particle multiplicity was already reported for pp~\cite{Chatrchyan:2012qb} and \pPb\ collisions~\cite{Abelev:2013haa} at lower energies.

It is interesting to notice that the ratio of the measured \pt\ spectra to the minimum bias one, shown in the bottom panel of 
\crefrange{fig:spectra-V0M}{fig:spectra-tracklets0815}, 
reaches a plateau for $\pt\gtrsim 4~\gevc$. This applies to all particle species and to all multiplicity estimators.
The trend at high-\pt\ is highlighted in Fig.~\ref{fig:yields-high-pt}, which shows the integrated yields for $\pt~>~4~\gevc$ as a function of the mid-rapidity multiplicity. Both the yields of strange hadrons and the charged particle multiplicity are self-normalised, i.e. they are divided by their average quoted on the \inelgtzero\ sample. The high-\pt\ yields of strange hadrons increase faster than the charged particle multiplicity. Despite the large uncertainties, the data also hint at the increase being non-linear.
The self-normalised yields of strange hadrons reach, at high multiplicity, larger values for the \NtrkEtaIn\ multiplicity selection as compared to the other estimators. For all multiplicity selections the self-normalised yields of baryons are higher than those of \pKzero\ mesons. The Monte Carlo models EPOS-LHC, PYTHIA 8 and PYTHIA 6, also shown in Fig.~\ref{fig:yields-high-pt}, reproduce the overall trend of strange hadrons seen in the data, with EPOS-LHC showing a clear difference between the \pKzero\ and the baryons, as observed in the data.
Indeed, a non-linear increase can be explained by the combined effect of multiplicity fluctuations and interactions between different MPIs, which produces a collective boost~\cite{Werner:2016nnf}. Both  colour reconnection effects implemented in PYTHIA and a collective hydrodynamic expansion can account for the non-linear increase pattern. 



\begin{figure}
  \begin{center}
    \includegraphics[width=0.47\textwidth]{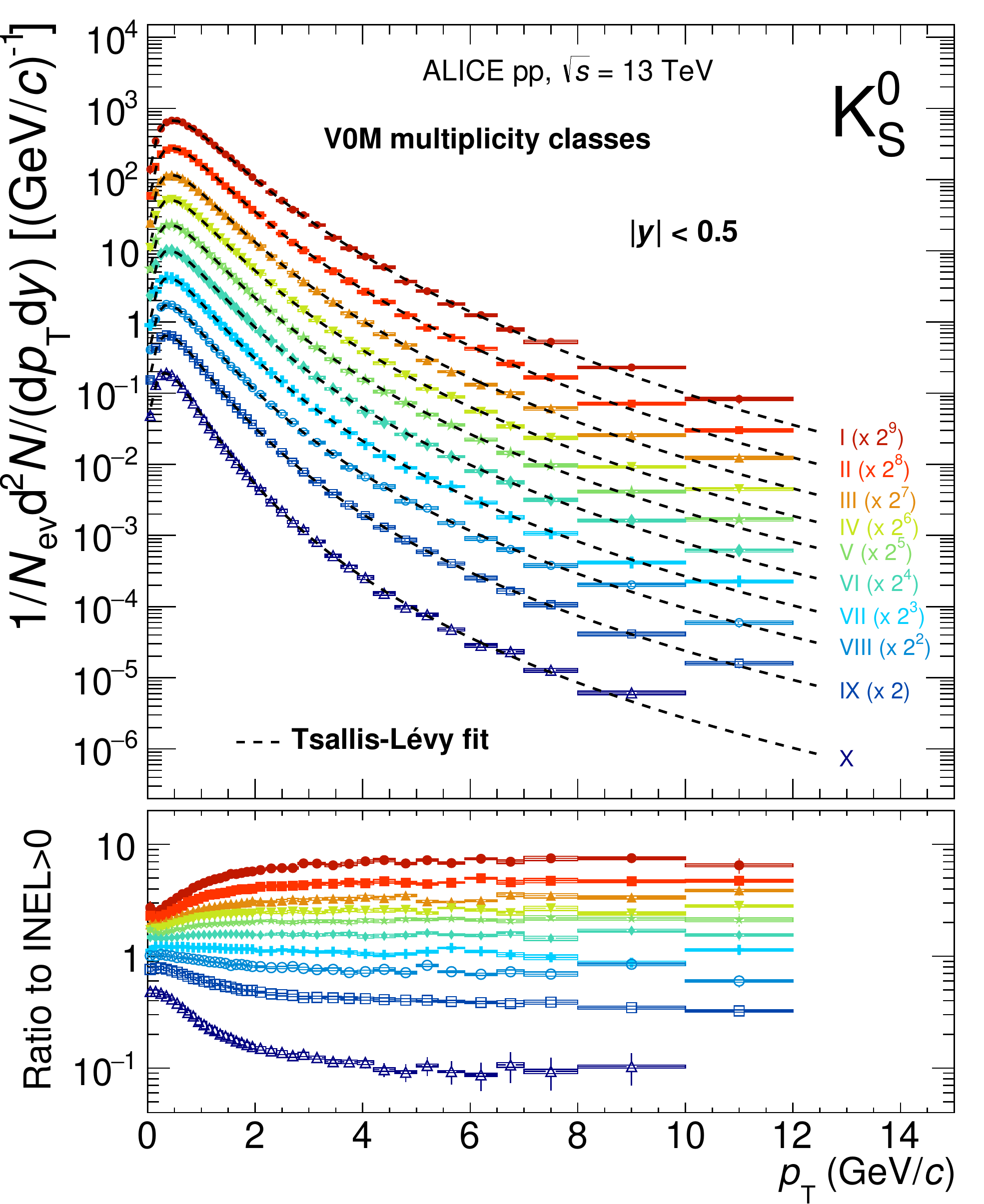}
    \includegraphics[width=0.47\textwidth]{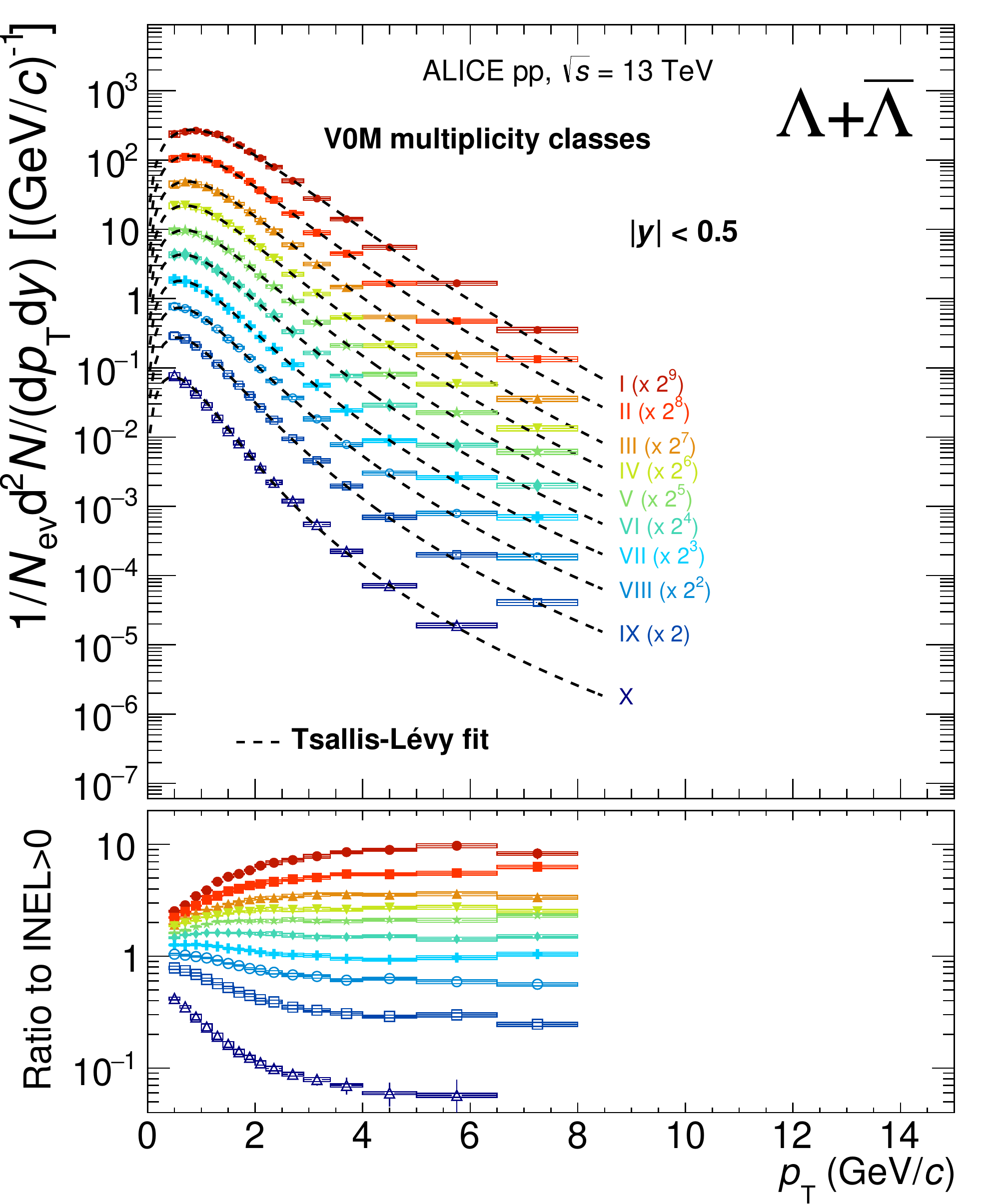}
    \includegraphics[width=0.47\textwidth]{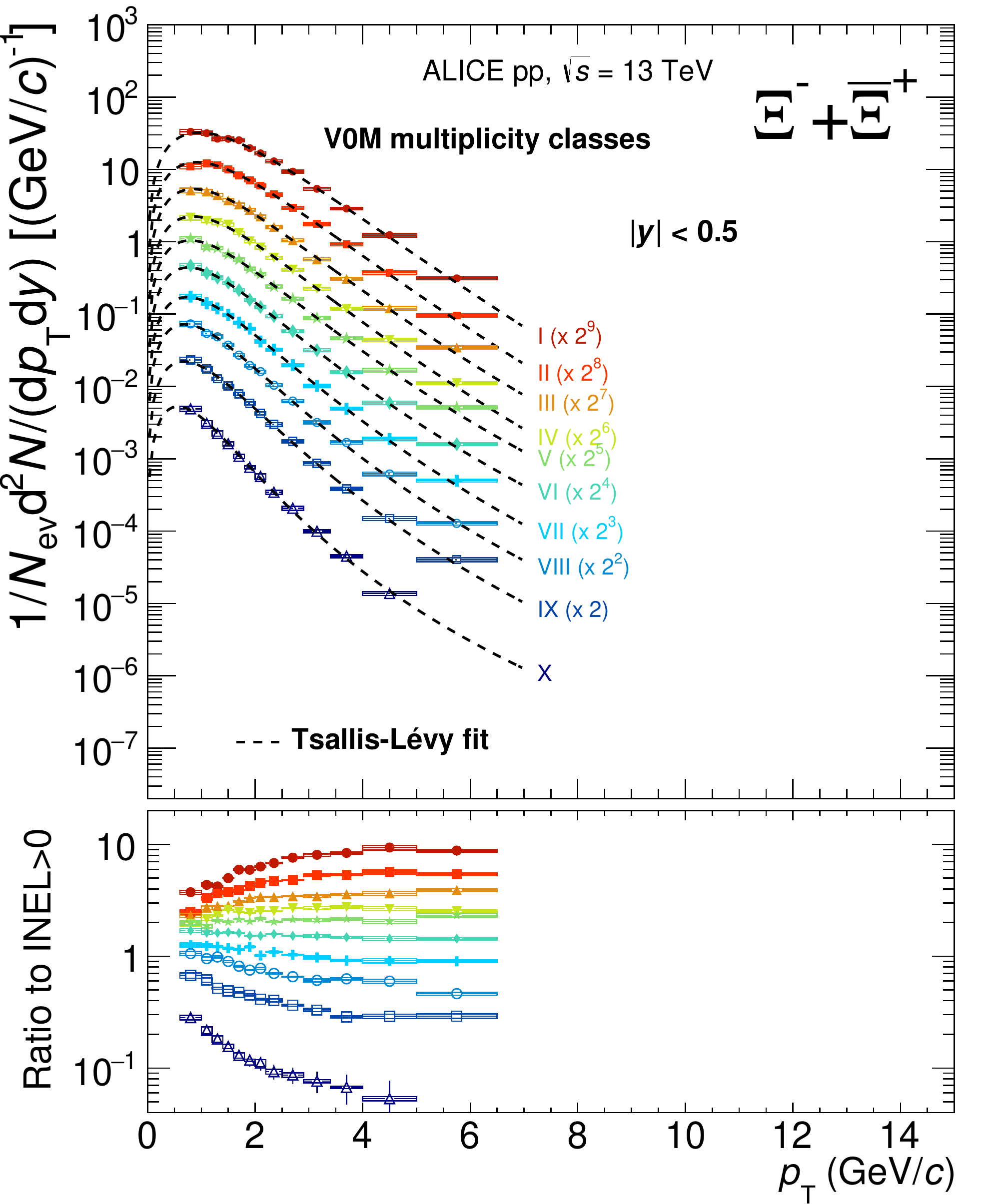}
    \includegraphics[width=0.47\textwidth]{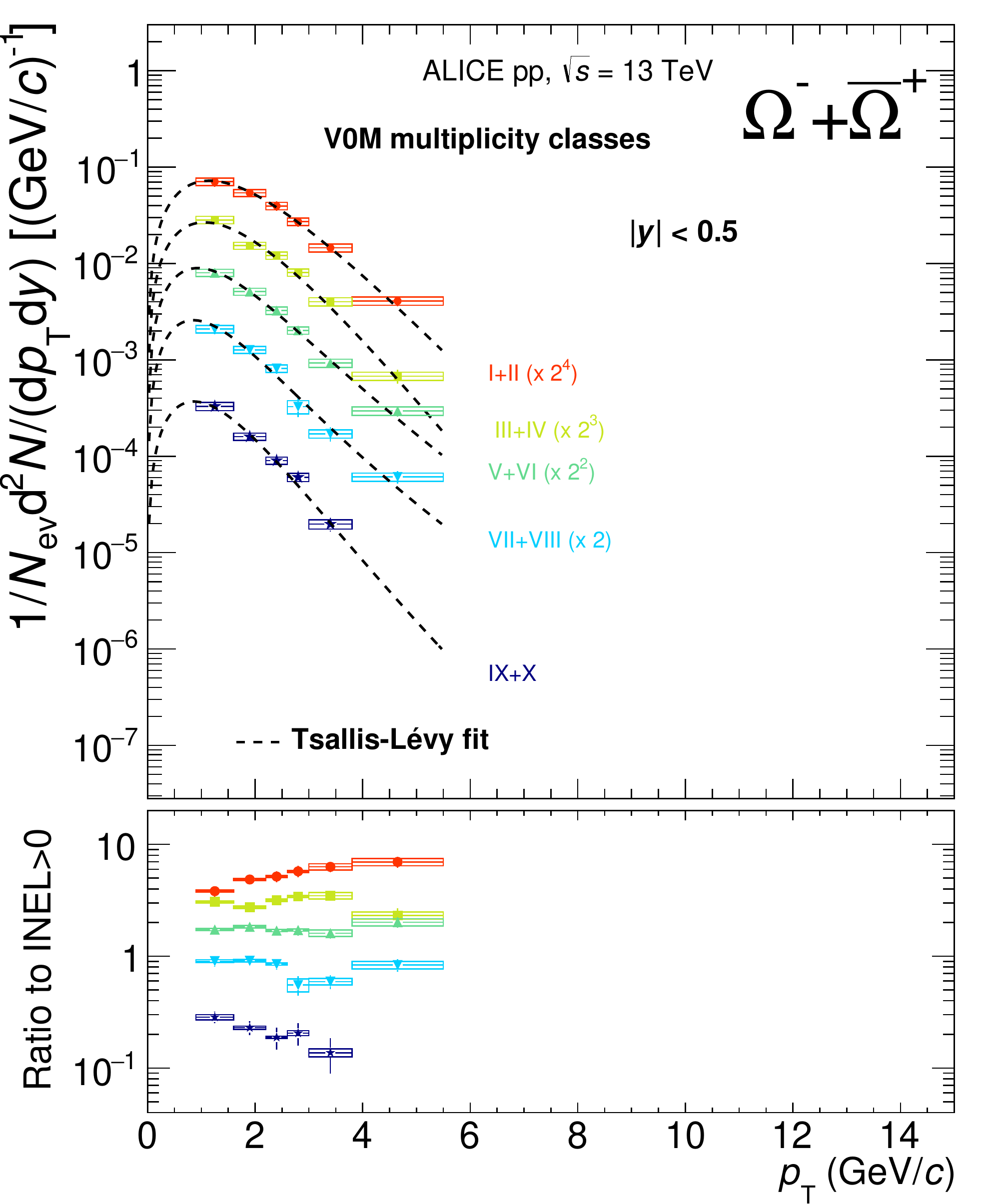}
  \end{center}
  \caption{\label{fig:spectra-V0M}Transverse momentum distribution of \pKzero, \sLambda, \sXi, and \sOmega, for multiplicity classes selected using the V0 detector. Statistical and total
systematic uncertainties are shown by error bars and boxes, respectively. In the bottom panels ratios of multiplicity dependent spectra to INEL $>$ 0 are shown.
The systematic uncertainties on the ratios are obtained by considering only contributions uncorrelated across multiplicity. The spectra are scaled
by different factors to improve the visibility. The dashed curves represent Tsallis-L\'evy fits to the measured spectra. }
\end{figure}

\begin{figure}
  \begin{center}
    \includegraphics[width=0.47\textwidth]{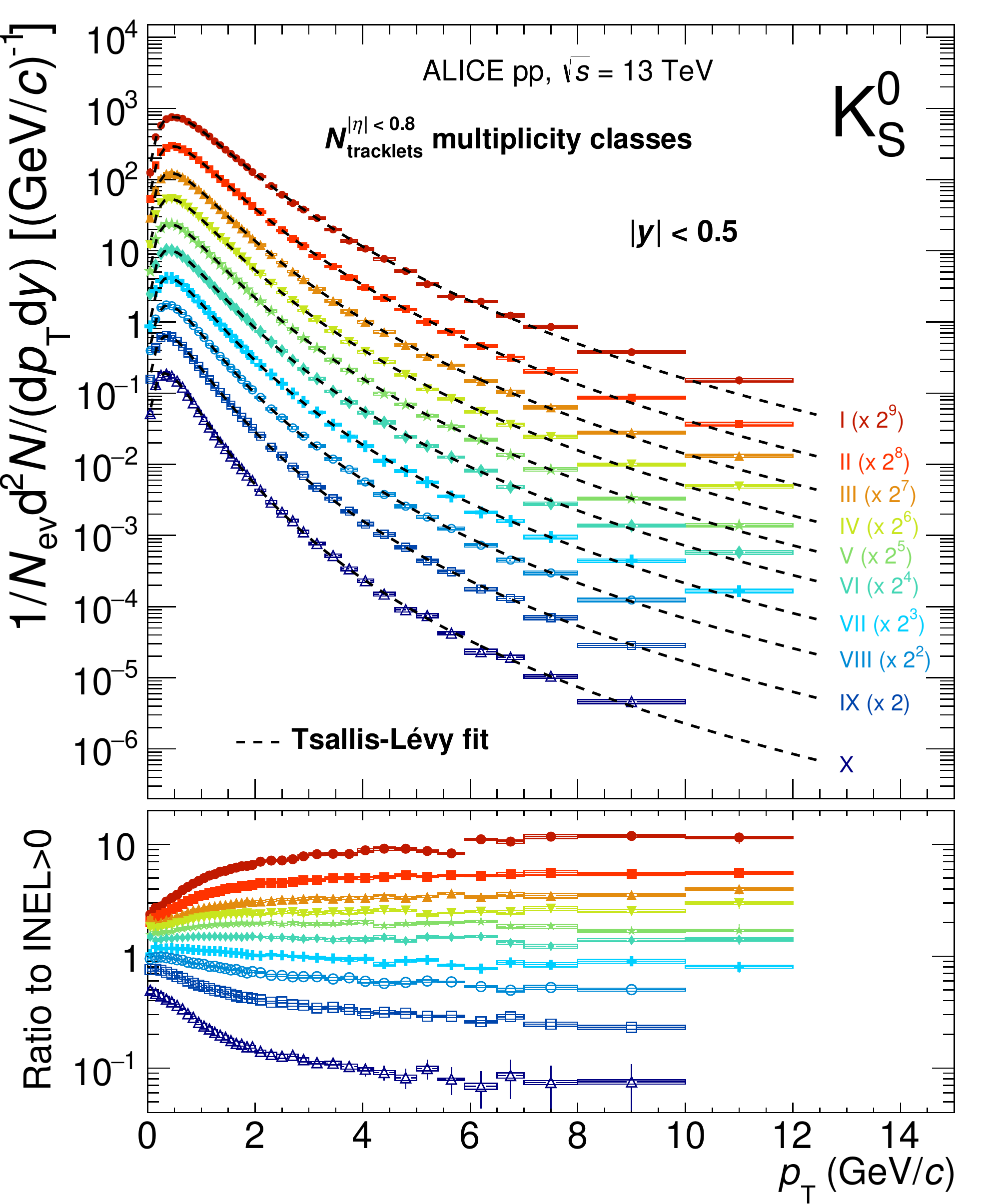}
    \includegraphics[width=0.47\textwidth]{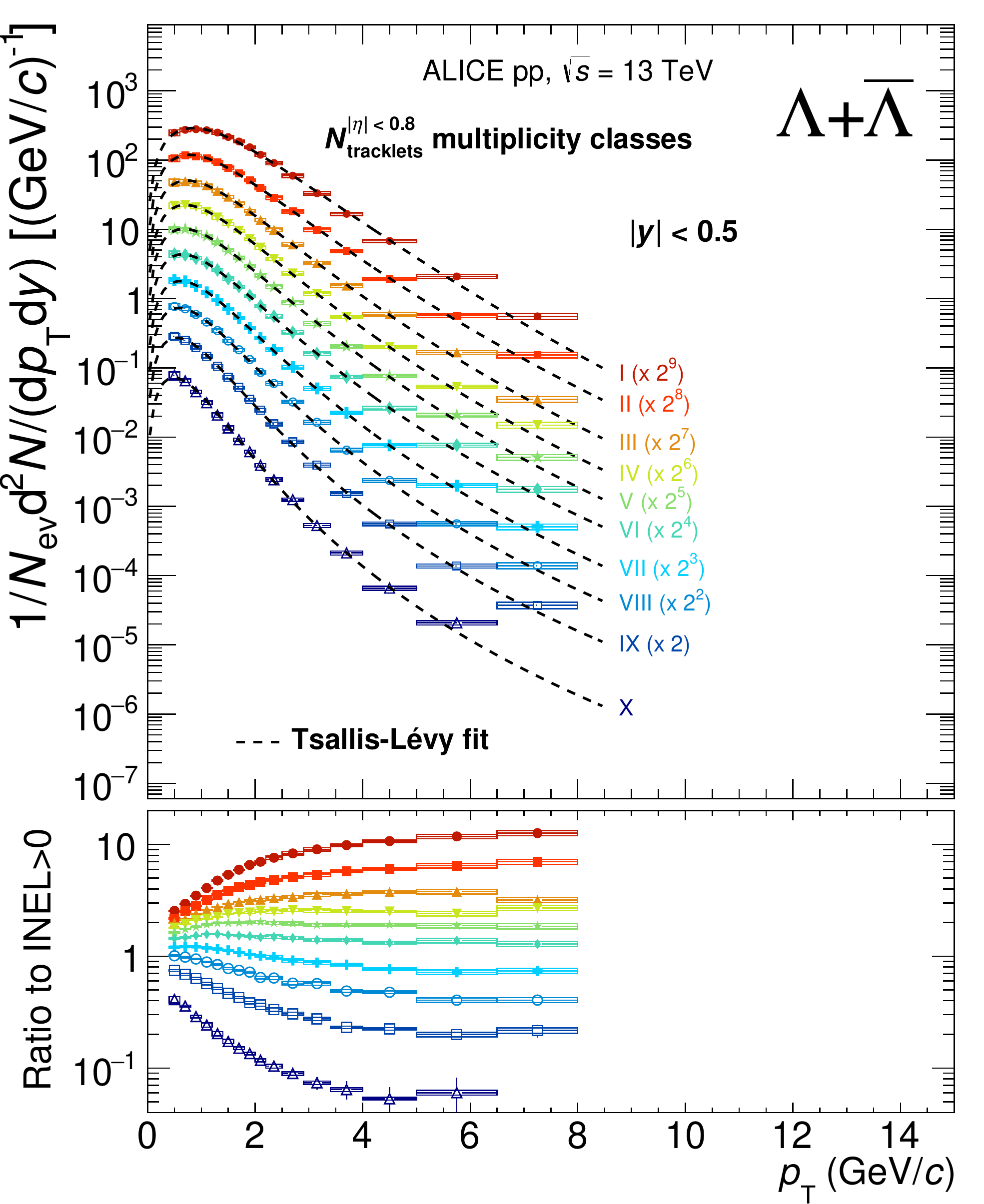}
    \includegraphics[width=0.47\textwidth]{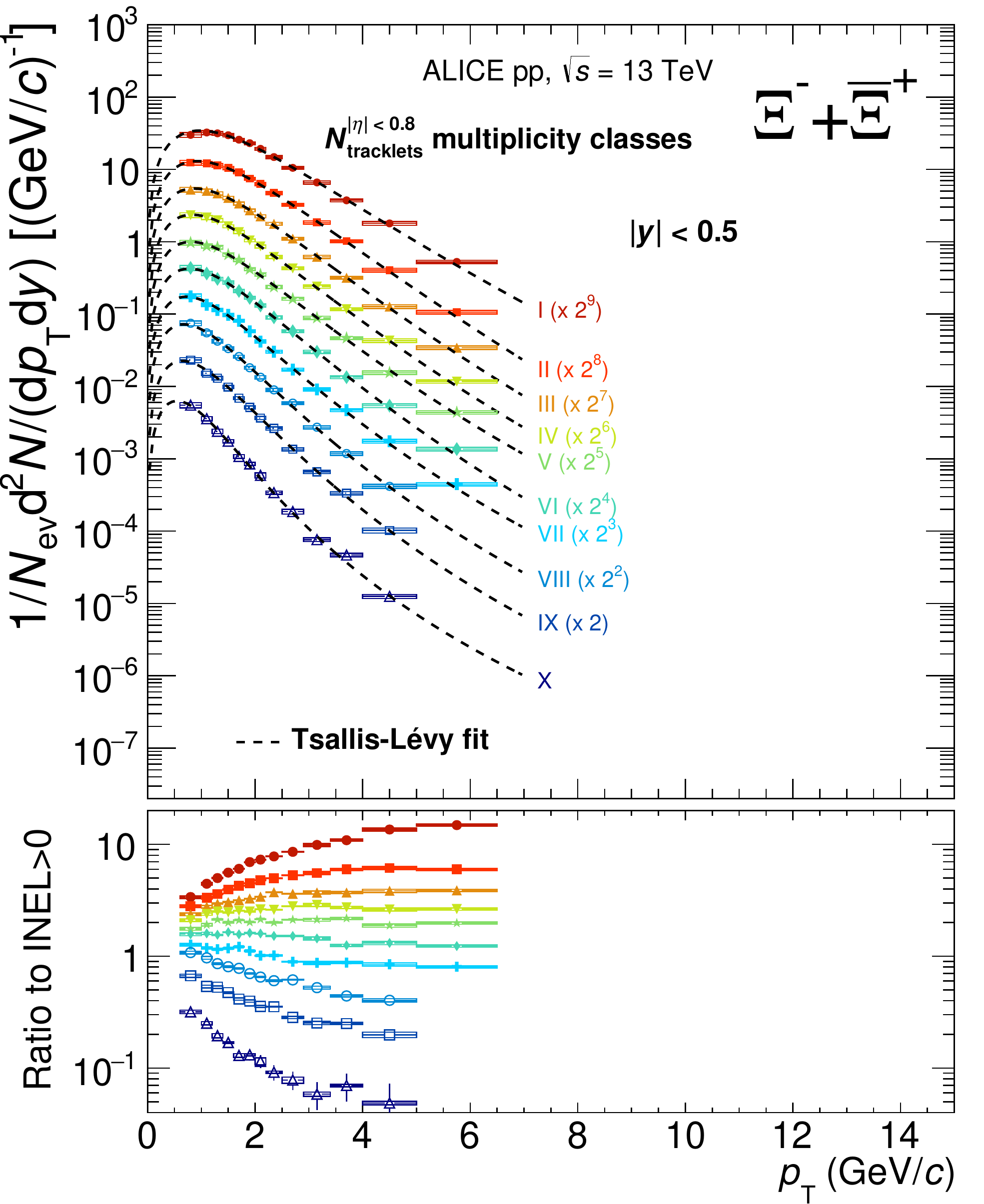}
    \includegraphics[width=0.47\textwidth]{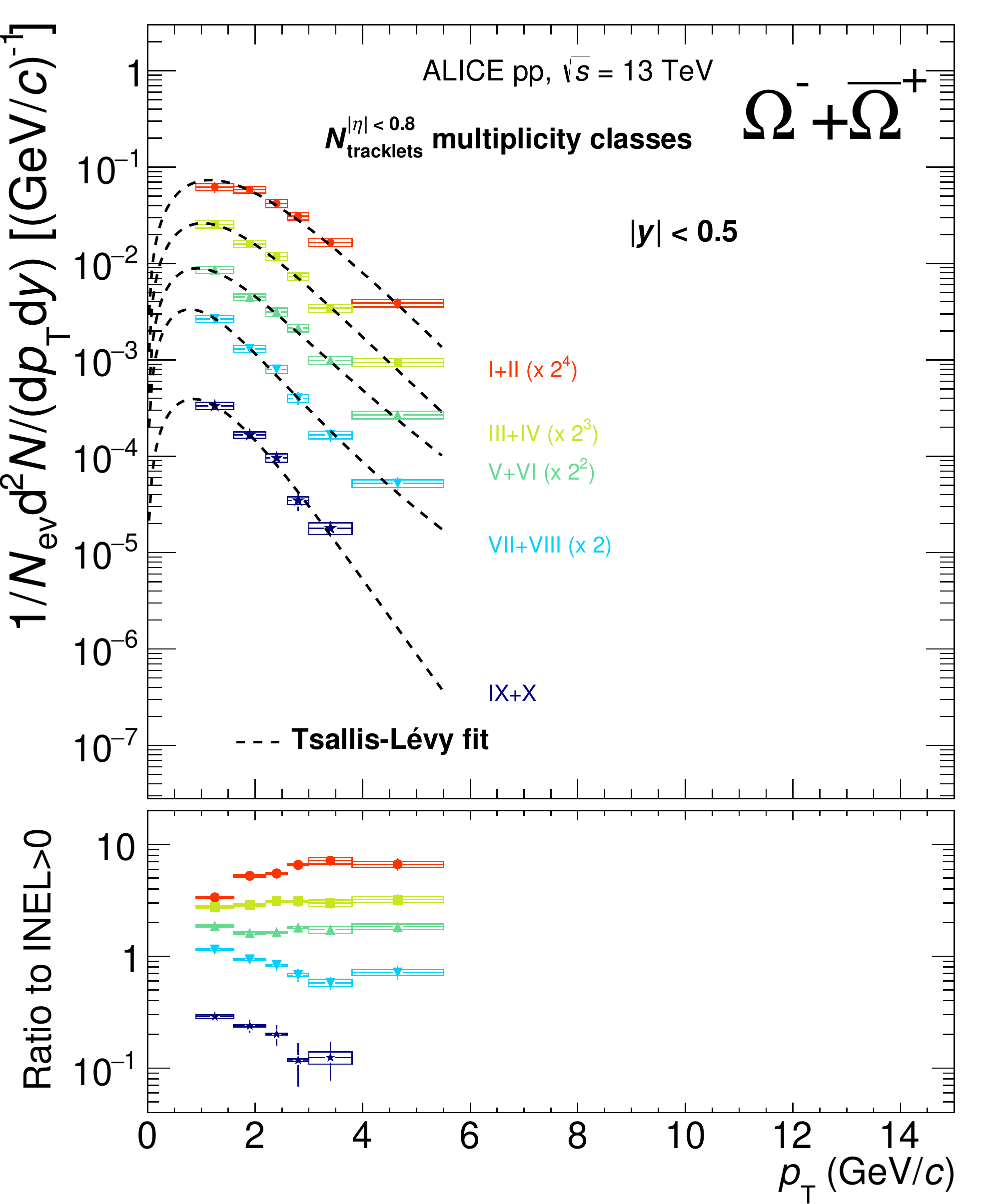}
  \end{center}
  \caption{\label{fig:spectra-tracklets08}Transverse momentum distribution of \pKzero, \sLambda, \sXi, and \sOmega, for multiplicity classes selected using the tracklets in \etaless{0.8}. Statistical and
total systematic uncertainties are shown by error bars and boxes, respectively. In the bottom panels ratios of multiplicity dependent spectra to INEL $>$ 0 are shown.
The systematic uncertainties on the ratios are obtained by considering only contributions uncorrelated across multiplicity. The spectra are scaled
by different factors to improve the visibility. The dashed curves represent Tsallis-L\'evy fits to the measured spectra. }
\end{figure}

\begin{figure}
  \begin{center}
    \includegraphics[width=0.47\textwidth]{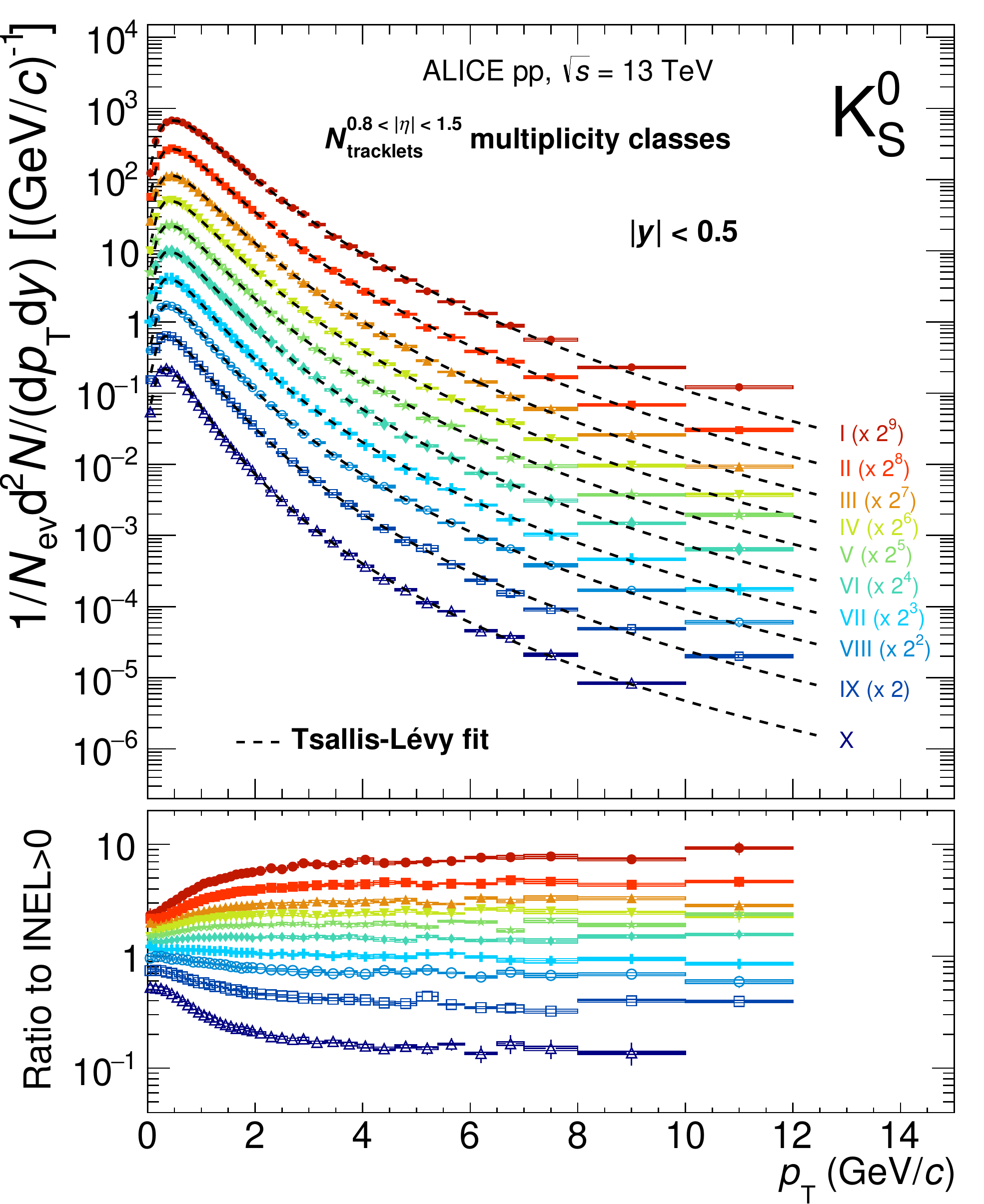}
    \includegraphics[width=0.47\textwidth]{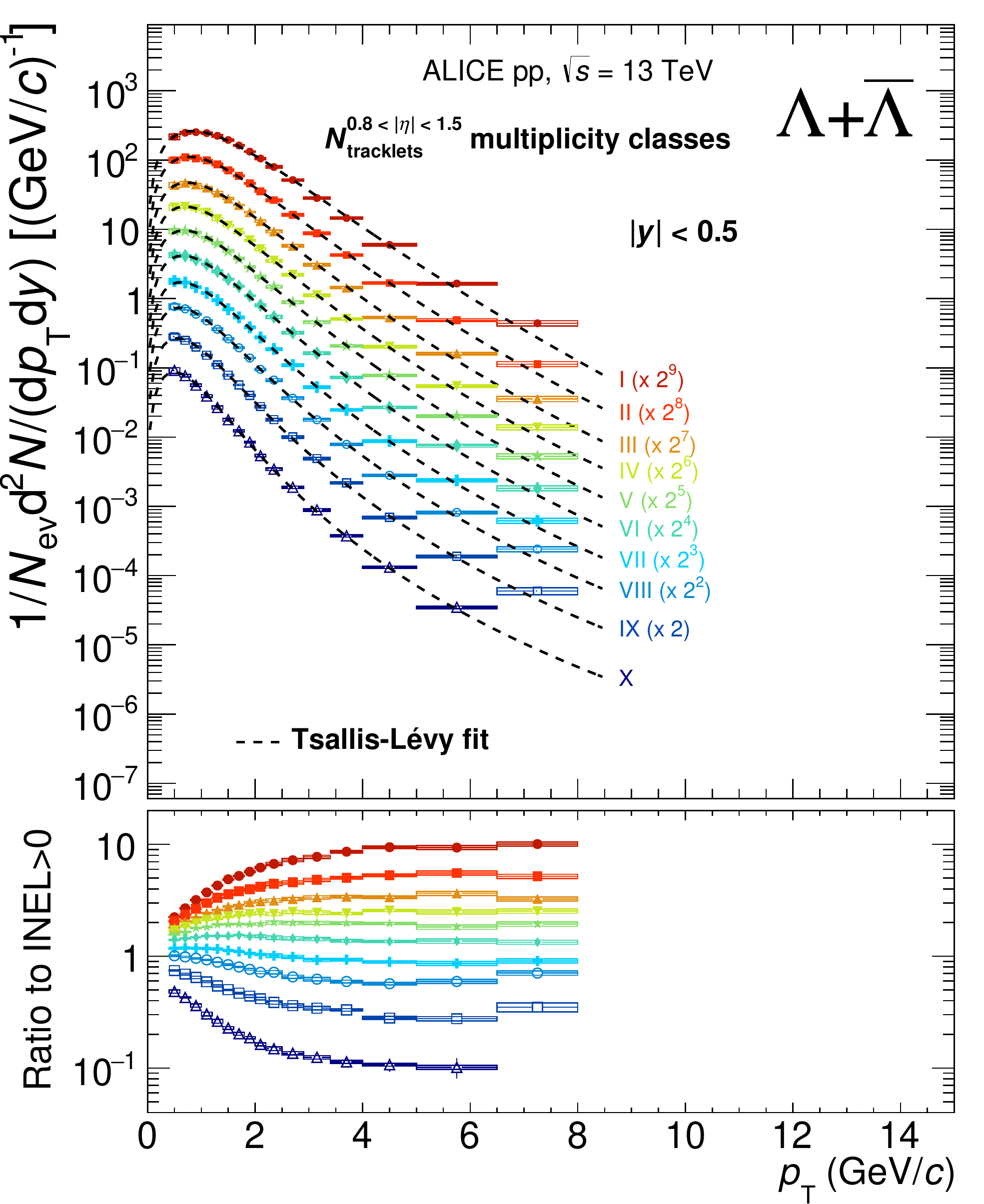}
    \includegraphics[width=0.47\textwidth]{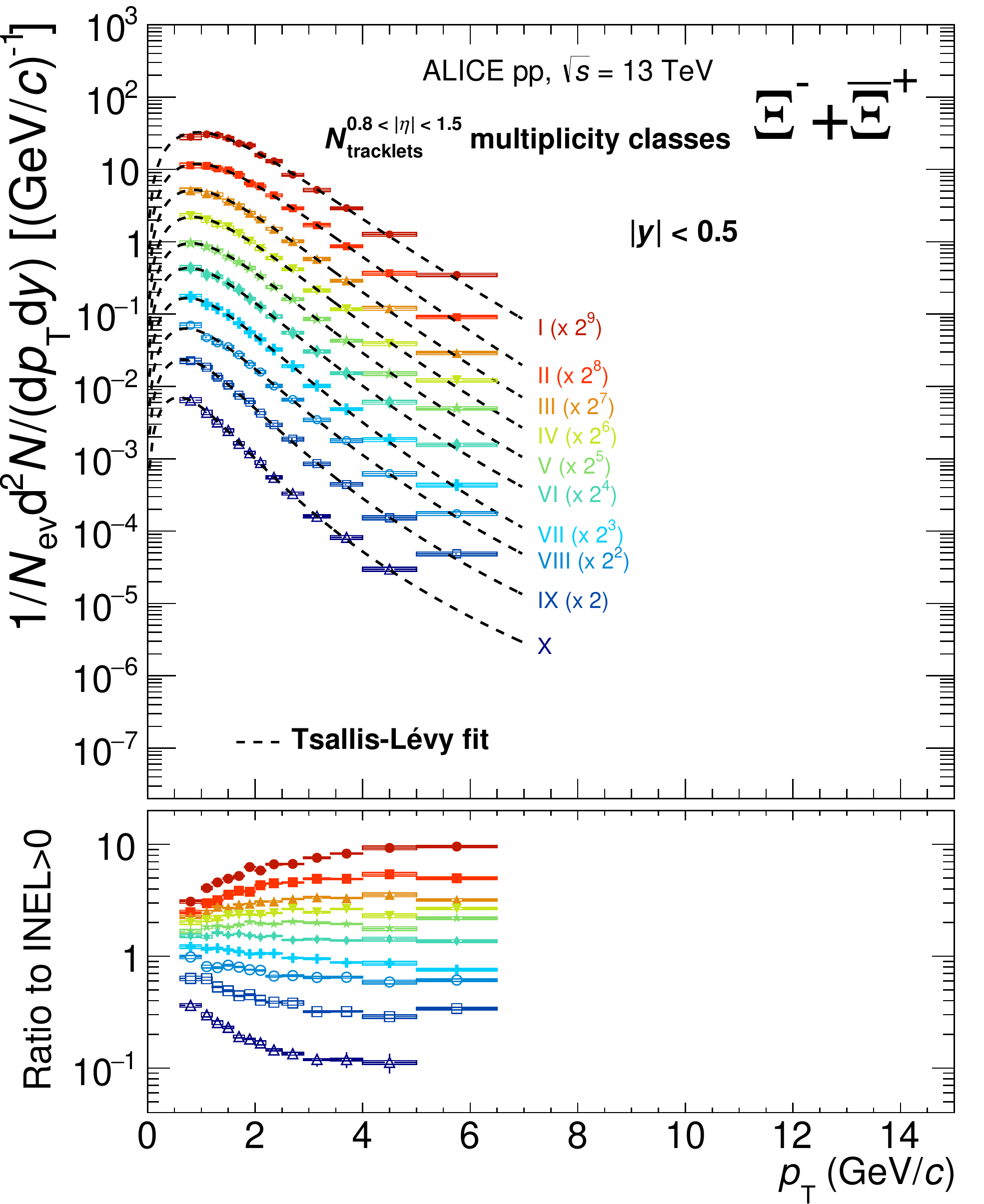}
    \includegraphics[width=0.47\textwidth]{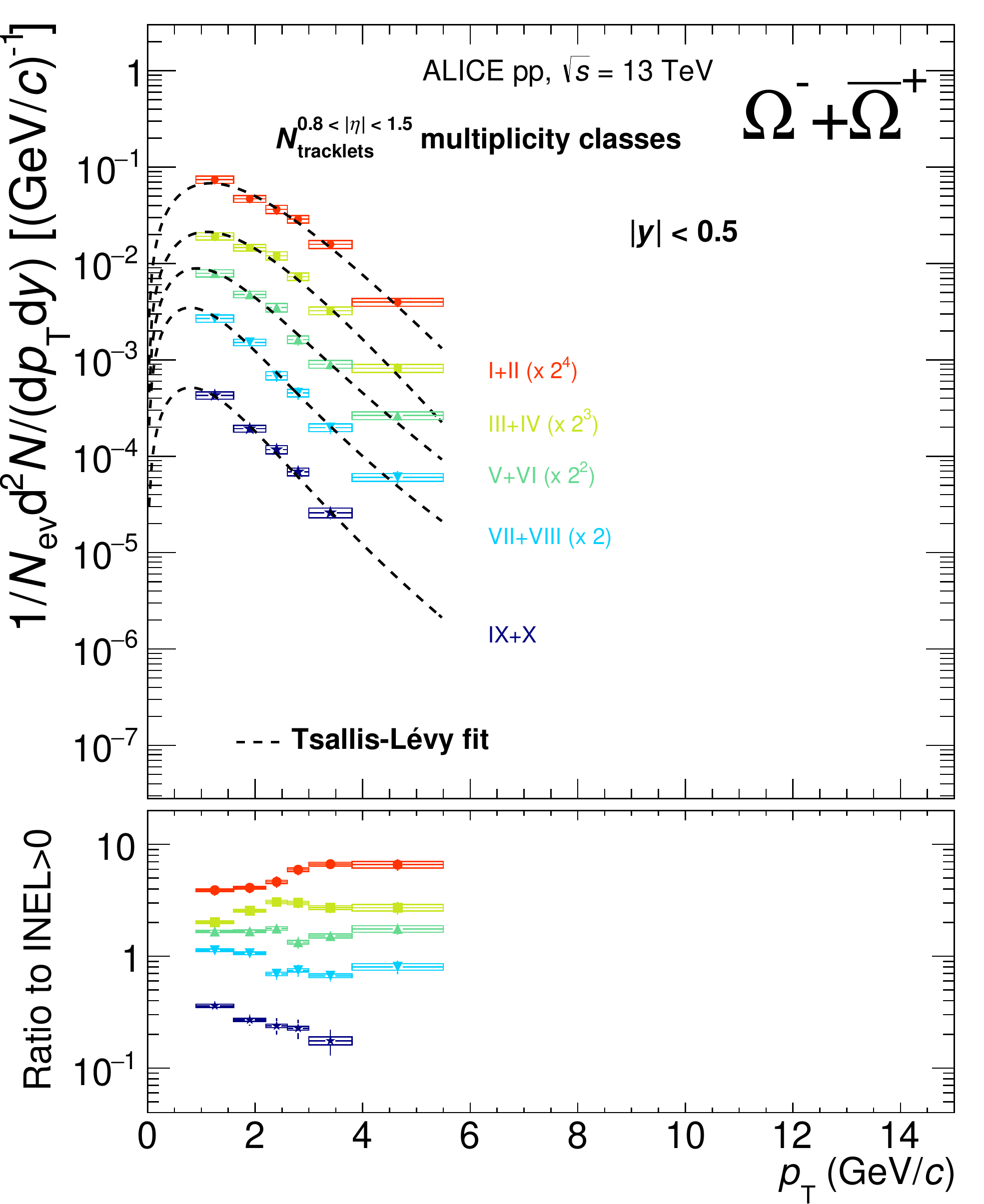}
  \end{center}
  \caption{\label{fig:spectra-tracklets0815}Transverse momentum distribution of \pKzero, \sLambda, \sXi, and \sOmega, for multiplicity classes selected using the tracklets  in $0.8 < \left| \eta \right| < 1.5$. Statistical and total
systematic uncertainties are shown by error bars and boxes, respectively. In the bottom panels ratios of multiplicity dependent spectra to INEL $>$ 0 are shown.
The systematic uncertainties on the ratios are obtained by considering only contributions uncorrelated across multiplicity. The spectra are scaled
by different factors to improve the visibility. The dashed curves represent Tsallis-L\'evy fits to the measured spectra. }
\end{figure}


\begin{figure}
  \begin{center}
  \includegraphics[width=0.84\textwidth]{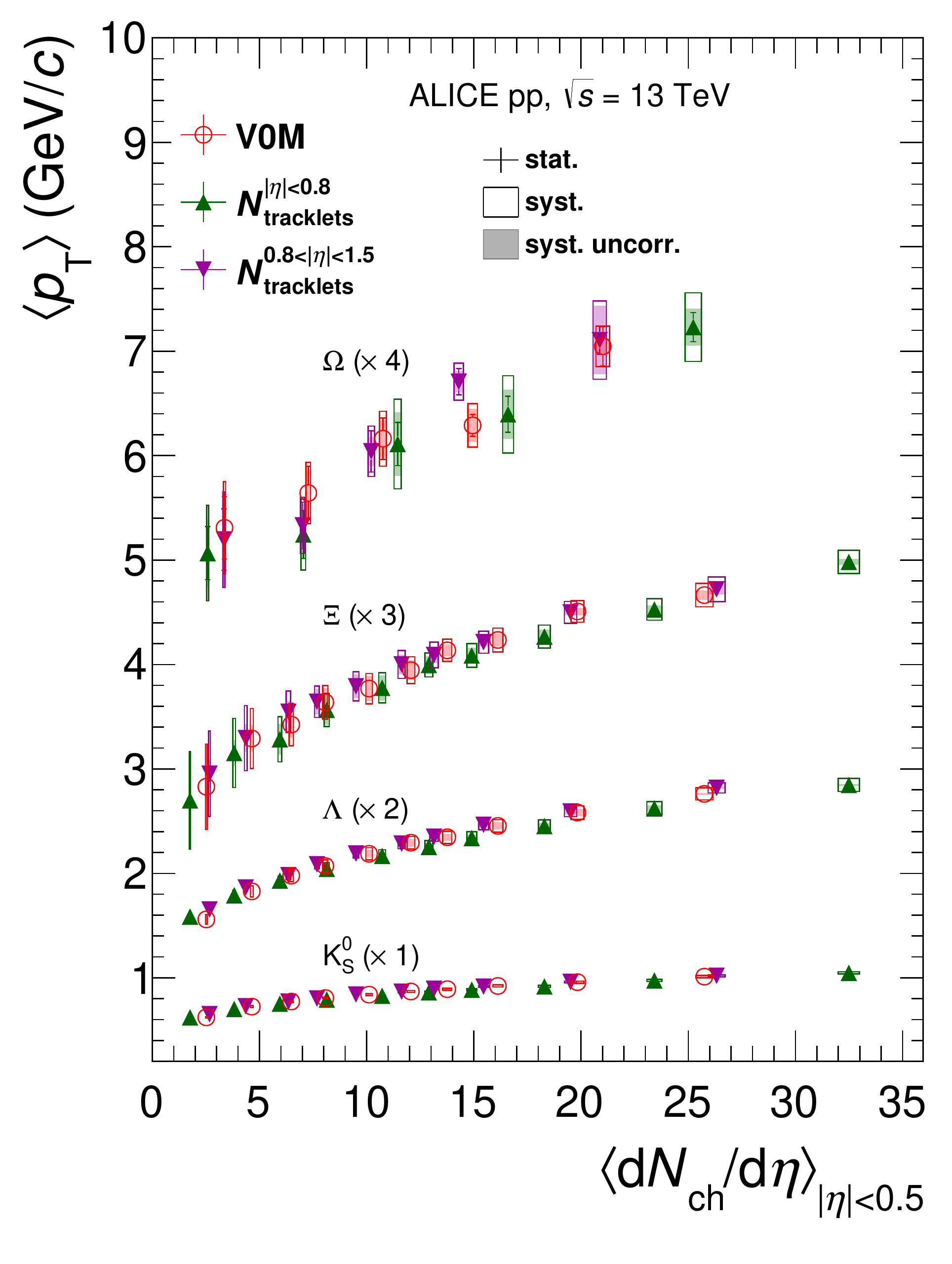}
  \end{center}
  \caption{\label{fig:avpt-vs-mult}\avpT\ of \pKzero, \sLambda, \sXi, and \sOmega in multiplicity event classes selected according to different estimators (see text for details).
Statistical and systematic uncertainties are shown by error bars and empty boxes, respectively. Shadowed boxes represent uncertainties uncorrelated across multiplicity.}
\end{figure}

\begin{figure}
  \centering
    \includegraphics[width=0.90\textwidth]{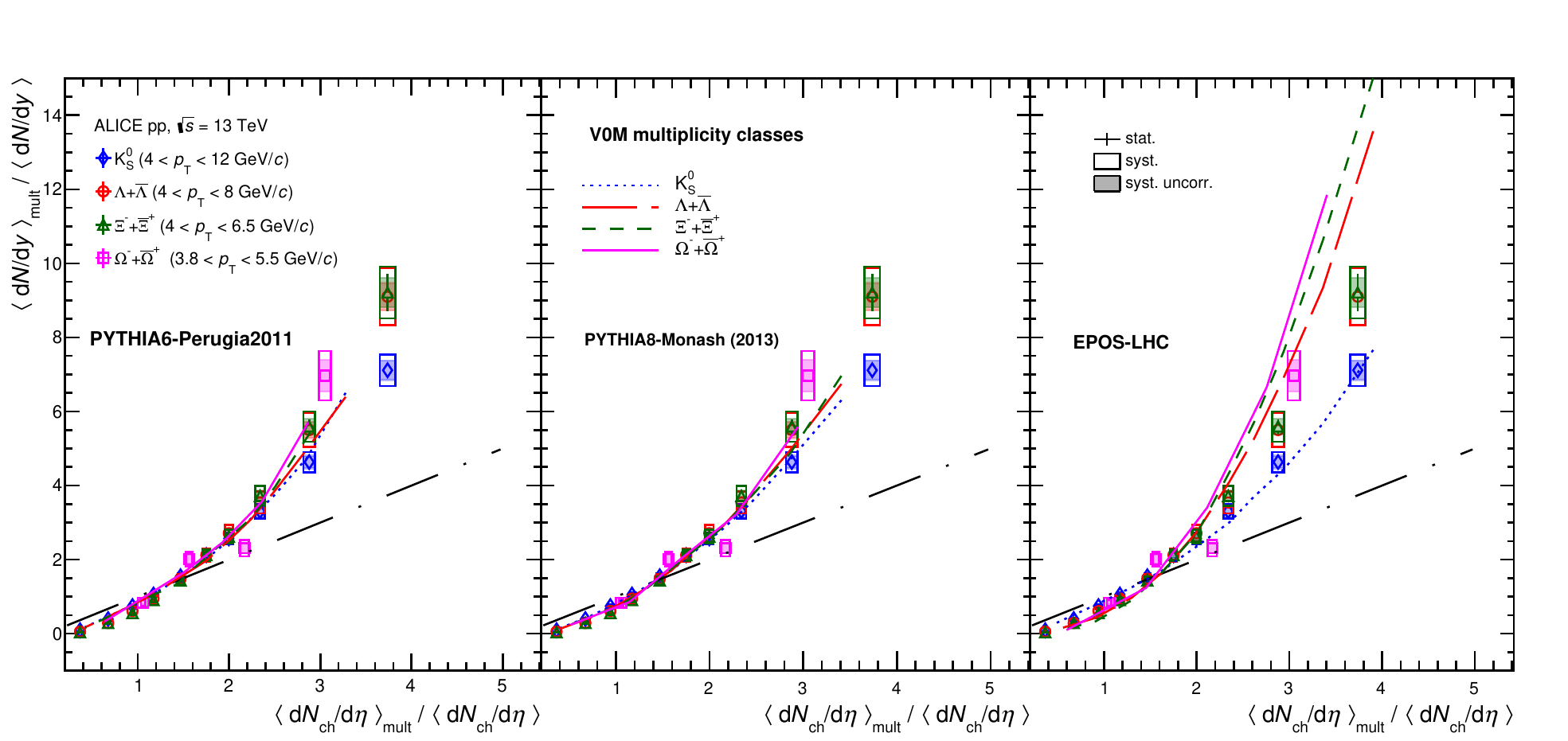} \\
    \includegraphics[width=0.90\textwidth]{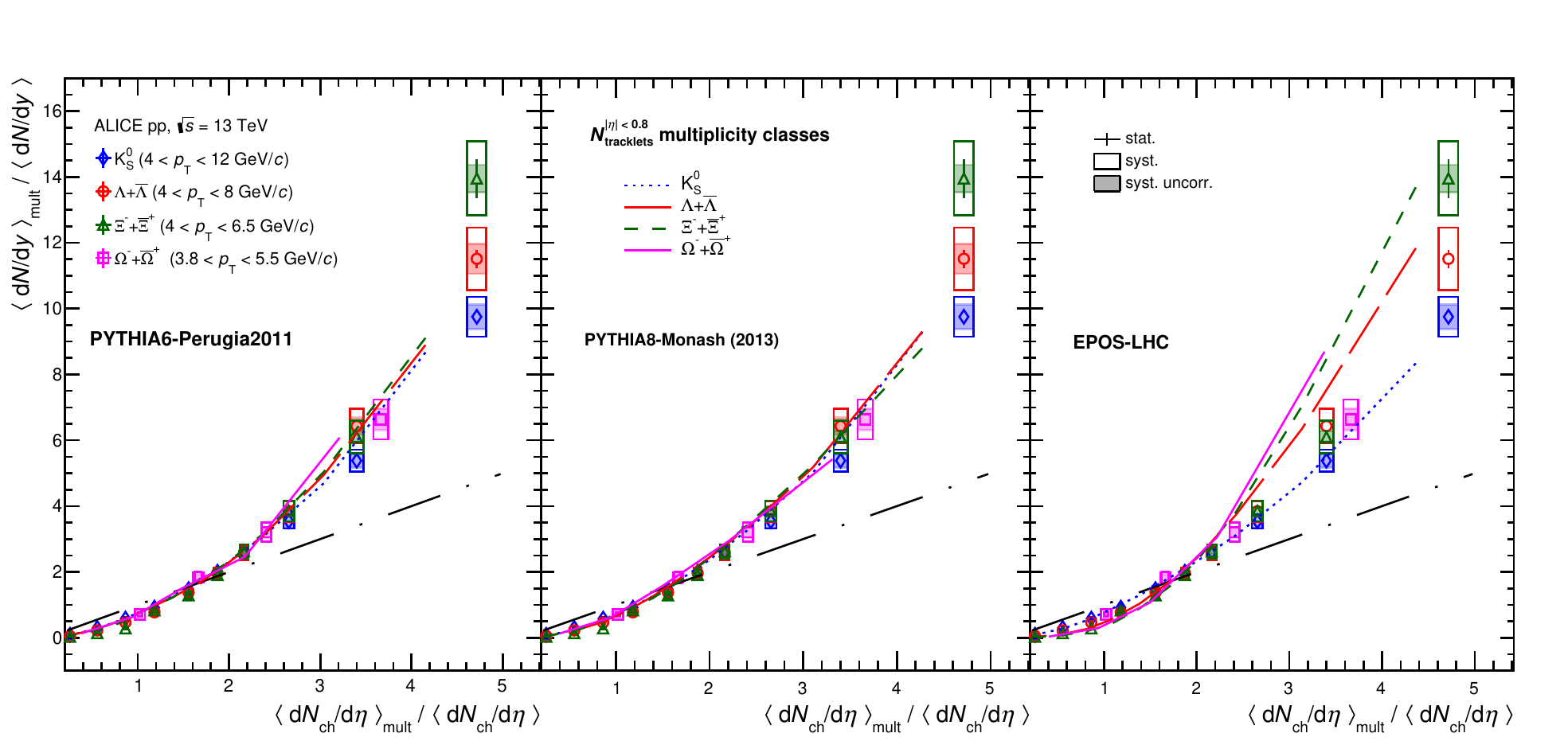} \\
    \includegraphics[width=0.90\textwidth]{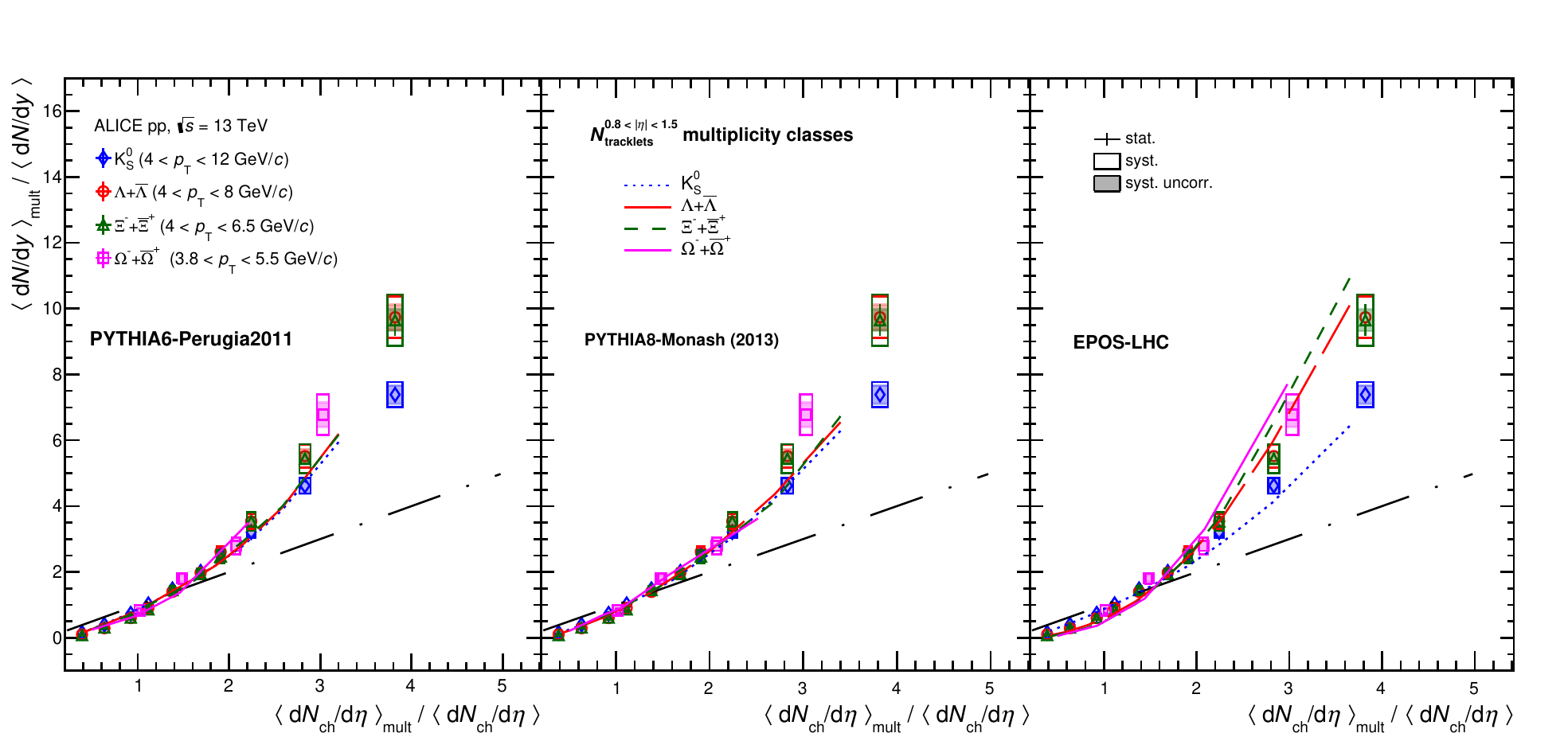}
  \caption{ Self-normalised yields of strange hadrons integrated for \pt $> 4$~\gevc\ (3.8 ~\gevc\ for the \sOmega) versus the self-normalised
    charged particle multiplicity quoted using different multiplicity estimators (see text for details). Statistical and systematic uncertainties are shown by error bars and empty boxes, respectively. Shadowed boxes represent uncertainties uncorrelated across multiplicity. Strange particle yields are compared to expectations from Monte Carlo models.}
    \label{fig:yields-high-pt}
\end{figure}

The left (right) panel of Fig.~\ref{fig:spactra-ratio-estimators} shows the ratio of the \pt\ spectra obtained with different estimators for 
multiplicity classes with comparable average $\dNdeta$ value ($\langle \dNdeta \rangle \simeq 26$);
the numerator refers to the \NtrkEtaOut\ (\NtrkEtaIn) estimator while the denominator refers to the V0M estimator. These selections correspond to the highest multiplicity class studied for the V0M and \NtrkEtaOut\ estimators. The  \NtrkEtaIn estimator, on the other hand, was re-adjusted to lead to the same \dNdeta\ value. These classes correspond to approximately 1\%, 1\% and 4\% of the \inelgtzero\ cross section for V0M, \NtrkEtaIn\ and \NtrkEtaOut, respectively. 
It is seen that the spectra are identical within uncertainties for the V0M and \NtrkEtaOut\ estimators. The comparison between the V0M and \NtrkEtaIn\ estimators shows that the bias introduced by the latter estimator does not depend on the particle species  and is more pronounced at low and intermediate \pt, although the uncertainties are large.


The \pt-integrated yields of strange hadrons are shown in Fig.~\ref{fig:yields-vs-mult} for the three estimators considered in this paper. For reference, Fig.~\ref{fig:yields-vs-mult-no-extrap} shows the results integrated in the measured \pt\ range with no extrapolation. These are characterised by a smaller uncertainty and can be useful for Monte Carlo tuning. In the rest of this section, we focus on the discussion of the fully extrapolated yields.
The results obtained with the V0M and \NtrkEtaOut estimators follow a similar linear trend, while the results obtained with \NtrkEtaIn\ tend to saturate, showing a lower \dNdy\ for a similar high-multiplicity class.
In order to gain insight on this difference, a generator-level PYTHIA (tune Perugia 2011) simulation study was performed. The abundance of strange hadrons in \yless{0.5} was studied as a function of the estimated mid-rapidity charged particle multiplicity for several event classes, selected using charged primary particles measured in the $\eta$ ranges corresponding to the experimental estimators presented in this paper:
this generator level study confirms the trend observed in the data, which can be understood in terms of a selection bias sensitive to fluctuations in particles yields. Indeed, an estimator based on charged tracks enhances charged primary particles over neutral particles or secondaries. If the multiplicity estimator and the observable under study are measured in the same $\eta$ region one expects primary charged-particles to be enhanced with respect to strange hadrons, which explains the saturation of yields observed in Fig.~\ref{fig:yields-vs-mult} for the \NtrkEtaIn\ estimator. As a further check of this interpretation, Fig.~\ref{fig:strangenes-vs-strangeness}
shows the correlations between the yields of different strange hadrons for the multiplicity estimators studied in this paper. The trend is linear, and similar for all estimators, confirming that the selection bias on primary charged-particles is stronger than that on strange hadrons. 


\begin{figure}
  \begin{center}
    \includegraphics[width=0.48\textwidth]{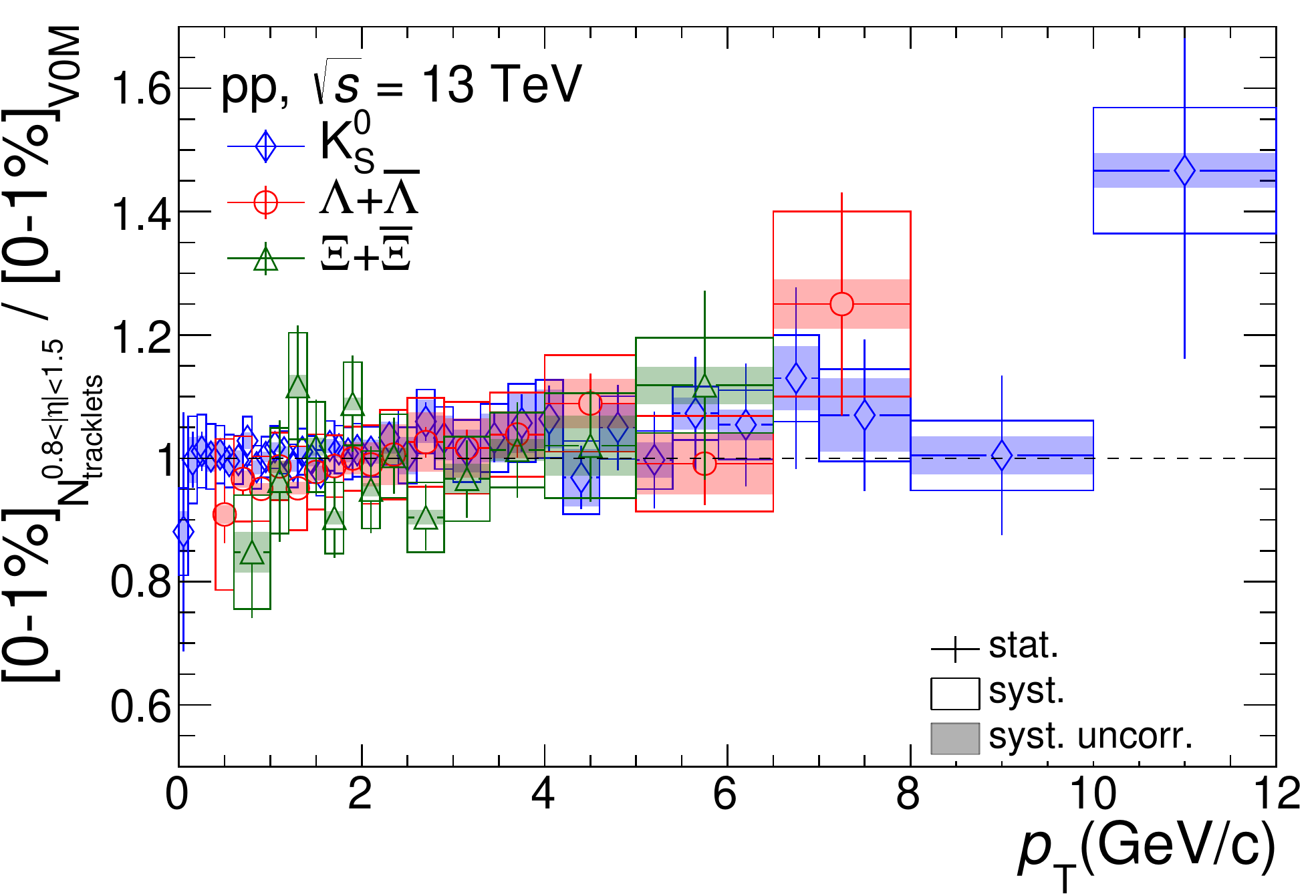}
    \includegraphics[width=0.48\textwidth]{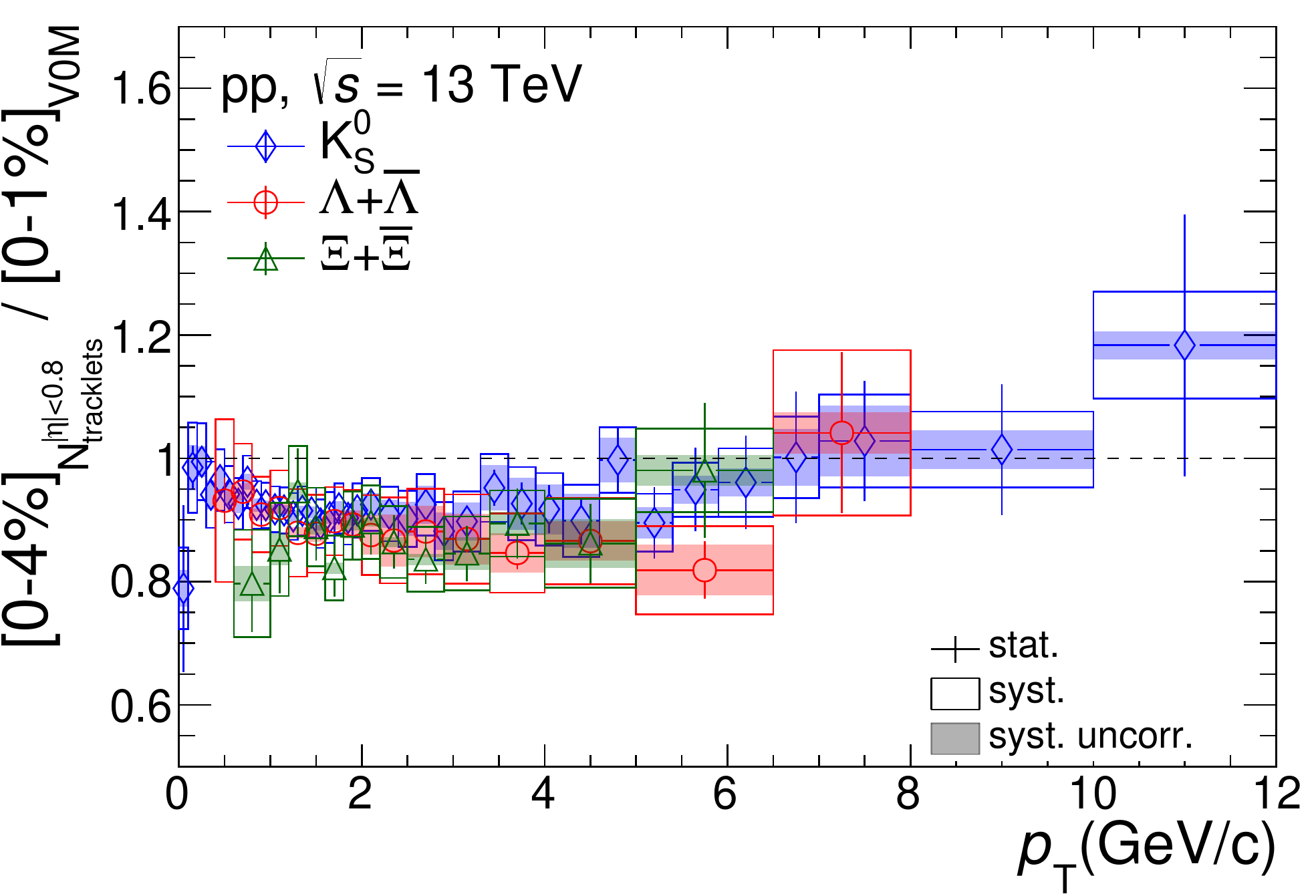}
  \end{center}
  \caption{\label{fig:spactra-ratio-estimators} Ratio of \pt\ spectra of \pKzero, \sLambda and \sXi obtained with two different estimators for 
  multiplicity classes with comparable average $\dNdeta$ value ($\langle \dNdeta \rangle \simeq 26$);
the numerator does refer to the \NtrkEtaOut\ (\NtrkEtaIn) estimator while the denominator refers to the V0M estimator. Statistical and systematic uncertainties are shown by error bars and empty boxes, respectively. Shadowed boxes represent uncertainties uncorrelated across multiplicity.}
\end{figure}

\begin{figure}
  \begin{center}
    \includegraphics[width=0.44\textwidth]{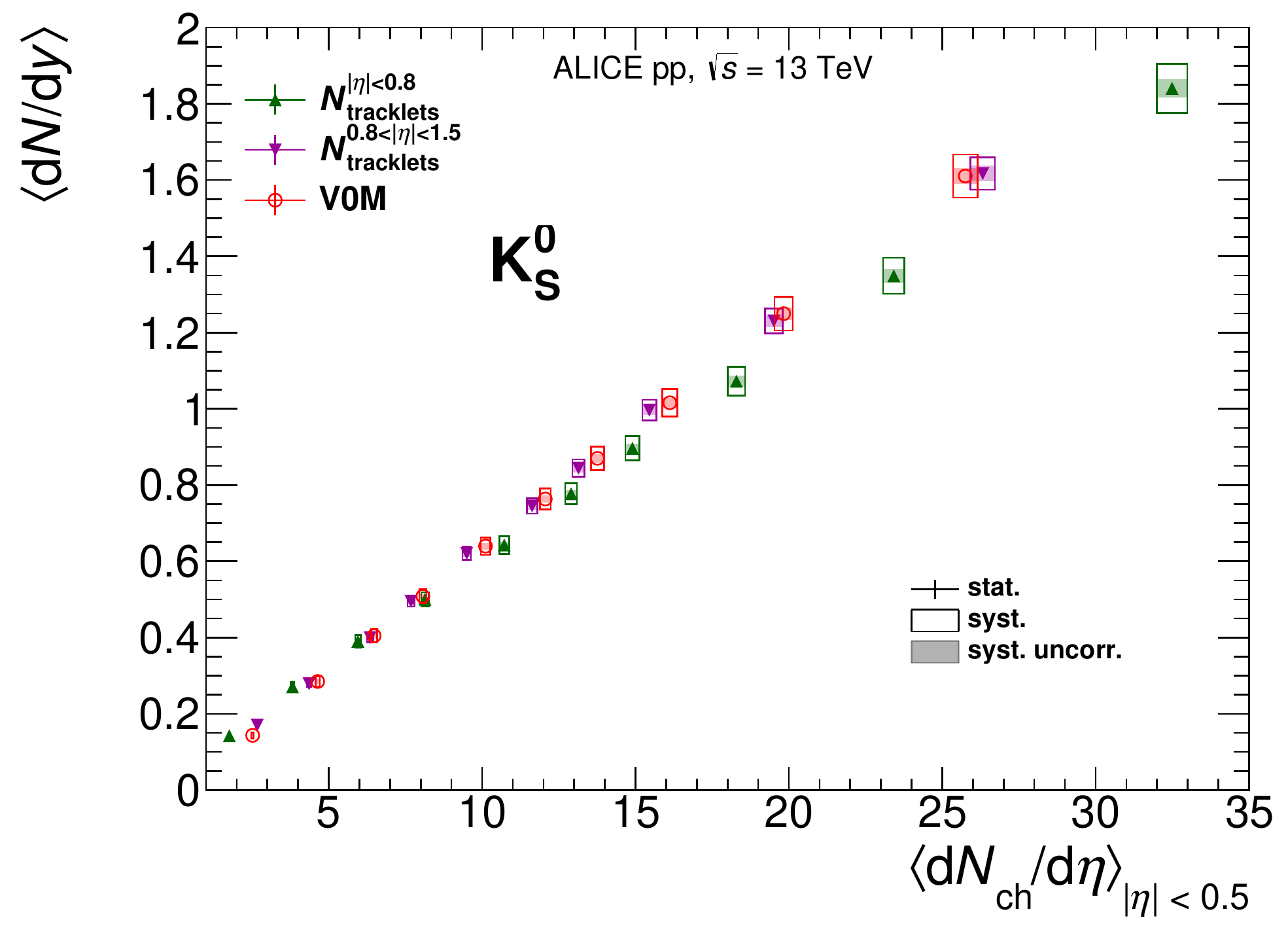}
    \includegraphics[width=0.44\textwidth]{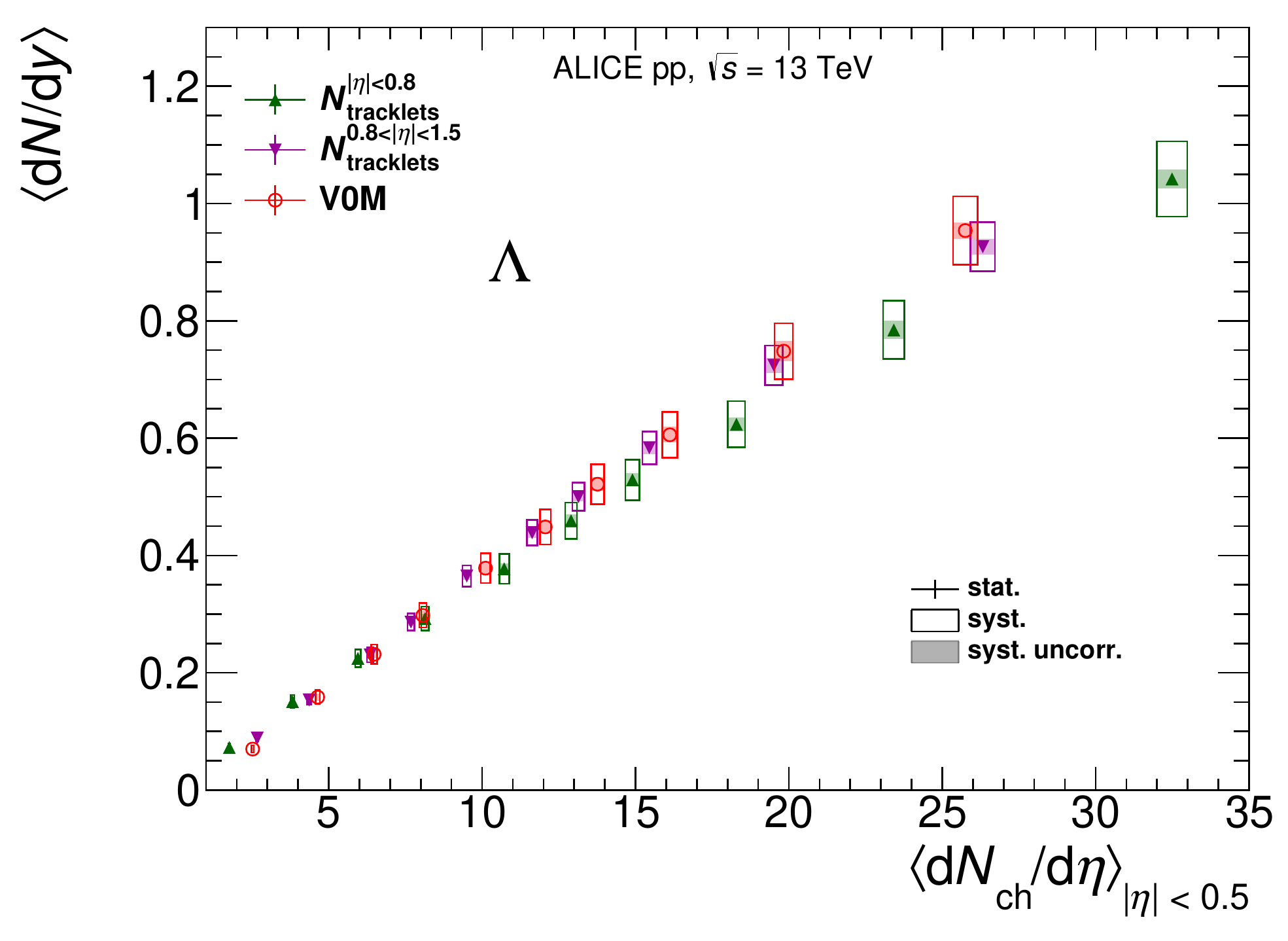}
    \includegraphics[width=0.44\textwidth]{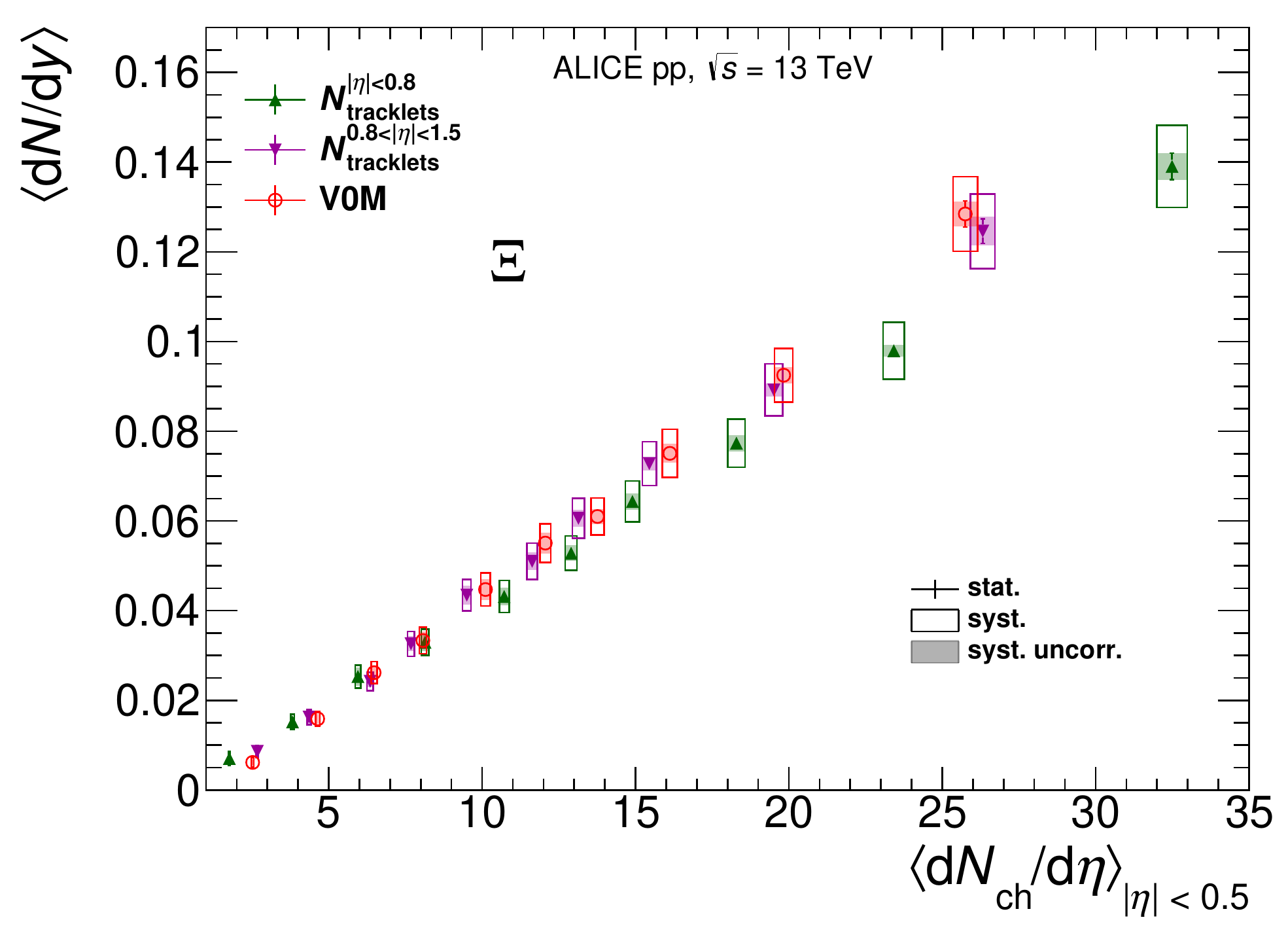}
    \includegraphics[width=0.44\textwidth]{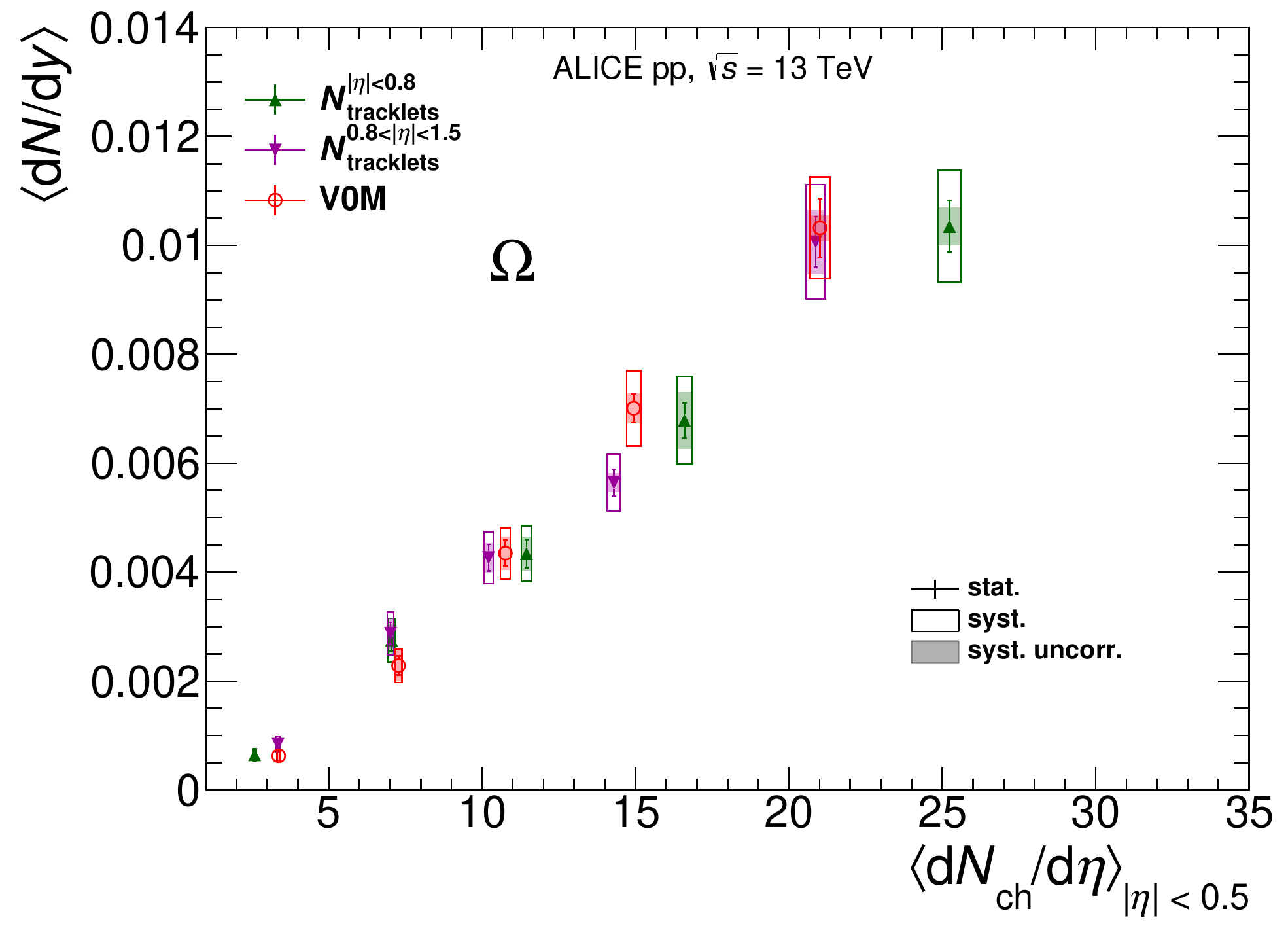}
  \end{center}
  \caption{\label{fig:yields-vs-mult}\dndy\ (integrated over the full \pt\ range) as a function of multiplicity for different strange particle species reported for different multiplicity classes (see text for details).
Error bars and boxes represent statistical and total systematic uncertainties, respectively. The bin-to-bin systematic uncertainties are shown by shadowed boxes.}
\end{figure}

\begin{figure}
  \begin{center}
    \includegraphics[width=0.44\textwidth]{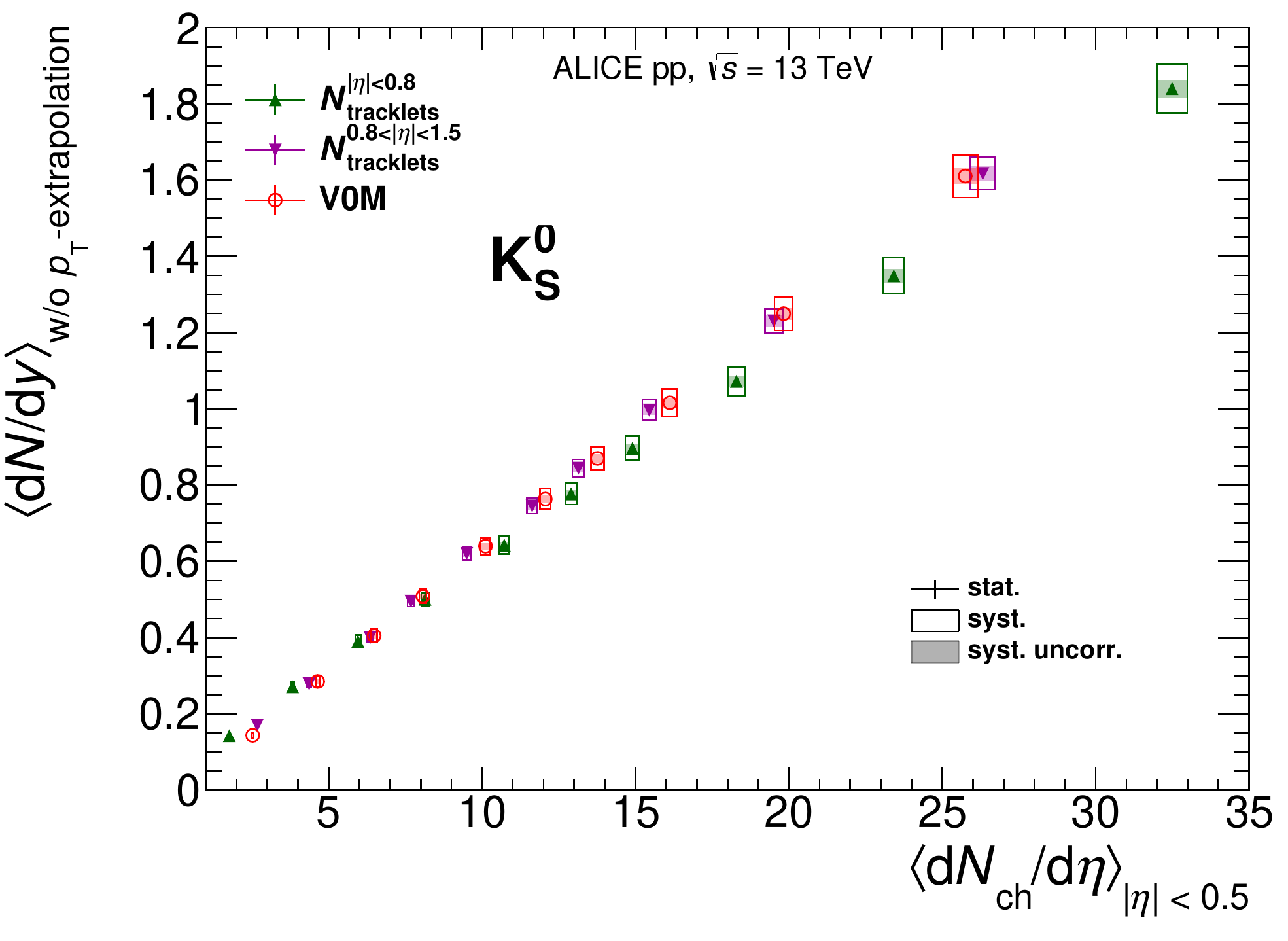}
    \includegraphics[width=0.44\textwidth]{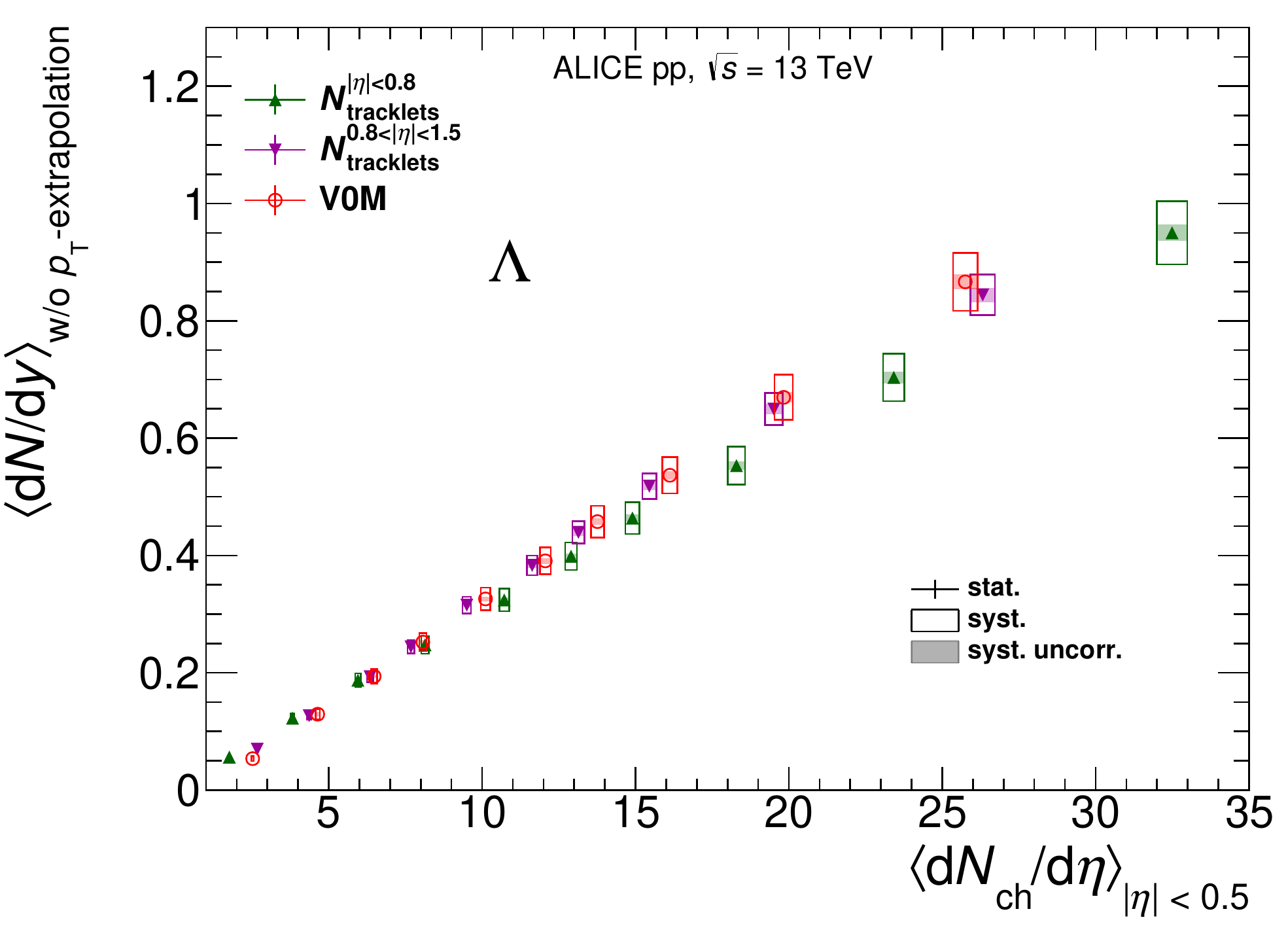}
    \includegraphics[width=0.44\textwidth]{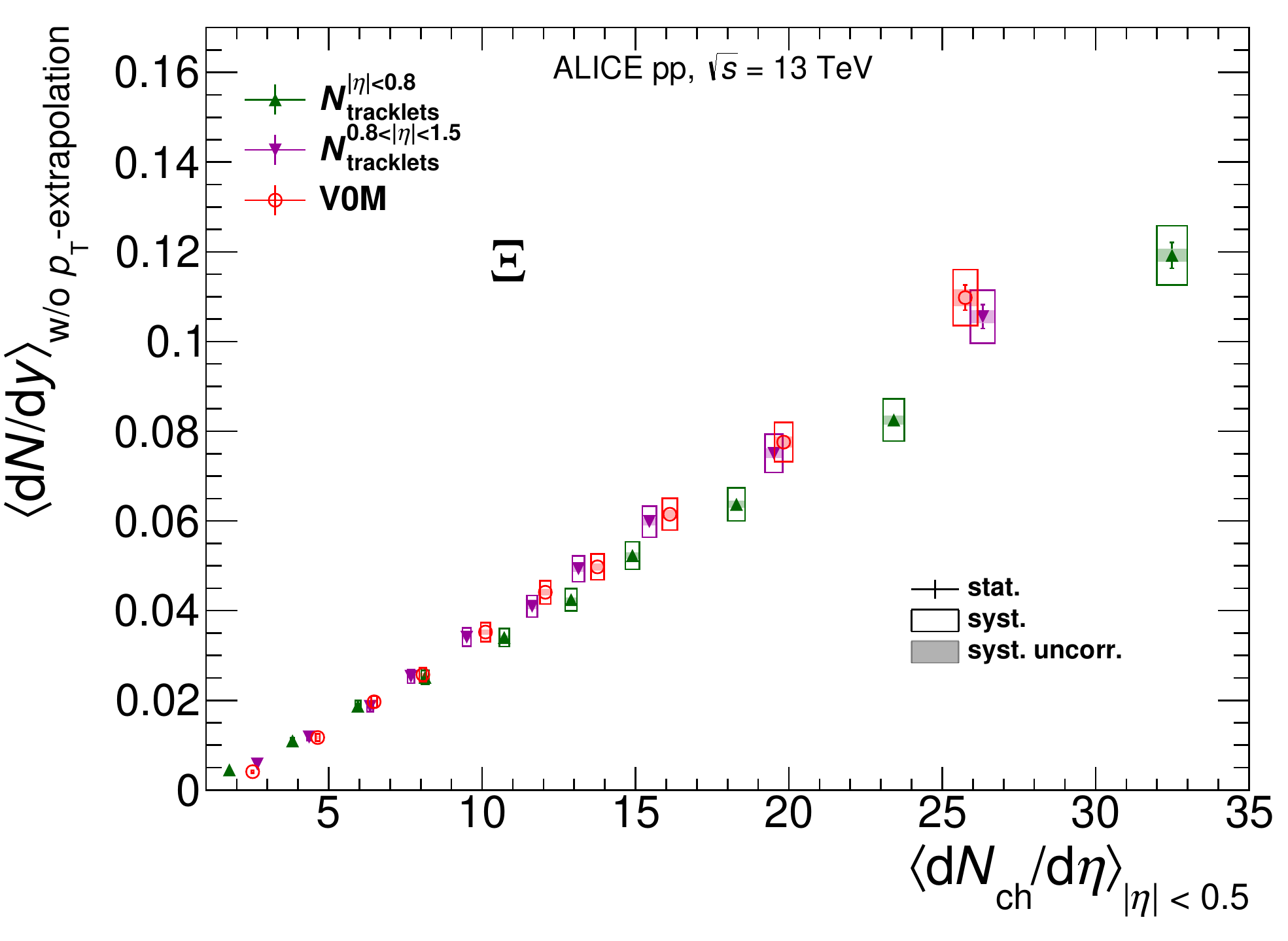}
    \includegraphics[width=0.44\textwidth]{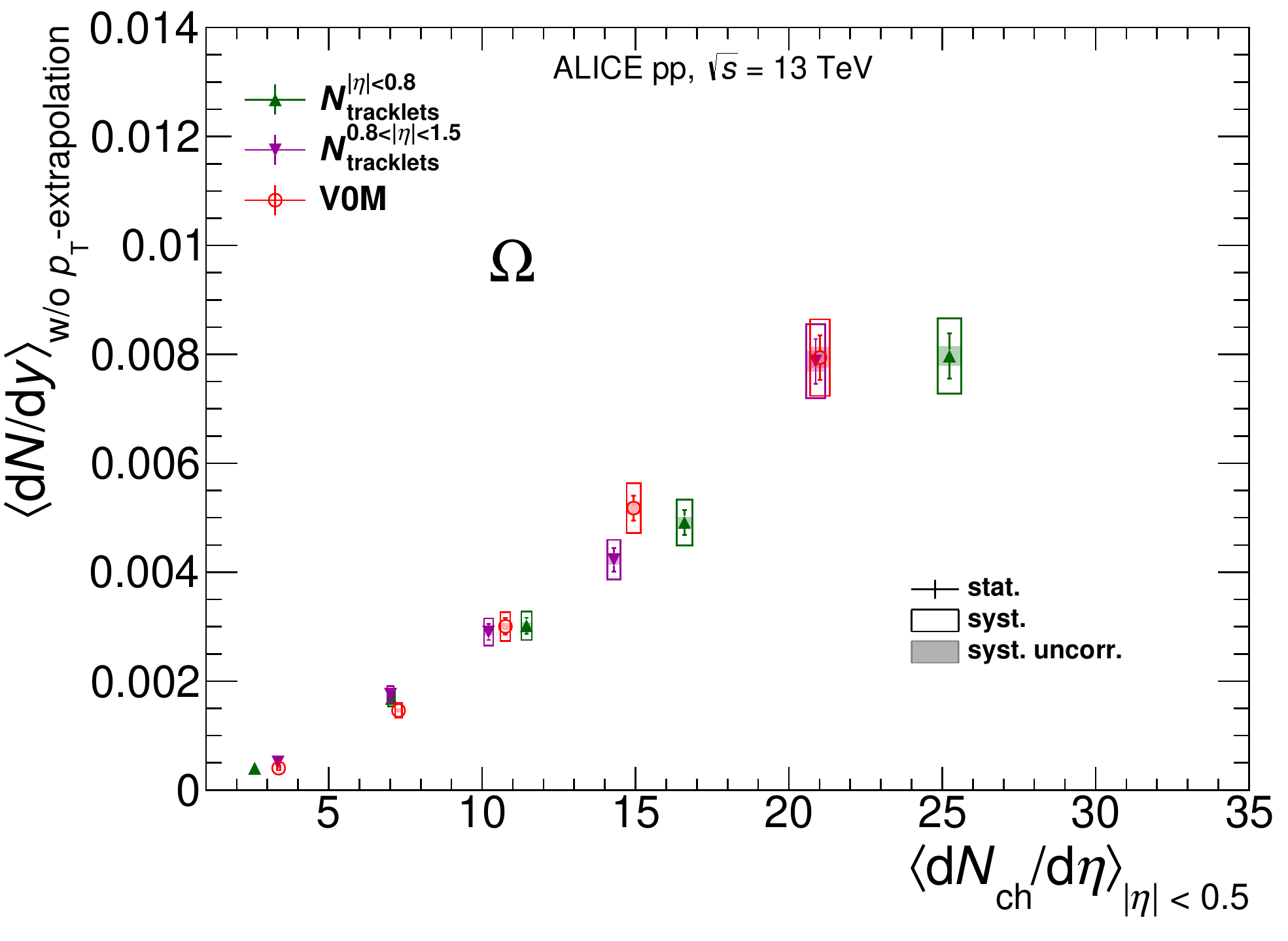}
  \end{center}
  \caption{\label{fig:yields-vs-mult-no-extrap}\dndy\ (integrated over the measured \pt\ ranges 0-12, 0.4-8, 0.6-6.5 and 0.9-5.5~\gevc\ for \pKzero, \pLambda, \pXi and \pOmega, respectively) as a function of multiplicity for different strange particle species reported for different multiplicity classes (see text for details).
Error bars and boxes represent statistical and total systematic uncertainties, respectively. The bin-to-bin systematic uncertainties are shown by shadowed boxes.}
\end{figure}

\begin{figure}
  \begin{center}
     \includegraphics[width=0.45\textwidth]{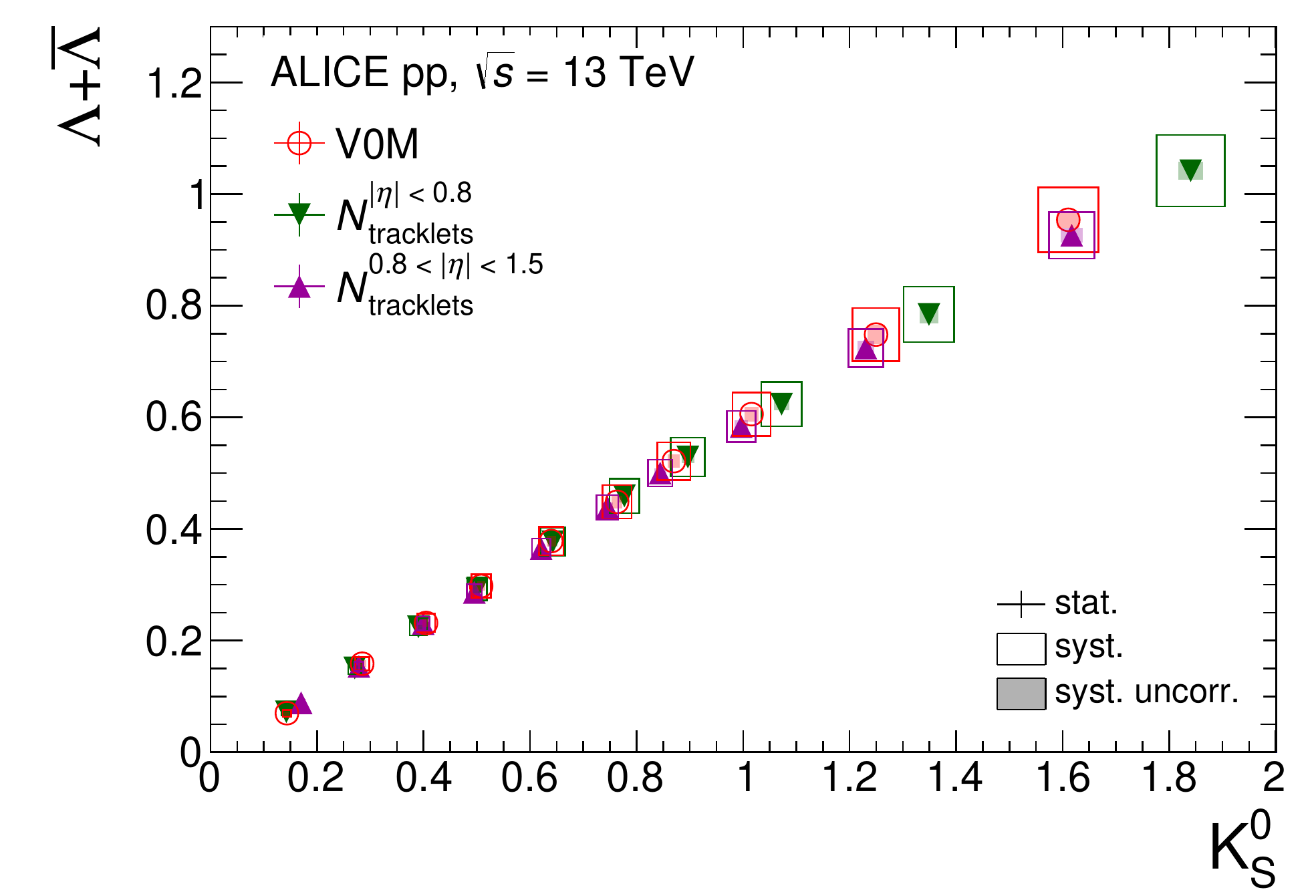}
    \includegraphics[width=0.45\textwidth]{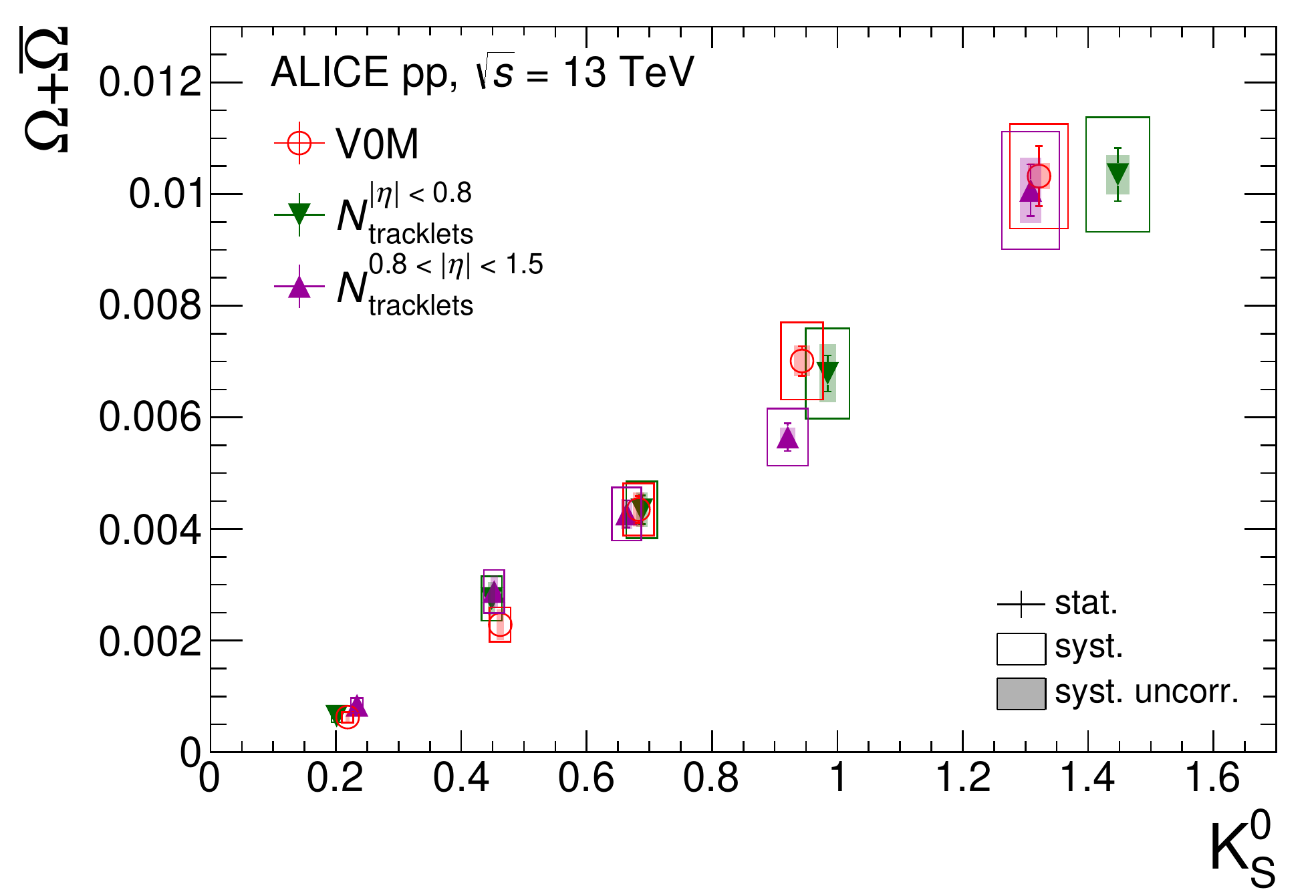}
    \includegraphics[width=0.45\textwidth]{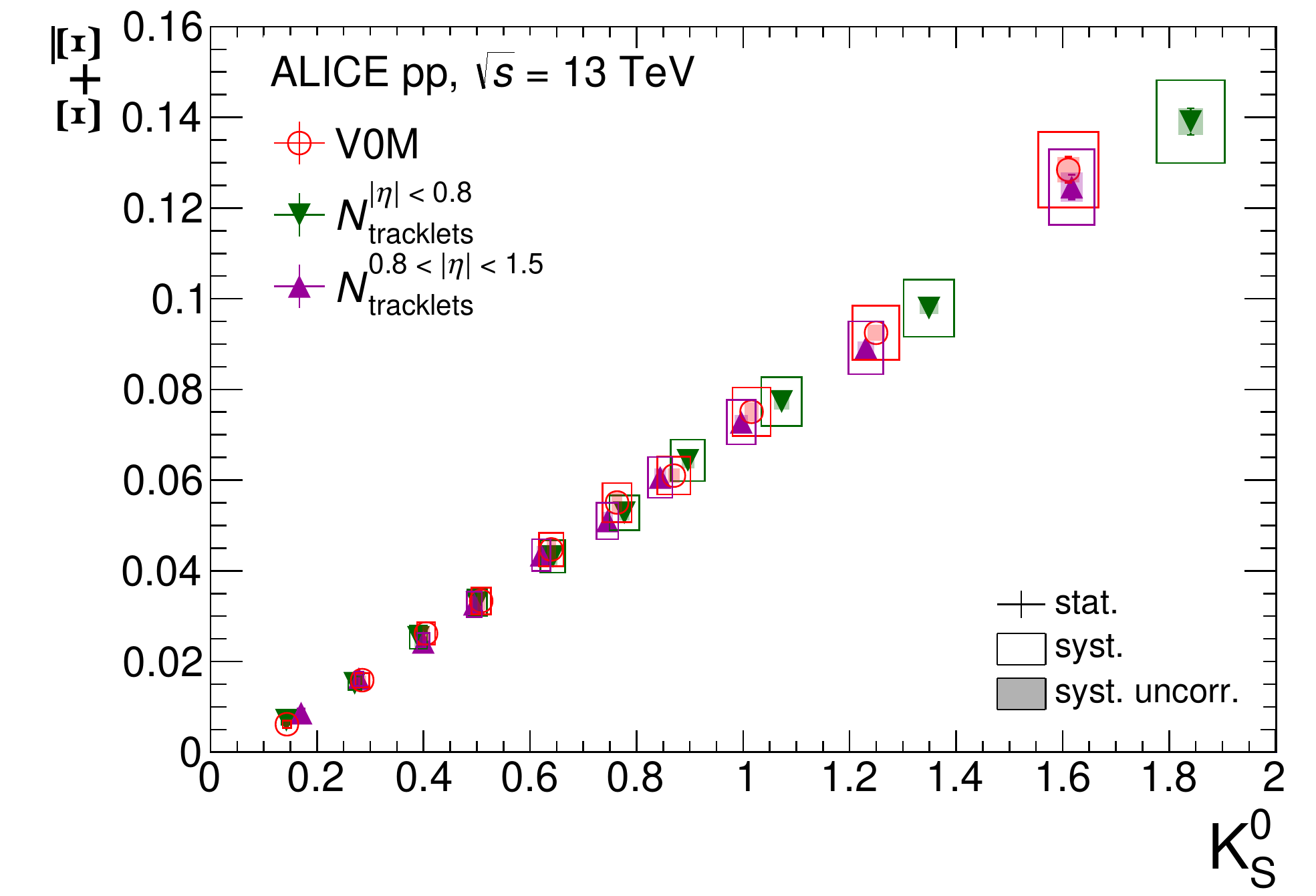}
    \includegraphics[width=0.45\textwidth]{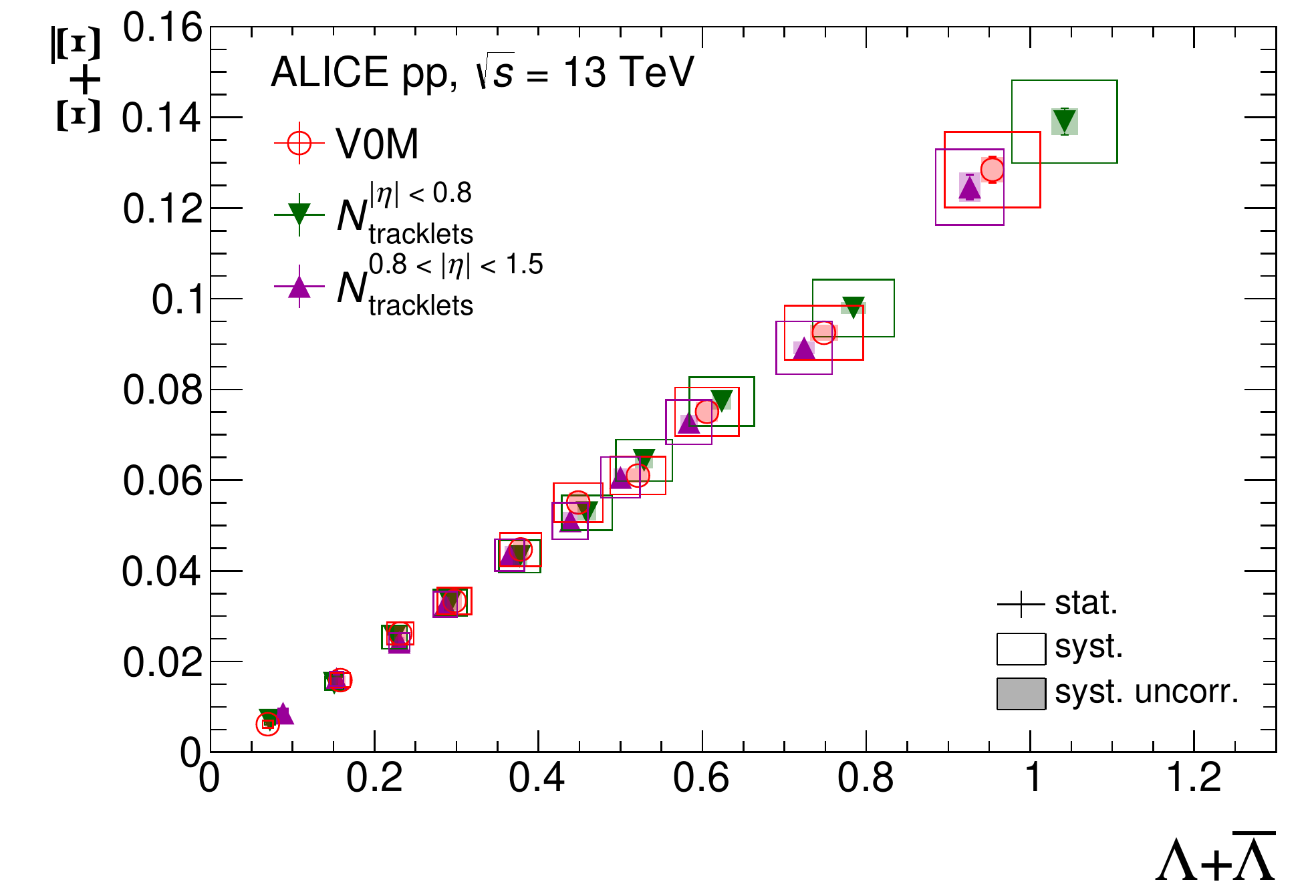}
  \end{center}
  \caption{\label{fig:strangenes-vs-strangeness}  Correlations between integrated yield of different strange hadrons in multiplicity classes selected according to different estimators
(see text for details). Statistical and systematic uncertainties are shown by error bars and empty boxes, respectively. Shadowed boxes represent uncertainties uncorrelated across multiplicity.} \warn{PB: the y scale is not clear. Density of particles per unit of pseudorapidity, right? -> yes, to be fixed}
\end{figure}


The energy dependence of the strangeness yields and \avpT\ versus the charged particle multiplicity at mid-rapidity is studied in 
\cref{fig:yields-vs-nch-energydep,fig:avpt-vs-nch-energydep},
where our results are compared to the previous pp measurements at \s~=~7~TeV~\cite{ALICE:2017jyt}. The minimum bias results in the \inelgtzero\ event class at \s~=~7 and 13~TeV~\cite{Abelev:2012jp} are also shown.

As can be seen in Fig.~\ref{fig:yields-vs-nch-energydep}, the yields of strange hadrons increase with the charged particle multiplicity following a power law behaviour, and the trend is the same at \s~=~7~and~13~TeV. The \inelgtzero\ results also follow the same trend at all the tested centre-of-mass energies. This result indicates that the abundance of strange hadrons depends on the local charged particle density and turns out to be invariant with the collision energy, i.e. an energy scaling property applies for the multiplicity-dependent yields of strange hadrons. 
It should also be noted that the yields of  particles with larger strange quark content increase faster as a function of multiplicity as already reported in~\cite{ALICE:2017jyt}. 
Figure~\ref{fig:ratios-to-K0s} shows the \pLambda/\pKzero (no \apLambda contribution considered here), \sXi/\pKzero\ and \sOmega/\pKzero\ ratios compared to calculations from grand-canonical thermal models~\cite{Andronic:2008gu,Wheaton:2004qb}, which were found to satisfactorily describe central Pb-Pb data at \snn~=~2.76~TeV.
In the context of a canonical thermal model for a gas of hadrons, an increase of the relative strangeness abundance depending on the strange quark content can be interpreted as a consequence of a change in the system volume, called canonical suppression. Indeed, it was recently shown in~\cite{Acharya:2018orn} that the existing ALICE data can be described within this framework, introducing an additional parameter to quantify the rapidity window over which strangeness is effectively correlated.
It was also suggested that strangeness follows a universal scaling behaviour in all colliding systems, when the transverse energy density~\cite{Cuautle:2016huw} or the multiplicity per transverse area~\cite{Castorina:2016eyx} are used as a scaling variable. While there are some caveats in the observation reported in these papers (due to the uncertainties on the transverse size of the system) this is a very intriguing observation, that could help to clarify the origin of strangeness enhancement.

The \avpT\ is seen in Fig.~\ref{fig:avpt-vs-nch-energydep} to be harder at 13 TeV than at 7 TeV for event classes with a similar \dNdeta. All the tested Monte Carlo models describe in a qualitative way the observed smooth rise of the \avpT\ with \dNdeta; from a quantitative point of view, EPOS-LHC provides a slightly better description of the \avpT\ multiplicity evolution, especially for what concerns the strange baryons.


\begin{figure}
  \begin{center}
    \includegraphics[width=0.84\textwidth]{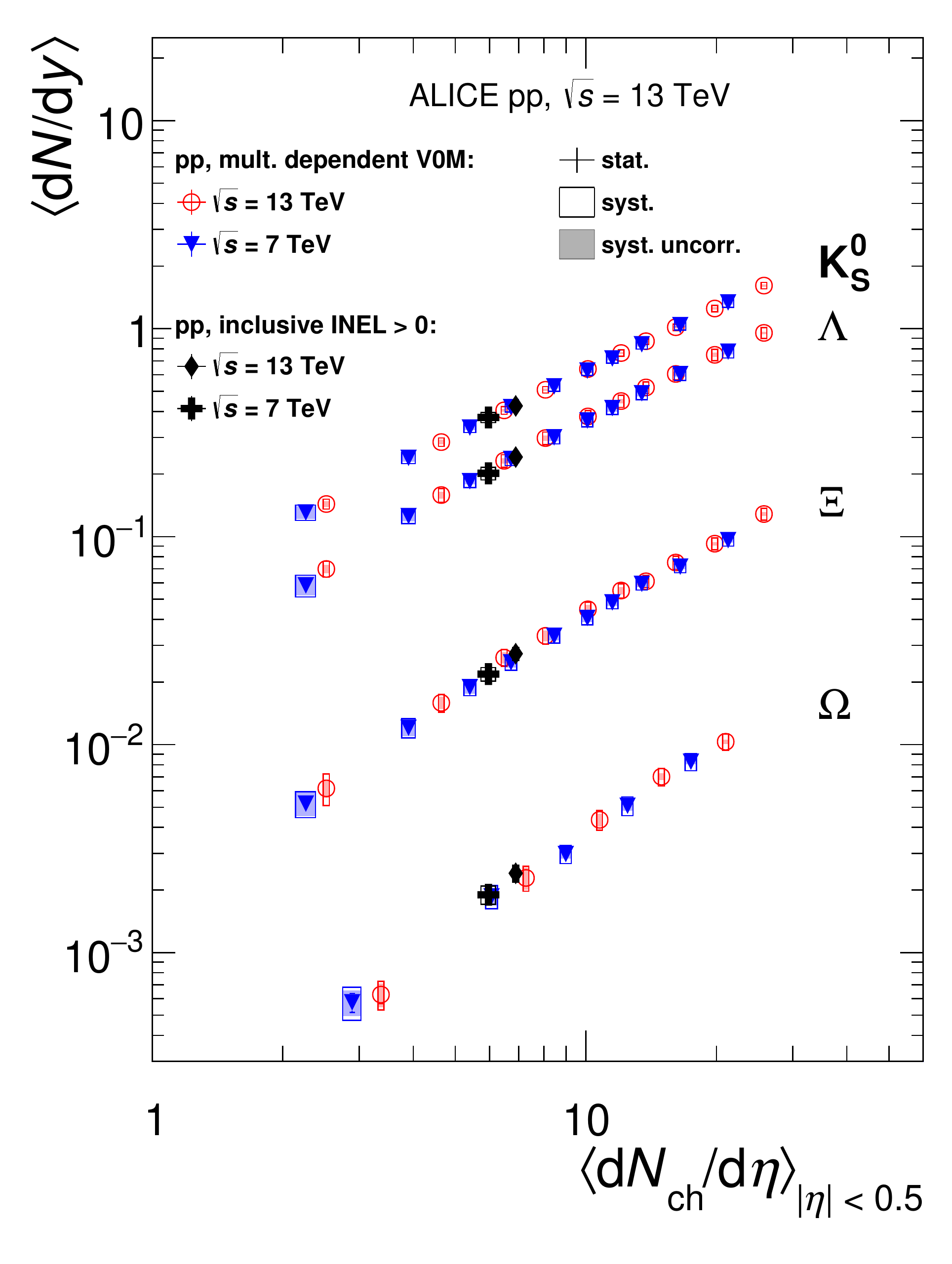}
  \end{center}
  \caption{\label{fig:yields-vs-nch-energydep}Integrated yields of \pKzero, \sLambda, \sXi, and \sOmega as a function of \dNdeta\ in V0M multiplicity
event classes at \s~=~7~and~13~TeV. Statistical and systematic uncertainties are shown by error bars and empty boxes, respectively. Shadowed boxes represent uncertainties uncorrelated across multiplicity.
The corresponding results obtained for \inelgtzero\ event class are also shown. }
\end{figure}

\begin{figure}
  \begin{center}
    \includegraphics[width=0.48\textwidth]{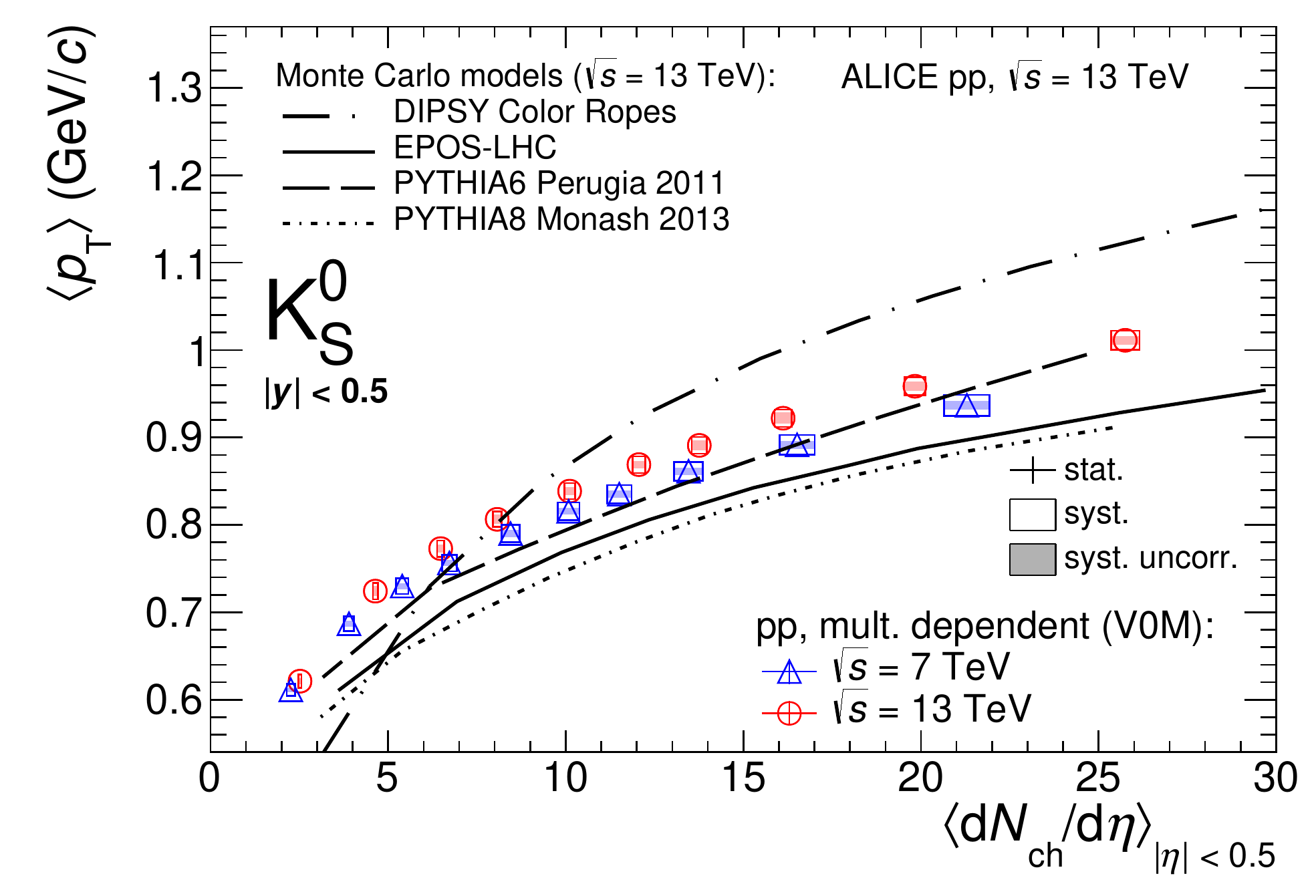}
    \includegraphics[width=0.48\textwidth]{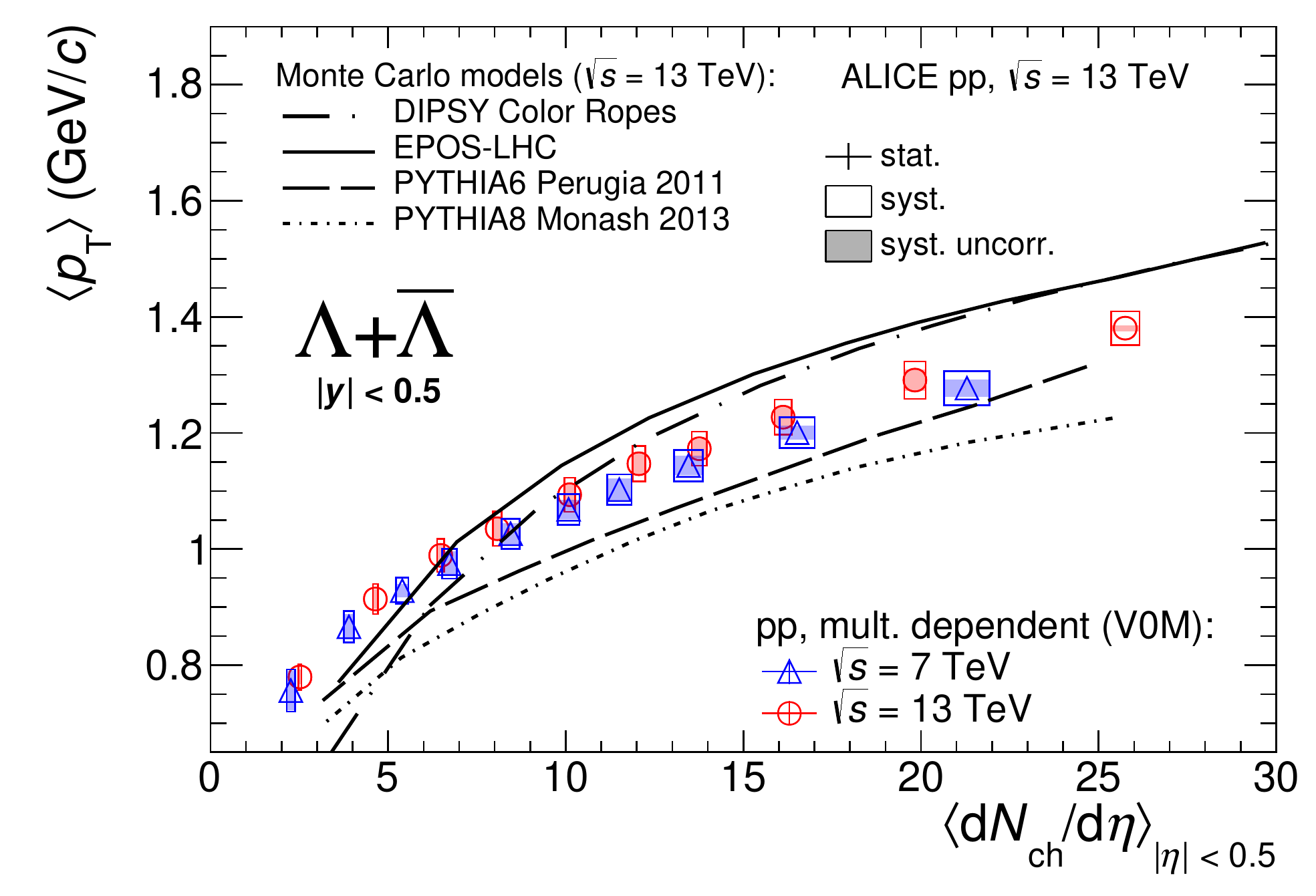}
    \includegraphics[width=0.48\textwidth]{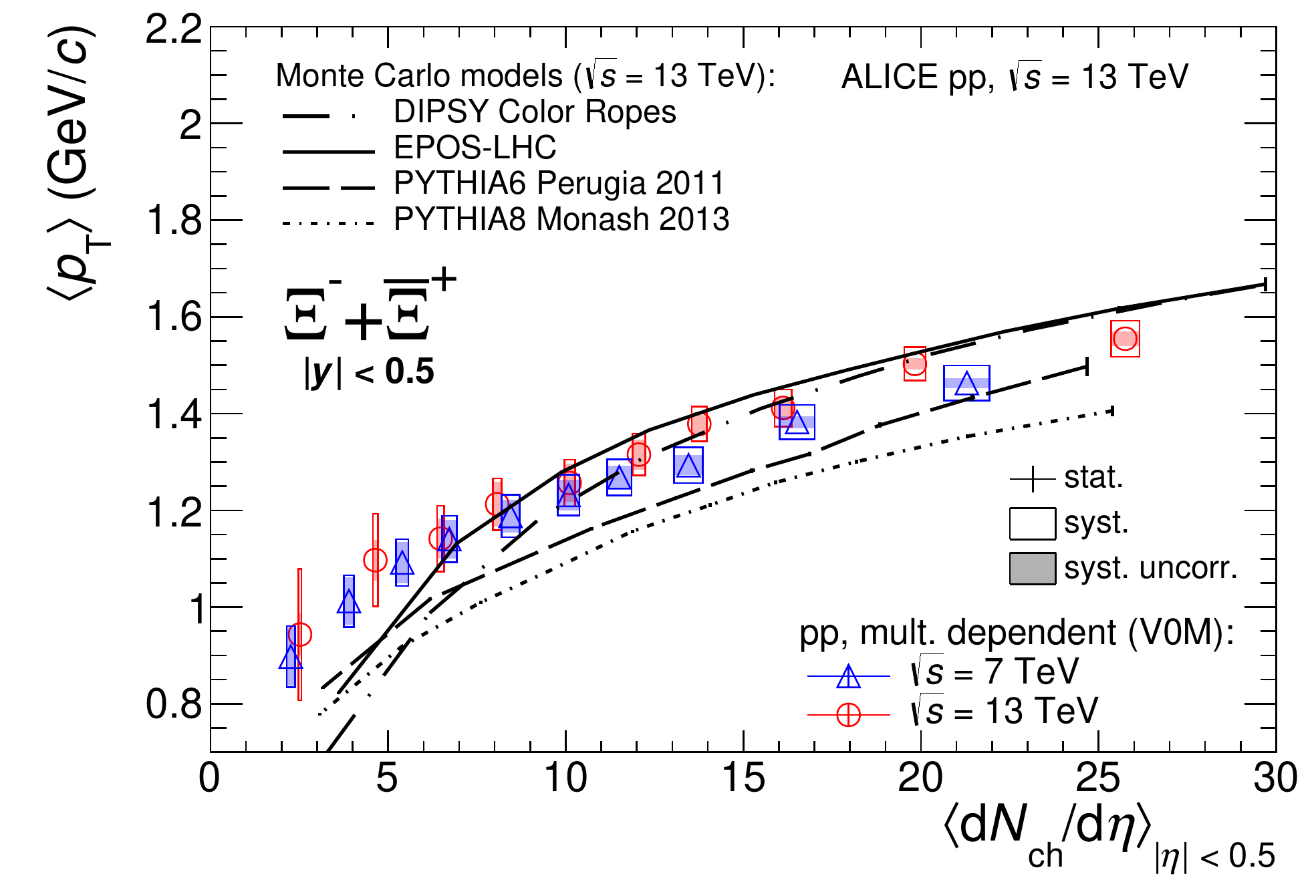}
    \includegraphics[width=0.48\textwidth]{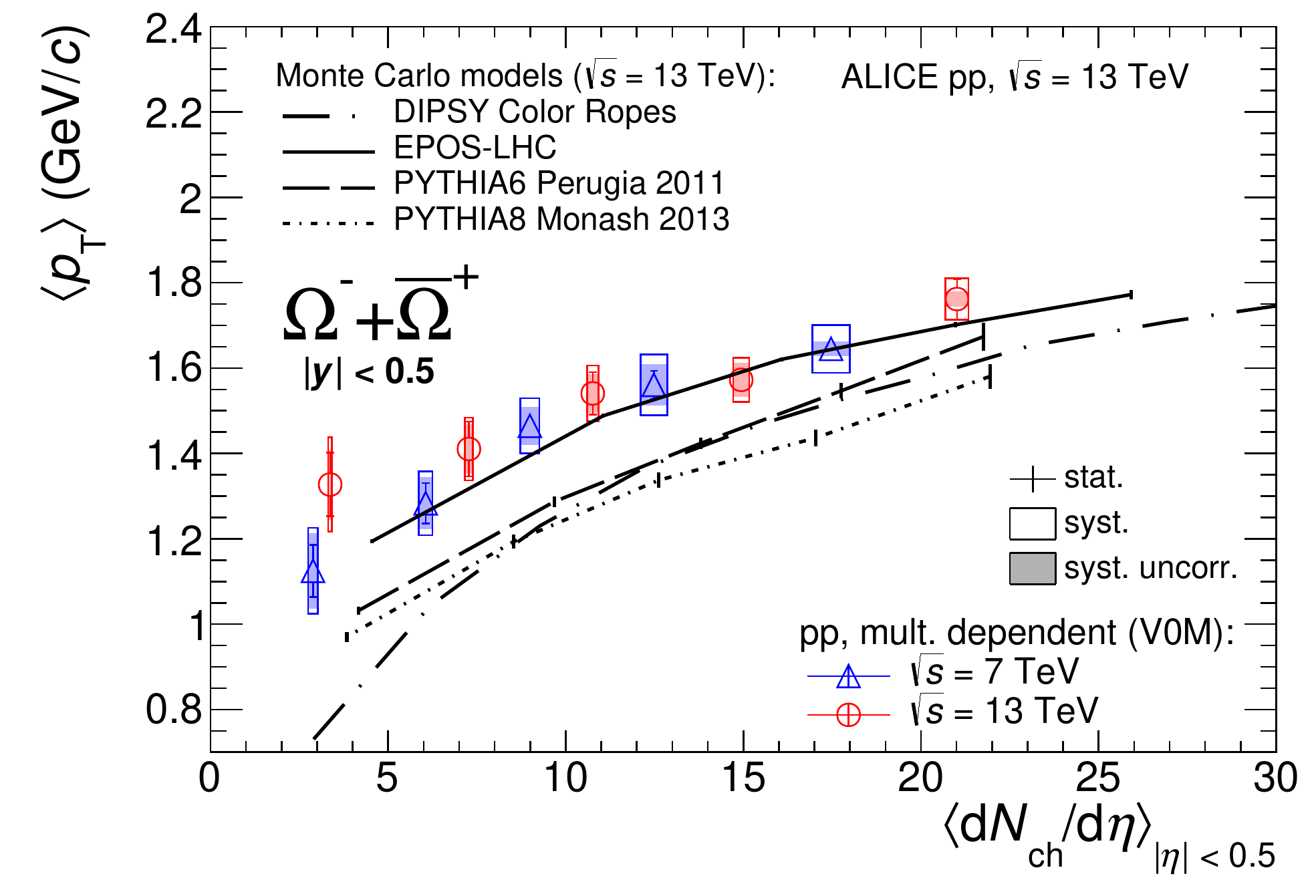}
  \end{center}
  \caption{\label{fig:avpt-vs-nch-energydep}\avpT\ of \pKzero, \sLambda, \sXi, and \sOmega as a function of \dNdeta\ in V0M multiplicity
event classes at \s~=~7~and~13~TeV. Statistical and systematic uncertainties are shown by error bars and empty boxes, respectively.
Shadowed boxes represent uncertainties uncorrelated across multiplicity. The results are compared to predictions from several Monte Carlo models.}
\end{figure}

\begin{figure}
  \begin{center}
    \includegraphics[width=0.60\textwidth]{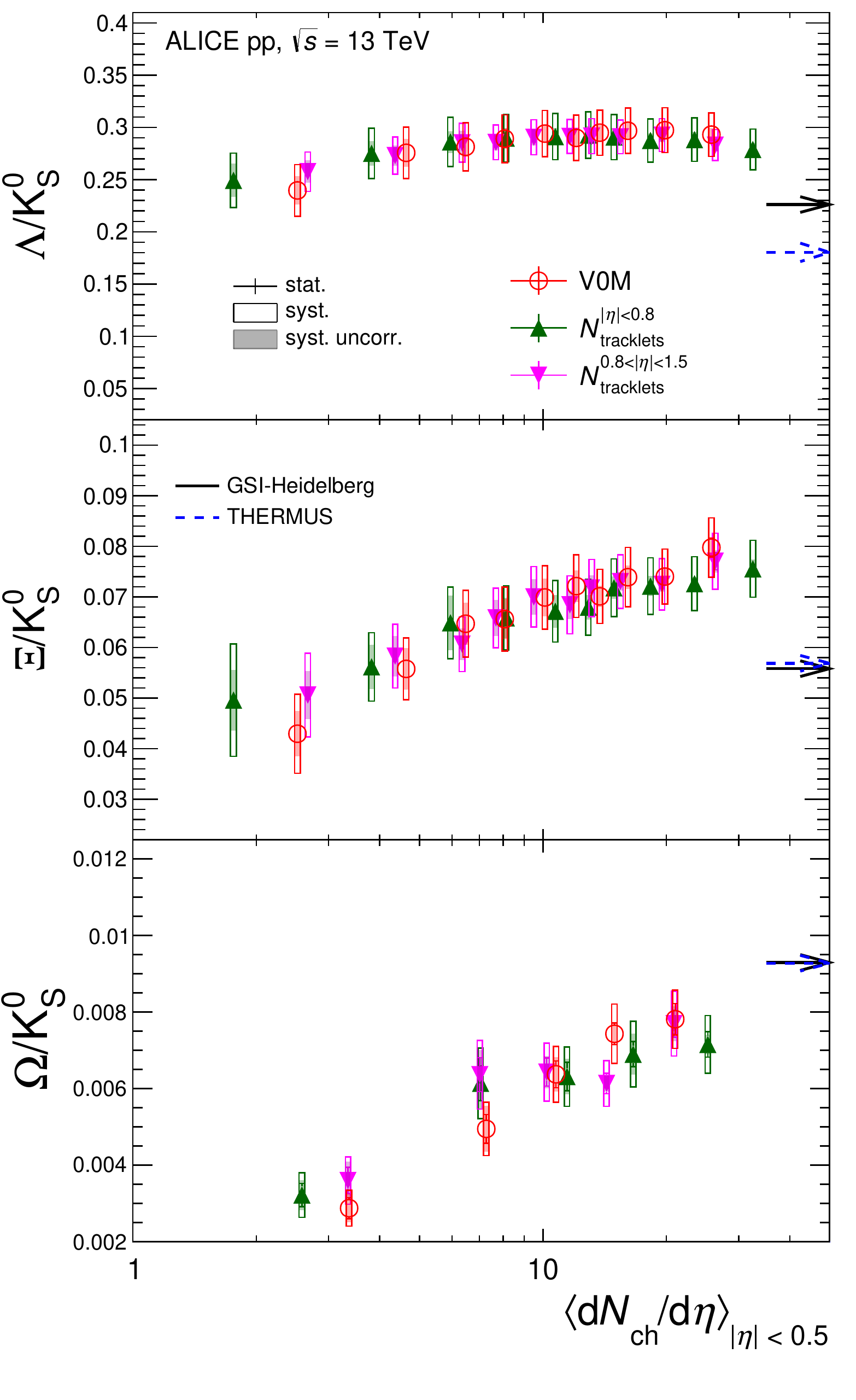}
  \end{center}
  \caption{\label{fig:ratios-to-K0s}Ratios of integrated yields of \sLambda (no \apLambda contribution considered here), \sXi, and \sOmega\ to \pKzero\ as a function of \dNdeta\ for different multiplicity
estimators (see text for details) in pp collisions at \s~=~13~TeV.  Statistical and systematic uncertainties are shown by error bars and empty boxes, respectively. Shadowed boxes represent uncertainties uncorrelated across multiplicity. The corresponding calculations from grand-canonical thermal models, which refer to most central Pb-Pb collisions at $\sqrt{s_{\rm NN}} = 2.76$ TeV, are shown.} 
\end{figure}


In Fig.~\ref{fig:dndy-comparison-to-mc} the results on the strange hadron yields as a function of the charged particle multiplicity at mid-rapidity are compared with some commonly-used general purpose QCD inspired models, focusing on the multiplicity classes defined by the V0M estimator.
The yields of all measured strange hadrons show an almost linear increase with multiplicity: a linear fit to the baryon trends would intercept the \dNdeta\ axis at positive values. 
Alternatively one may regard such qualitative observation as the trend of the points to have an enhanced (more than linear) increase for the hadrons with larger strangeness content. 


The PYTHIA 8.210 event generator (tune Monash 2013) gives a reasonable description of the \pKzero results, however it shows a less pronounced increase of the strange baryons yields versus the charged particle multiplicity than what is observed in the data. The description worsens with increasing strangeness content. 
Recent attempts to improve the PYTHIA colour-reconnection scheme to account for the observed strangeness enhancement are discussed in~\cite{Fischer:2016zzs}, however they cannot explain the reported experimental observations.

The DIPSY Monte Carlo~\cite{Avsar:2006jy,Flensburg:2011kk} is based on a BFKL inspired initial state dipole evolution model~\cite{Mueller:1994gb} interfaced to the Ariadne model \cite{Lonnblad:1992tz} for final state dipole evolution and to the PYTHIA hadronisation scheme, where the latter has been modified to take into account the high density of strings that occurs in events with several MPI. More specifically, depending on their transverse density, the strings are allowed to form ``colour-ropes'' leading to increased string tension, which leads to an increase of strangeness production stronger than in default PYTHIA~\cite{Bierlich:2014xba}. 
However, despite a qualitative improvement in the description of the data, the discrepancy for the \sOmega distributions remains large.
Moreover, as discussed in \cite{ALICE:2017jyt}, DIPSY also predicts an increase of protons normalised to pions as a function of multiplicity, which is not observed in our \s~=~7~TeV data. 

The Monte Carlo generators based on EPOS~\cite{Werner:2013yia} rely on the Parton Based Gribov Regge Theory (PBGRT)~\cite{Drescher:2000ha}. A common feature of all the EPOS versions is a collective evolution of matter in the secondary scattering stage in all reactions, from pp to AA, with a core-corona separation mechanism \cite{Werner:2007bf} which defines the initial conditions of the secondary interactions.
In EPOS, the initial parton scatterings create ``flux tubes'' that either escape the medium and hadronise as jets or contribute to the core, described in terms of hydrodynamics. The core is then hadronised in terms of a grand-canonical statistical model.\warn{double check this with Tanguy} The relative amount of multi-hadron production arising from the core grows with the number of MPIs. 
The EPOS-LHC model adopted in this analysis (distributed with the CRMC package 1.5.4)\warn{Does this need a reference?} shows a pronounced increase of the strangeness yields as a function of the charged particle multiplicity at mid-rapidity. 
Despite the differences in the underlying physics, the comparison to data leads to a similar conclusion for EPOS-LHC and DIPSY.

The comparison of the Monte Carlo model predictions to the data indicates that the origin of the strangeness enhancement in hadronic collisions and its relation to the QCD deconfinement phase transition are still open problems for models.
While the current versions of these models do not reproduce the data quantitatively, it is possible that the agreement can be improved with further tuning or with new implementations. 
A quantitative description of strangeness production and enhancement in a microscopic Monte Carlo model would represent a major step to understand flavour production in hadronic collisions at high energy. 

\warn{Add version of the plot with DIPSY.}
\warn{assicurarsi che ci sia un pistolotto sufficiente sui bias degli estimatori (overlapping region -> sensitive to fluctuations)}
 \warn{PB: in my opinion this is - again - due to the correlation between \NtrkEtaIn\ and the charged particle multiplicity at mid-rapidity. They are almost the same estimator! Adding forward multiplicity information allows to be more sensitive to the total multiplicity. Using only central information end-up in a sort of de-generation, i.e. we are no longer investigating yields in a 2D space, we are investigating  them in a 1D space.}

\begin{figure} 
  \centering
  \includegraphics[width=0.45\textwidth]{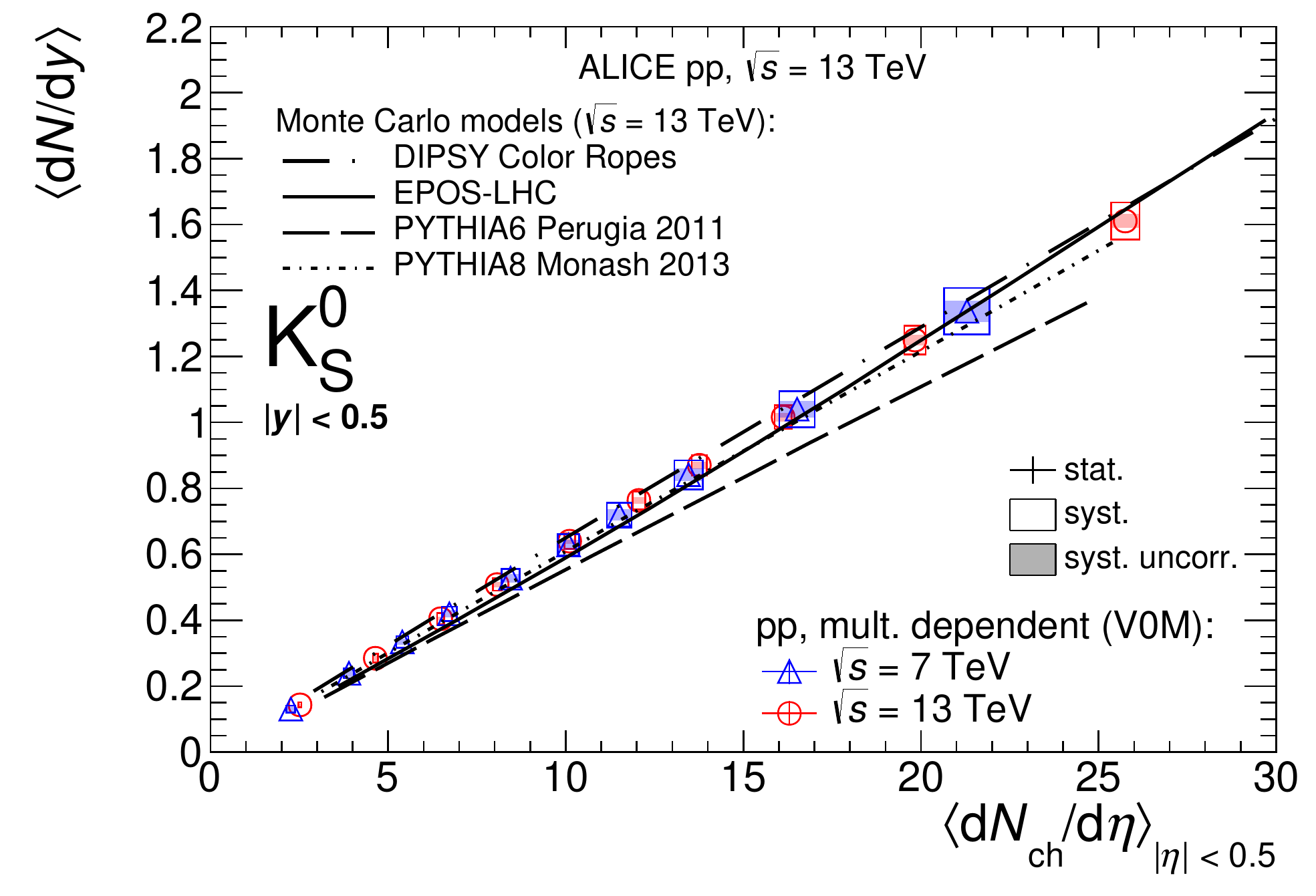}
  \includegraphics[width=0.45\textwidth]{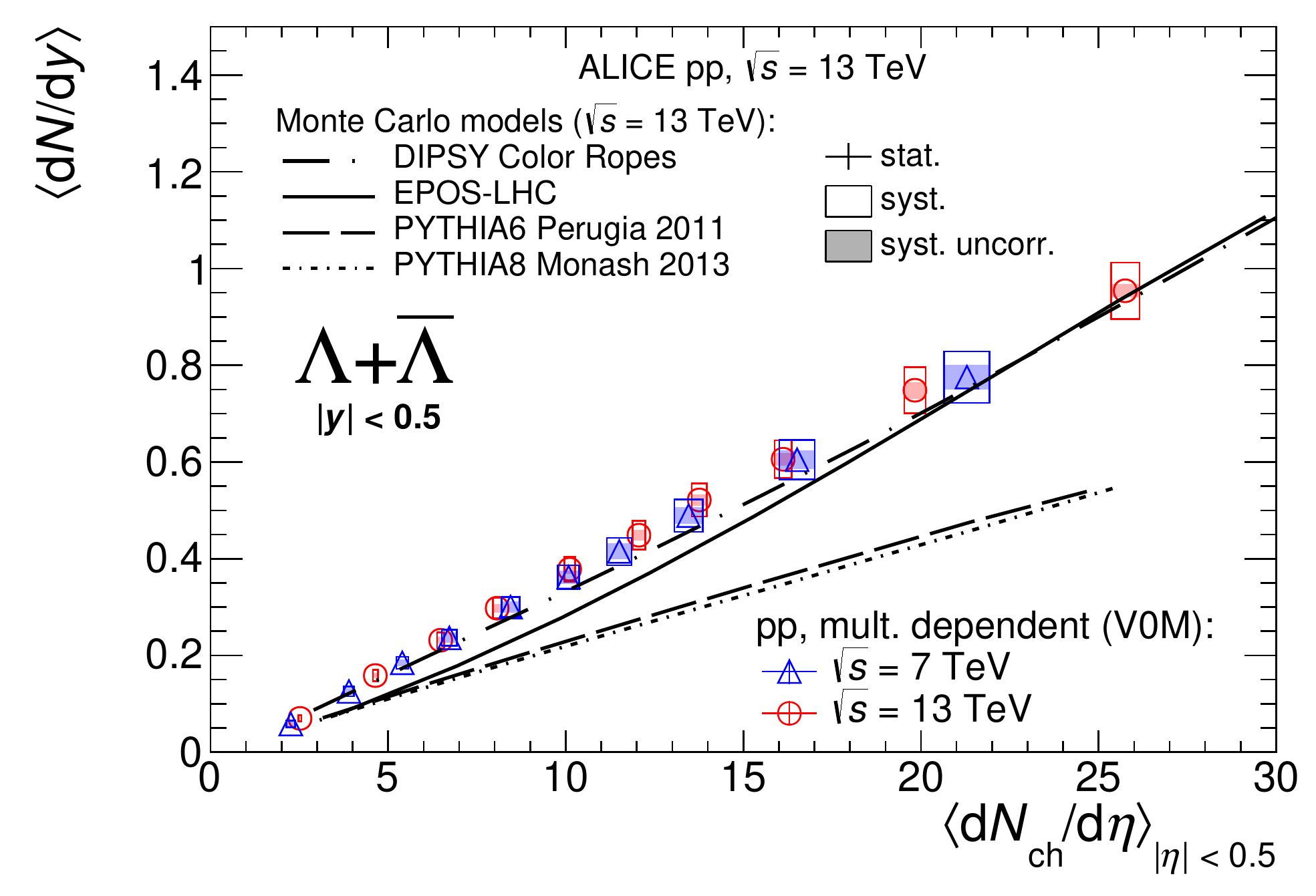}
  \includegraphics[width=0.45\textwidth]{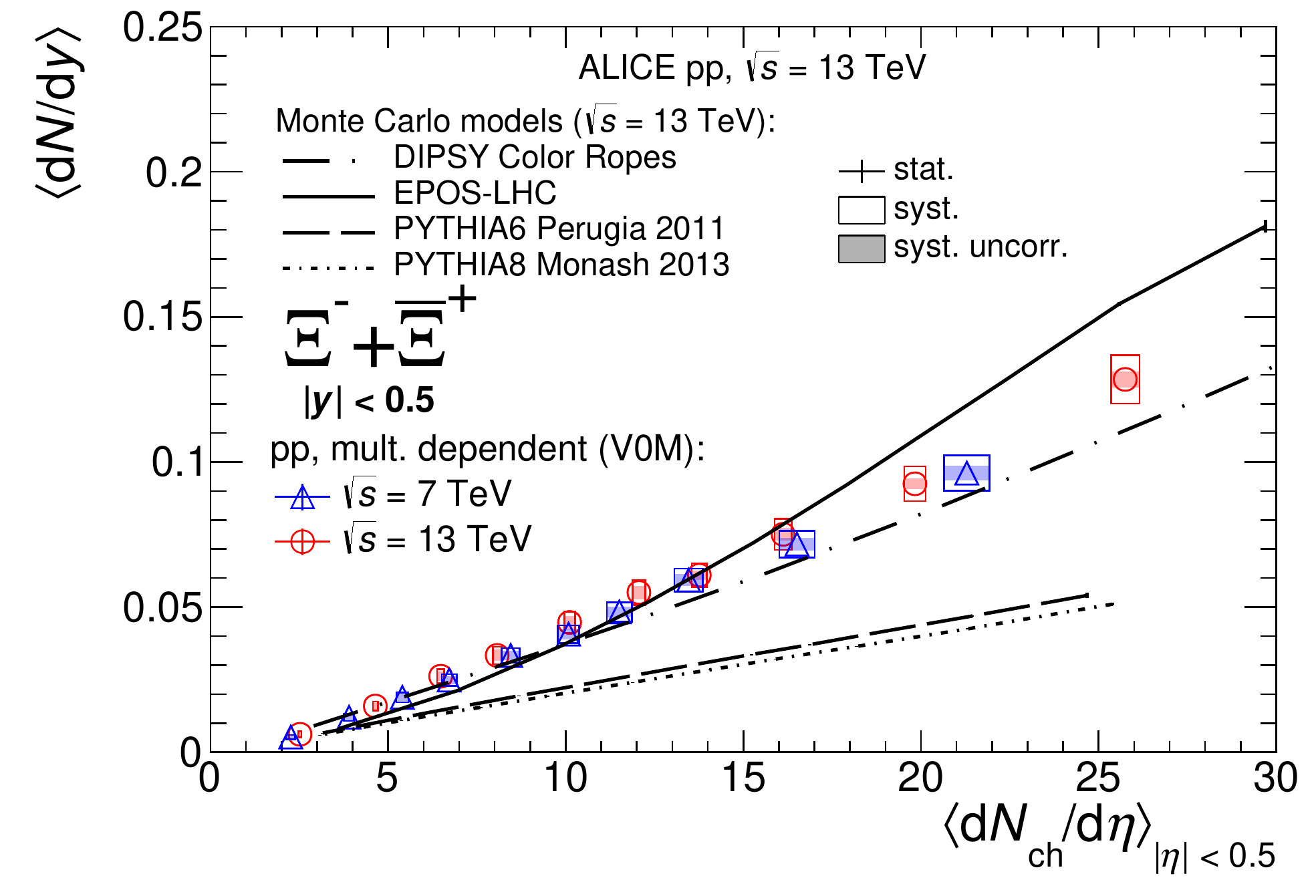}
  \includegraphics[width=0.45\textwidth]{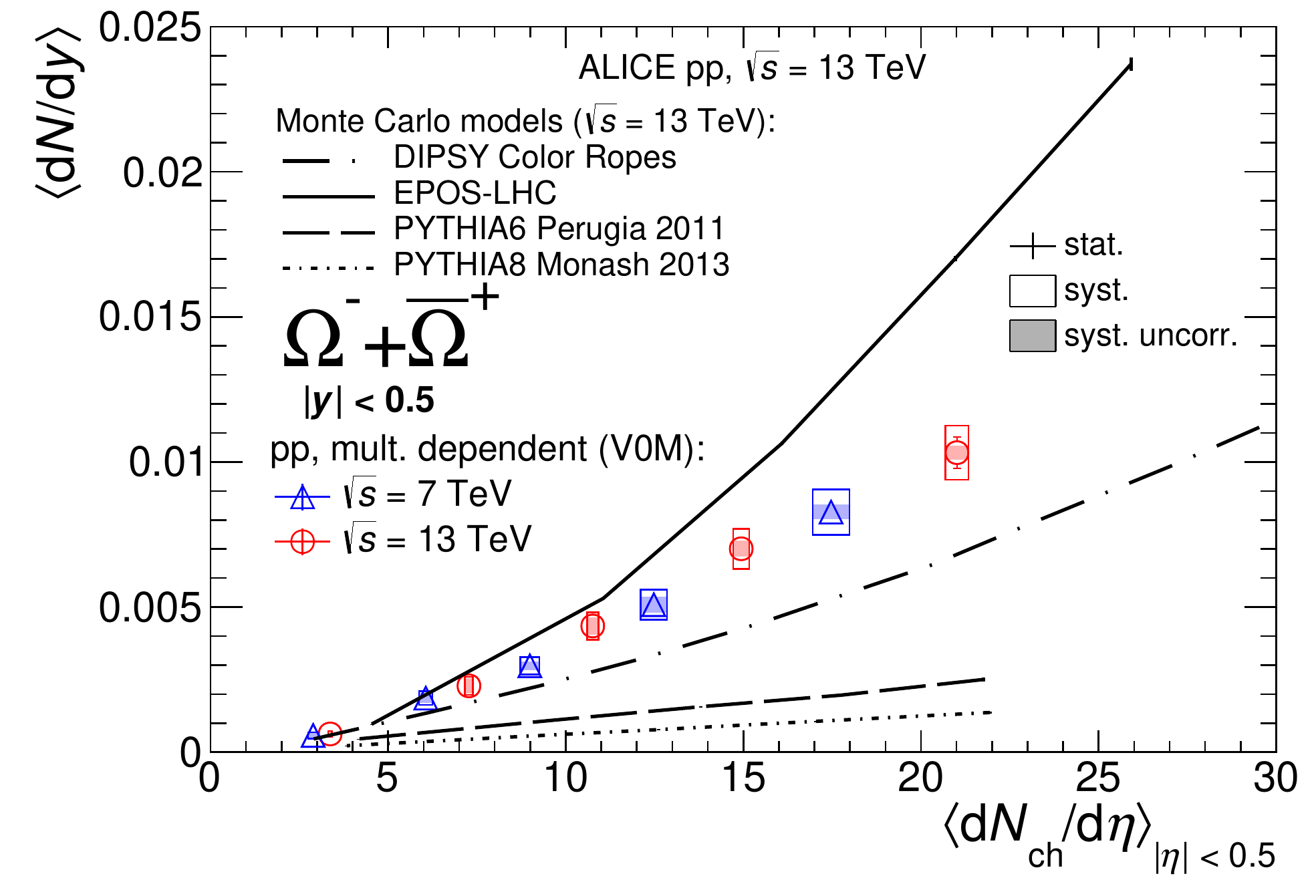}
  \caption{Integrated yields of \pKzero, \sLambda, \sXi, and \sOmega as a function of \dNdeta\ in V0M multiplicity
event classes at \s~=~7~and~13~TeV. Statistical and systematic uncertainties are shown by error bars and empty boxes, respectively.
Shadowed boxes represent uncertainties uncorrelated across multiplicity. The results are compared to predictions from several Monte Carlo models.}
  \label{fig:dndy-comparison-to-mc}
\end{figure}


\section{Summary}
\label{sec:summary}

We studied the production of primary strange and multi-strange hadrons at mid-rapidity in pp collisions at \s~=~13~TeV, focusing on the multiplicity dependence. The main feature of this analysis is the usage of multiplicity estimators defined in different pseudorapidity regions, providing a different sensitivity to the fragmentation and MPI components of particle production and allowing for a detailed study of the selection biases due to fluctuations. 

Hardening of the \pt\ spectra with the increase of the multiplicity is observed, as already reported for pp~\cite{Chatrchyan:2012qb} and \pPb\ collisions~\cite{Abelev:2013haa} at lower energies.  

The \pt-integrated yields of strange hadrons increase as a function of multiplicity faster than the ones of unidentified charged-particles. This behaviour is even more pronounced for particles with higher strangeness content, confirming the earlier observations in pp collisions at  \s~=~7~TeV~\cite{ALICE:2017jyt}. This leads to a multiplicity-dependent increase of the ratio of the strange baryons  \sXi and \sOmega\ over \pKzero, while the \sLambda\ over \pKzero ratio turns out to be constant within uncertainties. In the context of a canonical thermal model, an increase of the relative strangeness abundance depending on the strange quark content can be understood as a consequence of an increase in the system volume leading to a progressive removal of canonical suppression.

Comparing the 13 TeV results to the 7 TeV ones, the data exhibit an interesting scaling property: for multiplicity estimators selected in the forward region, the strange hadron yields turn out to be independent of the centre of mass energy, the \s\ increase resulting just in harder \pt\ spectra.

The use of high-multiplicity triggered data collected during the full Run 2 period will allow us to test these scaling ansatzes extending the measurement to higher multiplicities, comparable to peripheral Pb-Pb collisions at the LHC ($\dNdeta \sim 40-50$).

The colour reconnection effects implemented in PYTHIA 8 and DIPSY as well as the collective hydrodynamic expansion implemented in EPOS-LHC could not account quantitatively for all the reported results. Some of the qualitative features of the data are described by these models, including the indication of a non-linear increase of the yields of strange hadrons at high \pt, similar to what was previously reported in~\cite{Adam:2015ota} for \pJPsi\ and $D$ meson production.  Although none of the tested models can reproduce the data from a quantitative point of view, EPOS-LHC and DIPSY do provide a better qualitative description of the strangeness enhancement.




%
%

\newenvironment{acknowledgement}{\relax}{\relax}
\begin{acknowledgement}
\section*{Acknowledgements}

The ALICE Collaboration would like to thank all its engineers and technicians for their invaluable contributions to the construction of the experiment and the CERN accelerator teams for the outstanding performance of the LHC complex.
The ALICE Collaboration gratefully acknowledges the resources and support provided by all Grid centres and the Worldwide LHC Computing Grid (WLCG) collaboration.
The ALICE Collaboration acknowledges the following funding agencies for their support in building and running the ALICE detector:
A. I. Alikhanyan National Science Laboratory (Yerevan Physics Institute) Foundation (ANSL), State Committee of Science and World Federation of Scientists (WFS), Armenia;
Austrian Academy of Sciences, Austrian Science Fund (FWF): [M 2467-N36] and Nationalstiftung f\"{u}r Forschung, Technologie und Entwicklung, Austria;
Ministry of Communications and High Technologies, National Nuclear Research Center, Azerbaijan;
Conselho Nacional de Desenvolvimento Cient\'{\i}fico e Tecnol\'{o}gico (CNPq), Universidade Federal do Rio Grande do Sul (UFRGS), Financiadora de Estudos e Projetos (Finep) and Funda\c{c}\~{a}o de Amparo \`{a} Pesquisa do Estado de S\~{a}o Paulo (FAPESP), Brazil;
Ministry of Science \& Technology of China (MSTC), National Natural Science Foundation of China (NSFC) and Ministry of Education of China (MOEC) , China;
Croatian Science Foundation and Ministry of Science and Education, Croatia;
Centro de Aplicaciones Tecnol\'{o}gicas y Desarrollo Nuclear (CEADEN), Cubaenerg\'{\i}a, Cuba;
Ministry of Education, Youth and Sports of the Czech Republic, Czech Republic;
The Danish Council for Independent Research | Natural Sciences, the Carlsberg Foundation and Danish National Research Foundation (DNRF), Denmark;
Helsinki Institute of Physics (HIP), Finland;
Commissariat \`{a} l'Energie Atomique (CEA), Institut National de Physique Nucl\'{e}aire et de Physique des Particules (IN2P3) and Centre National de la Recherche Scientifique (CNRS) and R\'{e}gion des  Pays de la Loire, France;
Bundesministerium f\"{u}r Bildung und Forschung (BMBF) and GSI Helmholtzzentrum f\"{u}r Schwerionenforschung GmbH, Germany;
General Secretariat for Research and Technology, Ministry of Education, Research and Religions, Greece;
National Research, Development and Innovation Office, Hungary;
Department of Atomic Energy Government of India (DAE), Department of Science and Technology, Government of India (DST), University Grants Commission, Government of India (UGC) and Council of Scientific and Industrial Research (CSIR), India;
Indonesian Institute of Science, Indonesia;
Centro Fermi - Museo Storico della Fisica e Centro Studi e Ricerche Enrico Fermi and Istituto Nazionale di Fisica Nucleare (INFN), Italy;
Institute for Innovative Science and Technology , Nagasaki Institute of Applied Science (IIST), Japan Society for the Promotion of Science (JSPS) KAKENHI and Japanese Ministry of Education, Culture, Sports, Science and Technology (MEXT), Japan;
Consejo Nacional de Ciencia (CONACYT) y Tecnolog\'{i}a, through Fondo de Cooperaci\'{o}n Internacional en Ciencia y Tecnolog\'{i}a (FONCICYT) and Direcci\'{o}n General de Asuntos del Personal Academico (DGAPA), Mexico;
Nederlandse Organisatie voor Wetenschappelijk Onderzoek (NWO), Netherlands;
The Research Council of Norway, Norway;
Commission on Science and Technology for Sustainable Development in the South (COMSATS), Pakistan;
Pontificia Universidad Cat\'{o}lica del Per\'{u}, Peru;
Ministry of Science and Higher Education and National Science Centre, Poland;
Korea Institute of Science and Technology Information and National Research Foundation of Korea (NRF), Republic of Korea;
Ministry of Education and Scientific Research, Institute of Atomic Physics and Ministry of Research and Innovation and Institute of Atomic Physics, Romania;
Joint Institute for Nuclear Research (JINR), Ministry of Education and Science of the Russian Federation, National Research Centre Kurchatov Institute, Russian Science Foundation and Russian Foundation for Basic Research, Russia;
Ministry of Education, Science, Research and Sport of the Slovak Republic, Slovakia;
National Research Foundation of South Africa, South Africa;
Swedish Research Council (VR) and Knut \& Alice Wallenberg Foundation (KAW), Sweden;
European Organization for Nuclear Research, Switzerland;
National Science and Technology Development Agency (NSDTA), Suranaree University of Technology (SUT) and Office of the Higher Education Commission under NRU project of Thailand, Thailand;
Turkish Atomic Energy Agency (TAEK), Turkey;
National Academy of  Sciences of Ukraine, Ukraine;
Science and Technology Facilities Council (STFC), United Kingdom;
National Science Foundation of the United States of America (NSF) and United States Department of Energy, Office of Nuclear Physics (DOE NP), United States of America.    
\end{acknowledgement}

\bibliographystyle{utphys}   
\bibliography{biblio}

\providecommand{\href}[2]{#2}\begingroup\raggedright\begin{thebibliography}{10}

\bibitem{Forte:2013wc}
S.~Forte and G.~Watt, ``{Progress in the Determination of the Partonic
  Structure of the Proton},''
  \href{http://dx.doi.org/10.1146/annurev-nucl-102212-170607}{{\em Ann. Rev.
  Nucl. Part. Sci.} {\bfseries 63} (2013) 291--328},
\href{http://arxiv.org/abs/1301.6754}{{\ttfamily arXiv:1301.6754 [hep-ph]}}.

\bibitem{Nagy:2017ggp}
Z.~Nagy and D.~E. Soper, ``{What is a parton shower?},''
\href{http://arxiv.org/abs/1705.08093}{{\ttfamily arXiv:1705.08093 [hep-ph]}}.

\bibitem{Fischer:2016zzs}
N.~Fischer and T.~Sj{\"{o}}strand, ``{Thermodynamical String Fragmentation},''
  \href{http://dx.doi.org/10.1007/JHEP01(2017)140}{{\em JHEP} {\bfseries 01}
  (2017) 140},
\href{http://arxiv.org/abs/1610.09818}{{\ttfamily arXiv:1610.09818 [hep-ph]}}.

\bibitem{Abelev:2006cs}
{\bfseries STAR} Collaboration, B.~Abelev {\em et~al.}, ``{Strange particle
  production in pp collisions at $\sqrt{\it{s}}$ = 200~GeV},''
  \href{http://dx.doi.org/10.1103/PhysRevC.75.064901}{{\em Phys. Rev.}
  {\bfseries C75} (2007) 064901},
\href{http://arxiv.org/abs/nucl-ex/0607033}{{\ttfamily arXiv:nucl-ex/0607033
  [nucl-ex]}}.

\bibitem{Abelev:2012jp}
{\bfseries ALICE} Collaboration, B.~Abelev {\em et~al.}, ``{Multi-strange
  baryon production in $pp$ collisions at $\sqrt{\it{s}} = 7$ TeV with
  ALICE},'' \href{http://dx.doi.org/10.1016/j.physletb.2012.05.011}{{\em Phys.
  Lett.} {\bfseries B712} (2012) 309--318},
\href{http://arxiv.org/abs/1204.0282}{{\ttfamily arXiv:1204.0282 [nucl-ex]}}.

\bibitem{Aamodt:2011zza}
{\bfseries ALICE} Collaboration, K.~Aamodt {\em et~al.}, ``{Strange particle
  production in proton-proton collisions at $\sqrt{\it{s}}$ = 0.9 TeV with
  ALICE at the LHC},''
  \href{http://dx.doi.org/10.1140/epjc/s10052-011-1594-5}{{\em Eur. Phys. J.}
  {\bfseries C71} (2011) 1594},
\href{http://arxiv.org/abs/1012.3257}{{\ttfamily arXiv:1012.3257 [hep-ex]}}.

\bibitem{Rafelski:1982pu}
J.~Rafelski and B.~Muller, ``{Strangeness Production in the Quark - Gluon
  Plasma},''
{\em Phys. Rev. Lett.} {\bfseries 48} (1982) 1066.

\bibitem{Rafelski:1980rk}
J.~Rafelski and R.~Hagedorn, ``{From Hadron Gas to Quark Matter. 2},'' in {\em
  {International Symposium on Statistical Mechanics of Quarks and Hadrons
  Bielefeld, Germany, August 24-31, 1980}}, pp.~253--272.
\newblock 1980.
\newblock
\url{https://cds.cern.ch/record/126179}.
\newblock

\bibitem{Andersen:1998vu}
E.~Andersen {\em et~al.}, ``{Enhancement of central Lambda, Xi and Omega yields
  in Pb - Pb collisions at 158 A-GeV/c},''
\href{http://dx.doi.org/10.1016/S0370-2693(98)00689-3}{{\em Phys. Lett.}
  {\bfseries B433} (1998) 209--216}.

\bibitem{Abelev:2007xp}
{\bfseries STAR} Collaboration, B.~Abelev {\em et~al.}, ``{Enhanced strange
  baryon production in Au + Au collisions compared to pp at
  $\sqrt{\it{s}}_{\mathrm{NN}}$ = 200~GeV},''
  \href{http://dx.doi.org/10.1103/PhysRevC.77.044908}{{\em Phys. Rev.}
  {\bfseries C77} (2008) 044908},
\href{http://arxiv.org/abs/0705.2511}{{\ttfamily arXiv:0705.2511 [nucl-ex]}}.

\bibitem{ABELEV:2013zaa}
{\bfseries ALICE} Collaboration, B.~Abelev {\em et~al.}, ``{Multi-strange
  baryon production at mid-rapidity in Pb-Pb collisions at
  $\sqrt{\it{s}_{\mathrm{NN}}}$ = 2.76 TeV},''
  \href{http://dx.doi.org/10.1016/j.physletb.2013.11.048}{{\em Phys. Lett.}
  {\bfseries B728} (2014) 216--227},
  \href{http://arxiv.org/abs/1307.5543}{{\ttfamily arXiv:1307.5543 [nucl-ex]}}.
[Erratum: \href{https://doi.org/10.1016/j.physletb.2014.05.052}{ {\it Phys.
  Lett.} $\mathbf{B734}$ (2014) 409-410.}].

\bibitem{Koch:2017pda}
P.~Koch, B.~Muller, and J.~Rafelski, ``{From strangeness enhancement to
  quark-gluon plasma discovery},''
  \href{http://dx.doi.org/10.1142/S0217751X17300241}{{\em Int. J. Mod. Phys.}
  {\bfseries A32} no.~31, (2017) 1730024},
\href{http://arxiv.org/abs/1708.08115}{{\ttfamily arXiv:1708.08115 [nucl-th]}}.

\bibitem{Andronic:2008gu}
A.~Andronic, P.~Braun-Munzinger, and J.~Stachel, ``{Thermal hadron production
  in relativistic nuclear collisions: The Hadron mass spectrum, the horn, and
  the QCD phase transition},''
  \href{http://dx.doi.org/10.1016/j.physletb.2009.02.014}{{\em Phys. Lett.}
  {\bfseries B673} (2009) 142--145},
  \href{http://arxiv.org/abs/0812.1186}{{\ttfamily arXiv:0812.1186 [nucl-th]}}.
[Erratum: \href{https://doi.org/10.1016/j.physletb.2009.06.021}{ {\it Phys.
  Lett.} $\mathbf{B673}$ (2009) 142-145.}].

\bibitem{Wheaton:2004qb}
S.~Wheaton and J.~Cleymans, ``{THERMUS: A Thermal model package for ROOT},''
  \href{http://dx.doi.org/10.1016/j.cpc.2008.08.001}{{\em Comput. Phys.
  Commun.} {\bfseries 180} (2009) 84--106},
\href{http://arxiv.org/abs/hep-ph/0407174}{{\ttfamily arXiv:hep-ph/0407174
  [hep-ph]}}.

\bibitem{Tounsi:2001ck}
A.~Tounsi and K.~Redlich, ``{Strangeness enhancement and canonical
  suppression},''
\href{http://arxiv.org/abs/hep-ph/0111159}{{\ttfamily arXiv:hep-ph/0111159
  [hep-ph]}}.

\bibitem{Tounsi:2002nd}
A.~Tounsi, A.~Mischke, and K.~Redlich, ``{Canonical aspects of strangeness
  enhancement},'' \href{http://dx.doi.org/10.1016/S0375-9474(02)01524-5}{{\em
  Nucl. Phys.} {\bfseries A715} (2003) 565c--568c},
\href{http://arxiv.org/abs/hep-ph/0209284}{{\ttfamily arXiv:hep-ph/0209284
  [hep-ph]}}.

\bibitem{Becattini:2008yn}
F.~Becattini and J.~Manninen, ``{Strangeness production from SPS to LHC},''
  \href{http://dx.doi.org/10.1088/0954-3899/35/10/104013}{{\em J. Phys.}
  {\bfseries G35} (2008) 104013},
\href{http://arxiv.org/abs/0805.0098}{{\ttfamily arXiv:0805.0098 [nucl-th]}}.

\bibitem{Acharya:2018orn}
{\bfseries ALICE} Collaboration, S.~Acharya {\em et~al.}, ``{Multiplicity
  dependence of light-flavor hadron production in pp collisions at $\sqrt{s}$ =
  7 TeV},'' {\em Submitted to: Phys. Rev.} (2018) ,
\href{http://arxiv.org/abs/1807.11321}{{\ttfamily arXiv:1807.11321 [nucl-ex]}}.

\bibitem{Kraus:2007hf}
I.~Kraus, J.~Cleymans, H.~Oeschler, K.~Redlich, and S.~Wheaton, ``{Chemical
  Equilibrium in Collisions of Small Systems},''
  \href{http://dx.doi.org/10.1103/PhysRevC.76.064903}{{\em Phys. Rev.}
  {\bfseries C76} (2007) 064903},
\href{http://arxiv.org/abs/0707.3879}{{\ttfamily arXiv:0707.3879 [hep-ph]}}.

\bibitem{Kraus:2008fh}
I.~Kraus, J.~Cleymans, H.~Oeschler, and K.~Redlich, ``{Particle production in
  pp collisions and prediction for LHC energy},''
  \href{http://dx.doi.org/10.1103/PhysRevC.79.014901}{{\em Phys. Rev.}
  {\bfseries C79} (2009) 014901},
\href{http://arxiv.org/abs/0808.0611}{{\ttfamily arXiv:0808.0611 [hep-ph]}}.

\bibitem{ALICE:2017jyt}
{\bfseries ALICE} Collaboration, J.~Adam {\em et~al.}, ``{Enhanced production
  of multi-strange hadrons in high-multiplicity proton-proton collisions},''
  \href{http://dx.doi.org/10.1038/nphys4111}{{\em Nature Phys.} {\bfseries 13}
  (2017) 535--539},
\href{http://arxiv.org/abs/1606.07424}{{\ttfamily arXiv:1606.07424 [nucl-ex]}}.

\bibitem{Abelev:2013haa}
{\bfseries ALICE} Collaboration, B.~Abelev {\em et~al.}, ``{Multiplicity
  Dependence of Pion, Kaon, Proton and Lambda Production in p-Pb Collisions at
  $\sqrt{\it{s}_{\mathrm{NN}}}$ = 5.02 TeV},''
  \href{http://dx.doi.org/10.1016/j.physletb.2013.11.020}{{\em Phys. Lett.}
  {\bfseries B728} (2014) 25--38},
\href{http://arxiv.org/abs/1307.6796}{{\ttfamily arXiv:1307.6796 [nucl-ex]}}.

\bibitem{Adam:2015vsf}
{\bfseries ALICE} Collaboration, J.~Adam {\em et~al.}, ``{Multi-strange baryon
  production in p-Pb collisions at $\sqrt{\it{s}_\mathbf{NN}}=5.02$ TeV},''
  \href{http://dx.doi.org/10.1016/j.physletb.2016.05.027}{{\em Phys. Lett.}
  {\bfseries B758} (2016) 389--401},
\href{http://arxiv.org/abs/1512.07227}{{\ttfamily arXiv:1512.07227 [nucl-ex]}}.

\bibitem{Aamodt:2008zz}
{\bfseries ALICE} Collaboration, K.~Aamodt {\em et~al.}, ``{The ALICE
  experiment at the CERN LHC},''
\href{http://dx.doi.org/10.1088/1748-0221/3/08/S08002}{{\em JINST} {\bfseries
  3} (2008) S08002}.

\bibitem{Abelev:2014ffa}
{\bfseries ALICE} Collaboration, B.~Abelev {\em et~al.}, ``{Performance of the
  ALICE Experiment at the CERN LHC},''
  \href{http://dx.doi.org/10.1142/S0217751X14300440}{{\em Int. J. Mod. Phys.}
  {\bfseries A29} (2014) 1430044},
\href{http://arxiv.org/abs/1402.4476}{{\ttfamily arXiv:1402.4476 [nucl-ex]}}.

\bibitem{ALICE-PUBLIC-2017-005}
{\bfseries ALICE} Collaboration, ``{The ALICE definition of primary
  particles},''. \url{https://cds.cern.ch/record/2270008}.

\bibitem{ALICE:2012xs}
{\bfseries ALICE} Collaboration, B.~Abelev {\em et~al.}, ``{Pseudorapidity
  density of charged particles in $p$ + Pb collisions at
  $\sqrt{\it{s}_{\mathrm{NN}}}=5.02$ TeV},''
  \href{http://dx.doi.org/10.1103/PhysRevLett.110.032301}{{\em Phys. Rev.
  Lett.} {\bfseries 110} no.~3, (2013) 032301},
\href{http://arxiv.org/abs/1210.3615}{{\ttfamily arXiv:1210.3615 [nucl-ex]}}.

\bibitem{Tanabashi:2018oca}
{\bfseries Particle Data Group} Collaboration, M.~Tanabashi {\em et~al.},
  ``{Review of Particle Physics},''
\href{http://dx.doi.org/10.1103/PhysRevD.98.030001}{{\em Phys. Rev.} {\bfseries
  D98} no.~3, (2018) 030001}.

\bibitem{Sjostrand:2006za}
T.~Sj{\"{o}}strand, S.~Mrenna, and P.~Z. Skands, ``{PYTHIA 6.4 Physics and
  Manual},'' \href{http://dx.doi.org/10.1088/1126-6708/2006/05/026}{{\em JHEP}
  {\bfseries 05} (2006) 026},
\href{http://arxiv.org/abs/hep-ph/0603175}{{\ttfamily arXiv:hep-ph/0603175
  [hep-ph]}}.

\bibitem{Skands:2010ak}
P.~Z. Skands, ``{Tuning Monte Carlo Generators: The Perugia Tunes},''
  \href{http://dx.doi.org/10.1103/PhysRevD.82.074018}{{\em Phys. Rev.}
  {\bfseries D82} (2010) 074018},
\href{http://arxiv.org/abs/1005.3457}{{\ttfamily arXiv:1005.3457 [hep-ph]}}.

\bibitem{Brun:1994aa}
R.~Brun, F.~Bruyant, F.~Carminati, S.~Giani, M.~Maire, A.~McPherson,
  G.~Patrick, and L.~Urban, ``{GEANT Detector Description and Simulation
  Tool},''
\href{http://arxiv.org/abs/CERN-W5013, CERN-W-5013, W5013, W-5013}{{\ttfamily
  CERN-W5013, CERN-W-5013, W5013, W-5013}}.

\bibitem{Ferrari:898301}
A.~Ferrari, P.~R. Sala, A.~Fass$\grave{\mathrm{o}}$, and J.~Ranft, {\em {FLUKA:
  A multi-particle transport code (program version 2005)}}.
\newblock CERN Yellow Reports: Monographs. CERN, Geneva, 2005.
\newblock \url{https://cds.cern.ch/record/898301}.

\bibitem{BOHLEN2014211}
T.~B$\ddot{\mathrm{o}}$hlen, F.~Cerutti, M.~Chin, A.~Fass$\grave{\mathrm{o}}$,
  A.~Ferrari, P.~Ortega, A.~Mairani, P.~Sala, G.~Smirnov, and V.~Vlachoudis,
  ``The fluka code: Developments and challenges for high energy and medical
  applications,''
  \href{http://dx.doi.org/https://doi.org/10.1016/j.nds.2014.07.049}{{\em
  Nuclear Data Sheets} {\bfseries 120} (2014) 211 -- 214}.
  \url{http://www.sciencedirect.com/science/article/pii/S0090375214005018}.

\bibitem{Agostinelli:2002hh}
{\bfseries GEANT4} Collaboration, S.~Agostinelli {\em et~al.}, ``{GEANT4: A
  Simulation toolkit},''
\href{http://dx.doi.org/10.1016/S0168-9002(03)01368-8}{{\em Nucl. Instrum.
  Meth.} {\bfseries A506} (2003) 250--303}.

\bibitem{Sjostrand:2014zea}
T.~Sj{\"{o}}strand {\em et~al.}, ``{An Introduction to PYTHIA 8.2},''
  \href{http://dx.doi.org/10.1016/j.cpc.2015.01.024}{{\em Comput. Phys.
  Commun.} {\bfseries 191} (2015) 159--177},
\href{http://arxiv.org/abs/1410.3012}{{\ttfamily arXiv:1410.3012 [hep-ph]}}.

\bibitem{Skands:2014pea}
P.~Skands, S.~Carrazza, and J.~Rojo, ``{Tuning PYTHIA 8.1: the Monash 2013
  Tune},'' \href{http://dx.doi.org/10.1140/epjc/s10052-014-3024-y}{{\em Eur.
  Phys. J.} {\bfseries C74} no.~8, (2014) 3024},
\href{http://arxiv.org/abs/1404.5630}{{\ttfamily arXiv:1404.5630 [hep-ph]}}.

\bibitem{Pierog:2013ria}
T.~Pierog, I.~Karpenko, J.~M. Katzy, E.~Yatsenko, and K.~Werner, ``{EPOS LHC:
  Test of collective hadronization with data measured at the CERN Large Hadron
  Collider},'' \href{http://dx.doi.org/10.1103/PhysRevC.92.034906}{{\em Phys.
  Rev.} {\bfseries C92} no.~3, (2015) 034906},
\href{http://arxiv.org/abs/1306.0121}{{\ttfamily arXiv:1306.0121 [hep-ph]}}.

\bibitem{Prato:1999jj}
D.~Prato and C.~Tsallis, ``{Nonextensive foundation of L{\'e}vy
  distributions},''
\href{http://dx.doi.org/10.1103/PhysRevE.60.2398}{{\em Phys. Rev.} {\bfseries
  E60} (1999) 2398}.

\bibitem{Barlow:2002yb}
R.~Barlow, ``{Systematic errors: Facts and fictions},'' in {\em {Advanced
  Statistical Techniques in Particle Physics. Proceedings, Conference, Durham,
  UK, March 18-22, 2002}}, pp.~134--144.
\newblock 2002.
\newblock \href{http://arxiv.org/abs/hep-ex/0207026}{{\ttfamily
  arXiv:hep-ex/0207026 [hep-ex]}}.
\newblock
\url{http://www.ippp.dur.ac.uk/Workshops/02/statistics/proceedings//barlow.pdf}.
\newblock

\bibitem{Chatrchyan:2012qb}
{\bfseries CMS} Collaboration, S.~Chatrchyan {\em et~al.}, ``{Study of the
  inclusive production of charged pions, kaons, and protons in $pp$ collisions
  at $\sqrt{\it{s}}=0.9$, 2.76, and 7 TeV},''
  \href{http://dx.doi.org/10.1140/epjc/s10052-012-2164-1}{{\em Eur. Phys. J.}
  {\bfseries C72} (2012) 2164},
\href{http://arxiv.org/abs/1207.4724}{{\ttfamily arXiv:1207.4724 [hep-ex]}}.

\bibitem{Werner:2016nnf}
K.~Werner, B.~Guiot, I.~Karpenko, T.~Pierog, and G.~Sophys, ``{Charm production
  in high multiplicity pp events},''
\href{http://dx.doi.org/10.1088/1742-6596/736/1/012009}{{\em J. Phys. Conf.
  Ser.} {\bfseries 736} no.~1, (2016) 012009}.

\bibitem{Cuautle:2016huw}
E.~Cuautle and G.~Pai\'c, ``{The energy density representation of the
  strangeness enhancement from pp to Pb+Pb},''
\href{http://arxiv.org/abs/1608.02101}{{\ttfamily arXiv:1608.02101 [hep-ph]}}.

\bibitem{Castorina:2016eyx}
P.~Castorina, S.~Plumari, and H.~Satz, ``{Universal Strangeness Production in
  Hadronic and Nuclear Collisions},''
  \href{http://dx.doi.org/10.1142/S0218301316500580}{{\em Int. J. Mod. Phys.}
  {\bfseries E25} no.~08, (2016) 1650058},
\href{http://arxiv.org/abs/1603.06529}{{\ttfamily arXiv:1603.06529 [hep-ph]}}.

\bibitem{Avsar:2006jy}
E.~Avsar, G.~Gustafson, and L.~Lonnblad, ``{Small-x dipole evolution beyond the
  large-N(c) imit},''
  \href{http://dx.doi.org/10.1088/1126-6708/2007/01/012}{{\em JHEP} {\bfseries
  01} (2007) 012},
\href{http://arxiv.org/abs/hep-ph/0610157}{{\ttfamily arXiv:hep-ph/0610157
  [hep-ph]}}.

\bibitem{Flensburg:2011kk}
C.~Flensburg, G.~Gustafson, and L.~Lonnblad, ``{Inclusive and Exclusive
  Observables from Dipoles in High Energy Collisions},''
  \href{http://dx.doi.org/10.1007/JHEP08(2011)103}{{\em JHEP} {\bfseries 08}
  (2011) 103},
\href{http://arxiv.org/abs/1103.4321}{{\ttfamily arXiv:1103.4321 [hep-ph]}}.

\bibitem{Mueller:1994gb}
A.~H. Mueller, ``{Unitarity and the BFKL pomeron},''
  \href{http://dx.doi.org/10.1016/0550-3213(94)00480-3}{{\em Nucl. Phys.}
  {\bfseries B437} (1995) 107--126},
\href{http://arxiv.org/abs/hep-ph/9408245}{{\ttfamily arXiv:hep-ph/9408245
  [hep-ph]}}.

\bibitem{Lonnblad:1992tz}
L.~Lonnblad, ``{ARIADNE version 4: A Program for simulation of QCD cascades
  implementing the color dipole model},''
\href{http://dx.doi.org/10.1016/0010-4655(92)90068-A}{{\em Comput. Phys.
  Commun.} {\bfseries 71} (1992) 15--31}.

\bibitem{Bierlich:2014xba}
C.~Bierlich, G.~Gustafson, L.~Lonnblad, and A.~Tarasov, ``{Effects of
  Overlapping Strings in pp Collisions},''
  \href{http://dx.doi.org/10.1007/JHEP03(2015)148}{{\em JHEP} {\bfseries 03}
  (2015) 148},
\href{http://arxiv.org/abs/1412.6259}{{\ttfamily arXiv:1412.6259 [hep-ph]}}.

\bibitem{Werner:2013yia}
K.~Werner, L.~Karpenko, M.~Bleicher, and T.~Pierog, ``{The Physics of EPOS},''
\href{http://dx.doi.org/10.1051/epjconf/20125205001}{{\em EPJ Web Conf.}
  {\bfseries 52} (2013) 05001}.

\bibitem{Drescher:2000ha}
H.~J. Drescher, M.~Hladik, S.~Ostapchenko, T.~Pierog, and K.~Werner, ``{Parton
  based Gribov-Regge theory},''
  \href{http://dx.doi.org/10.1016/S0370-1573(00)00122-8}{{\em Phys. Rept.}
  {\bfseries 350} (2001) 93--289},
\href{http://arxiv.org/abs/hep-ph/0007198}{{\ttfamily arXiv:hep-ph/0007198
  [hep-ph]}}.

\bibitem{Werner:2007bf}
K.~Werner, ``{Core-corona separation in ultra-relativistic heavy ion
  collisions},'' \href{http://dx.doi.org/10.1103/PhysRevLett.98.152301}{{\em
  Phys. Rev. Lett.} {\bfseries 98} (2007) 152301},
\href{http://arxiv.org/abs/0704.1270}{{\ttfamily arXiv:0704.1270 [nucl-th]}}.

\bibitem{Adam:2015ota}
{\bfseries ALICE} Collaboration, J.~Adam {\em et~al.}, ``{Measurement of charm
  and beauty production at central rapidity versus charged-particle
  multiplicity in proton-proton collisions at $ \sqrt{\it{s}}=7 $ TeV},''
  \href{http://dx.doi.org/10.1007/JHEP09(2015)148}{{\em JHEP} {\bfseries 09}
  (2015) 148},
\href{http://arxiv.org/abs/1505.00664}{{\ttfamily arXiv:1505.00664 [nucl-ex]}}.

\end{thebibliography}\endgroup

\newpage
\appendix
\section{The ALICE Collaboration}
\label{app:collab}

\begingroup
\small
\begin{flushleft}
S.~Acharya\Irefn{org141}\And 
D.~Adamov\'{a}\Irefn{org93}\And 
S.P.~Adhya\Irefn{org141}\And 
A.~Adler\Irefn{org73}\And 
J.~Adolfsson\Irefn{org79}\And 
M.M.~Aggarwal\Irefn{org98}\And 
G.~Aglieri Rinella\Irefn{org34}\And 
M.~Agnello\Irefn{org31}\And 
N.~Agrawal\Irefn{org10}\textsuperscript{,}\Irefn{org48}\textsuperscript{,}\Irefn{org53}\And 
Z.~Ahammed\Irefn{org141}
% ...
\And
\renewcommand\labelenumi{\textsuperscript{\theenumi}~}

\section*{Affiliation notes}
\renewcommand\theenumi{\roman{enumi}}
\begin{Authlist}
\item \Adef{org*}Deceased
\item \Adef{orgI}Dipartimento DET del Politecnico di Torino, Turin, Italy
\item \Adef{orgII}M.V. Lomonosov Moscow State University, D.V. Skobeltsyn Institute of Nuclear, Physics, Moscow, Russia
\item \Adef{orgIII}Department of Applied Physics, Aligarh Muslim University, Aligarh, India
\item \Adef{orgIV}Institute of Theoretical Physics, University of Wroclaw, Poland
\end{Authlist}

\section*{Collaboration Institutes}
\renewcommand\theenumi{\arabic{enumi}~}
\begin{Authlist}
\item \Idef{org1}A.I. Alikhanyan National Science Laboratory (Yerevan Physics Institute) Foundation, Yerevan, Armenia
\item \Idef{org2}Bogolyubov Institute for Theoretical Physics, National Academy of Sciences of Ukraine, Kiev, Ukraine
\item \Idef{org3}Bose Institute, Department of Physics  and Centre for Astroparticle Physics and Space Science (CAPSS), Kolkata, India
\item \Idef{org4}Budker Institute for Nuclear Physics, Novosibirsk, Russia
\item \Idef{org5}California Polytechnic State University, San Luis Obispo, California, United States
\item \Idef{org6}Central China Normal University, Wuhan, China
\item \Idef{org7}Centre de Calcul de l'IN2P3, Villeurbanne, Lyon, France
\item \Idef{org8}Centro de Aplicaciones Tecnol\'{o}gicas y Desarrollo Nuclear (CEADEN), Havana, Cuba
\item \Idef{org9}Centro de Investigaci\'{o}n y de Estudios Avanzados (CINVESTAV), Mexico City and M\'{e}rida, Mexico
\item \Idef{org10}Centro Fermi - Museo Storico della Fisica e Centro Studi e Ricerche ``Enrico Fermi', Rome, Italy
\item \Idef{org11}Chicago State University, Chicago, Illinois, United States
\item \Idef{org12}China Institute of Atomic Energy, Beijing, China
\item \Idef{org13}Chonbuk National University, Jeonju, Republic of Korea
\item \Idef{org14}Comenius University Bratislava, Faculty of Mathematics, Physics and Informatics, Bratislava, Slovakia
\item \Idef{org15}COMSATS University Islamabad, Islamabad, Pakistan
\item \Idef{org16}Creighton University, Omaha, Nebraska, United States
\item \Idef{org17}Department of Physics, Aligarh Muslim University, Aligarh, India
\item \Idef{org18}Department of Physics, Pusan National University, Pusan, Republic of Korea
\item \Idef{org19}Department of Physics, Sejong University, Seoul, Republic of Korea
\item \Idef{org20}Department of Physics, University of California, Berkeley, California, United States
\item \Idef{org21}Department of Physics, University of Oslo, Oslo, Norway
\item \Idef{org22}Department of Physics and Technology, University of Bergen, Bergen, Norway
\item \Idef{org23}Dipartimento di Fisica dell'Universit\`{a} 'La Sapienza' and Sezione INFN, Rome, Italy
\item \Idef{org24}Dipartimento di Fisica dell'Universit\`{a} and Sezione INFN, Cagliari, Italy
\item \Idef{org25}Dipartimento di Fisica dell'Universit\`{a} and Sezione INFN, Trieste, Italy
\item \Idef{org26}Dipartimento di Fisica dell'Universit\`{a} and Sezione INFN, Turin, Italy
\item \Idef{org27}Dipartimento di Fisica e Astronomia dell'Universit\`{a} and Sezione INFN, Bologna, Italy
\item \Idef{org28}Dipartimento di Fisica e Astronomia dell'Universit\`{a} and Sezione INFN, Catania, Italy
\item \Idef{org29}Dipartimento di Fisica e Astronomia dell'Universit\`{a} and Sezione INFN, Padova, Italy
\item \Idef{org30}Dipartimento di Fisica `E.R.~Caianiello' dell'Universit\`{a} and Gruppo Collegato INFN, Salerno, Italy
\item \Idef{org31}Dipartimento DISAT del Politecnico and Sezione INFN, Turin, Italy
\item \Idef{org32}Dipartimento di Scienze e Innovazione Tecnologica dell'Universit\`{a} del Piemonte Orientale and INFN Sezione di Torino, Alessandria, Italy
\item \Idef{org33}Dipartimento Interateneo di Fisica `M.~Merlin' and Sezione INFN, Bari, Italy
\item \Idef{org34}European Organization for Nuclear Research (CERN), Geneva, Switzerland
\item \Idef{org35}Faculty of Electrical Engineering, Mechanical Engineering and Naval Architecture, University of Split, Split, Croatia
\item \Idef{org36}Faculty of Engineering and Science, Western Norway University of Applied Sciences, Bergen, Norway
\item \Idef{org37}Faculty of Nuclear Sciences and Physical Engineering, Czech Technical University in Prague, Prague, Czech Republic
\item \Idef{org38}Faculty of Science, P.J.~\v{S}af\'{a}rik University, Ko\v{s}ice, Slovakia
\item \Idef{org39}Frankfurt Institute for Advanced Studies, Johann Wolfgang Goethe-Universit\"{a}t Frankfurt, Frankfurt, Germany
\item \Idef{org40}Gangneung-Wonju National University, Gangneung, Republic of Korea
\item \Idef{org41}Gauhati University, Department of Physics, Guwahati, India
\item \Idef{org42}Helmholtz-Institut f\"{u}r Strahlen- und Kernphysik, Rheinische Friedrich-Wilhelms-Universit\"{a}t Bonn, Bonn, Germany
\item \Idef{org43}Helsinki Institute of Physics (HIP), Helsinki, Finland
\item \Idef{org44}High Energy Physics Group,  Universidad Aut\'{o}noma de Puebla, Puebla, Mexico
\item \Idef{org45}Hiroshima University, Hiroshima, Japan
\item \Idef{org46}Hochschule Worms, Zentrum  f\"{u}r Technologietransfer und Telekommunikation (ZTT), Worms, Germany
\item \Idef{org47}Horia Hulubei National Institute of Physics and Nuclear Engineering, Bucharest, Romania
\item \Idef{org48}Indian Institute of Technology Bombay (IIT), Mumbai, India
\item \Idef{org49}Indian Institute of Technology Indore, Indore, India
\item \Idef{org50}Indonesian Institute of Sciences, Jakarta, Indonesia
\item \Idef{org51}INFN, Laboratori Nazionali di Frascati, Frascati, Italy
\item \Idef{org52}INFN, Sezione di Bari, Bari, Italy
\item \Idef{org53}INFN, Sezione di Bologna, Bologna, Italy
\item \Idef{org54}INFN, Sezione di Cagliari, Cagliari, Italy
\item \Idef{org55}INFN, Sezione di Catania, Catania, Italy
\item \Idef{org56}INFN, Sezione di Padova, Padova, Italy
\item \Idef{org57}INFN, Sezione di Roma, Rome, Italy
\item \Idef{org58}INFN, Sezione di Torino, Turin, Italy
\item \Idef{org59}INFN, Sezione di Trieste, Trieste, Italy
\item \Idef{org60}Inha University, Incheon, Republic of Korea
\item \Idef{org61}Institut de Physique Nucl\'{e}aire d'Orsay (IPNO), Institut National de Physique Nucl\'{e}aire et de Physique des Particules (IN2P3/CNRS), Universit\'{e} de Paris-Sud, Universit\'{e} Paris-Saclay, Orsay, France
\item \Idef{org62}Institute for Nuclear Research, Academy of Sciences, Moscow, Russia
\item \Idef{org63}Institute for Subatomic Physics, Utrecht University/Nikhef, Utrecht, Netherlands
\item \Idef{org64}Institute of Experimental Physics, Slovak Academy of Sciences, Ko\v{s}ice, Slovakia
\item \Idef{org65}Institute of Physics, Homi Bhabha National Institute, Bhubaneswar, India
\item \Idef{org66}Institute of Physics of the Czech Academy of Sciences, Prague, Czech Republic
\item \Idef{org67}Institute of Space Science (ISS), Bucharest, Romania
\item \Idef{org68}Institut f\"{u}r Kernphysik, Johann Wolfgang Goethe-Universit\"{a}t Frankfurt, Frankfurt, Germany
\item \Idef{org69}Instituto de Ciencias Nucleares, Universidad Nacional Aut\'{o}noma de M\'{e}xico, Mexico City, Mexico
\item \Idef{org70}Instituto de F\'{i}sica, Universidade Federal do Rio Grande do Sul (UFRGS), Porto Alegre, Brazil
\item \Idef{org71}Instituto de F\'{\i}sica, Universidad Nacional Aut\'{o}noma de M\'{e}xico, Mexico City, Mexico
\item \Idef{org72}iThemba LABS, National Research Foundation, Somerset West, South Africa
\item \Idef{org73}Johann-Wolfgang-Goethe Universit\"{a}t Frankfurt Institut f\"{u}r Informatik, Fachbereich Informatik und Mathematik, Frankfurt, Germany
\item \Idef{org74}Joint Institute for Nuclear Research (JINR), Dubna, Russia
\item \Idef{org75}Korea Institute of Science and Technology Information, Daejeon, Republic of Korea
\item \Idef{org76}KTO Karatay University, Konya, Turkey
\item \Idef{org77}Laboratoire de Physique Subatomique et de Cosmologie, Universit\'{e} Grenoble-Alpes, CNRS-IN2P3, Grenoble, France
\item \Idef{org78}Lawrence Berkeley National Laboratory, Berkeley, California, United States
\item \Idef{org79}Lund University Department of Physics, Division of Particle Physics, Lund, Sweden
\item \Idef{org80}Nagasaki Institute of Applied Science, Nagasaki, Japan
\item \Idef{org81}Nara Women{'}s University (NWU), Nara, Japan
\item \Idef{org82}National and Kapodistrian University of Athens, School of Science, Department of Physics , Athens, Greece
\item \Idef{org83}National Centre for Nuclear Research, Warsaw, Poland
\item \Idef{org84}National Institute of Science Education and Research, Homi Bhabha National Institute, Jatni, India
\item \Idef{org85}National Nuclear Research Center, Baku, Azerbaijan
\item \Idef{org86}National Research Centre Kurchatov Institute, Moscow, Russia
\item \Idef{org87}Niels Bohr Institute, University of Copenhagen, Copenhagen, Denmark
\item \Idef{org88}Nikhef, National institute for subatomic physics, Amsterdam, Netherlands
\item \Idef{org89}NRC Kurchatov Institute IHEP, Protvino, Russia
\item \Idef{org90}NRC «Kurchatov Institute»  - ITEP, Moscow, Russia
\item \Idef{org91}NRNU Moscow Engineering Physics Institute, Moscow, Russia
\item \Idef{org92}Nuclear Physics Group, STFC Daresbury Laboratory, Daresbury, United Kingdom
\item \Idef{org93}Nuclear Physics Institute of the Czech Academy of Sciences, \v{R}e\v{z} u Prahy, Czech Republic
\item \Idef{org94}Oak Ridge National Laboratory, Oak Ridge, Tennessee, United States
\item \Idef{org95}Ohio State University, Columbus, Ohio, United States
\item \Idef{org96}Petersburg Nuclear Physics Institute, Gatchina, Russia
\item \Idef{org97}Physics department, Faculty of science, University of Zagreb, Zagreb, Croatia
\item \Idef{org98}Physics Department, Panjab University, Chandigarh, India
\item \Idef{org99}Physics Department, University of Jammu, Jammu, India
\item \Idef{org100}Physics Department, University of Rajasthan, Jaipur, India
\item \Idef{org101}Physikalisches Institut, Eberhard-Karls-Universit\"{a}t T\"{u}bingen, T\"{u}bingen, Germany
\item \Idef{org102}Physikalisches Institut, Ruprecht-Karls-Universit\"{a}t Heidelberg, Heidelberg, Germany
\item \Idef{org103}Physik Department, Technische Universit\"{a}t M\"{u}nchen, Munich, Germany
\item \Idef{org104}Politecnico di Bari, Bari, Italy
\item \Idef{org105}Research Division and ExtreMe Matter Institute EMMI, GSI Helmholtzzentrum f\"ur Schwerionenforschung GmbH, Darmstadt, Germany
\item \Idef{org106}Rudjer Bo\v{s}kovi\'{c} Institute, Zagreb, Croatia
\item \Idef{org107}Russian Federal Nuclear Center (VNIIEF), Sarov, Russia
\item \Idef{org108}Saha Institute of Nuclear Physics, Homi Bhabha National Institute, Kolkata, India
\item \Idef{org109}School of Physics and Astronomy, University of Birmingham, Birmingham, United Kingdom
\item \Idef{org110}Secci\'{o}n F\'{\i}sica, Departamento de Ciencias, Pontificia Universidad Cat\'{o}lica del Per\'{u}, Lima, Peru
\item \Idef{org111}Shanghai Institute of Applied Physics, Shanghai, China
\item \Idef{org112}St. Petersburg State University, St. Petersburg, Russia
\item \Idef{org113}Stefan Meyer Institut f\"{u}r Subatomare Physik (SMI), Vienna, Austria
\item \Idef{org114}SUBATECH, IMT Atlantique, Universit\'{e} de Nantes, CNRS-IN2P3, Nantes, France
\item \Idef{org115}Suranaree University of Technology, Nakhon Ratchasima, Thailand
\item \Idef{org116}Technical University of Ko\v{s}ice, Ko\v{s}ice, Slovakia
\item \Idef{org117}Technische Universit\"{a}t M\"{u}nchen, Excellence Cluster 'Universe', Munich, Germany
\item \Idef{org118}The Henryk Niewodniczanski Institute of Nuclear Physics, Polish Academy of Sciences, Cracow, Poland
\item \Idef{org119}The University of Texas at Austin, Austin, Texas, United States
\item \Idef{org120}Universidad Aut\'{o}noma de Sinaloa, Culiac\'{a}n, Mexico
\item \Idef{org121}Universidade de S\~{a}o Paulo (USP), S\~{a}o Paulo, Brazil
\item \Idef{org122}Universidade Estadual de Campinas (UNICAMP), Campinas, Brazil
\item \Idef{org123}Universidade Federal do ABC, Santo Andre, Brazil
\item \Idef{org124}University of Cape Town, Cape Town, South Africa
\item \Idef{org125}University of Houston, Houston, Texas, United States
\item \Idef{org126}University of Jyv\"{a}skyl\"{a}, Jyv\"{a}skyl\"{a}, Finland
\item \Idef{org127}University of Liverpool, Liverpool, United Kingdom
\item \Idef{org128}University of Science and Techonology of China, Hefei, China
\item \Idef{org129}University of South-Eastern Norway, Tonsberg, Norway
\item \Idef{org130}University of Tennessee, Knoxville, Tennessee, United States
\item \Idef{org131}University of the Witwatersrand, Johannesburg, South Africa
\item \Idef{org132}University of Tokyo, Tokyo, Japan
\item \Idef{org133}University of Tsukuba, Tsukuba, Japan
\item \Idef{org134}Universit\'{e} Clermont Auvergne, CNRS/IN2P3, LPC, Clermont-Ferrand, France
\item \Idef{org135}Universit\'{e} de Lyon, Universit\'{e} Lyon 1, CNRS/IN2P3, IPN-Lyon, Villeurbanne, Lyon, France
\item \Idef{org136}Universit\'{e} de Strasbourg, CNRS, IPHC UMR 7178, F-67000 Strasbourg, France, Strasbourg, France
\item \Idef{org137}Universit\'{e} Paris-Saclay Centre d'Etudes de Saclay (CEA), IRFU, D\'{e}partment de Physique Nucl\'{e}aire (DPhN), Saclay, France
\item \Idef{org138}Universit\`{a} degli Studi di Foggia, Foggia, Italy
\item \Idef{org139}Universit\`{a} degli Studi di Pavia, Pavia, Italy
\item \Idef{org140}Universit\`{a} di Brescia, Brescia, Italy
\item \Idef{org141}Variable Energy Cyclotron Centre, Homi Bhabha National Institute, Kolkata, India
\item \Idef{org142}Warsaw University of Technology, Warsaw, Poland
\item \Idef{org143}Wayne State University, Detroit, Michigan, United States
\item \Idef{org144}Westf\"{a}lische Wilhelms-Universit\"{a}t M\"{u}nster, Institut f\"{u}r Kernphysik, M\"{u}nster, Germany
\item \Idef{org145}Wigner Research Centre for Physics, Hungarian Academy of Sciences, Budapest, Hungary
\item \Idef{org146}Yale University, New Haven, Connecticut, United States
\item \Idef{org147}Yonsei University, Seoul, Republic of Korea
\end{Authlist}
\endgroup
\end{document}